\documentclass[12pt, preprint]{aastex}

\shorttitle{Emission Line Properties of Seyfert Galaxies in the 12 Micron Sample}
\shortauthors{Malkan et al.}

\usepackage{amsmath}

\usepackage{epstopdf}

\begin{document}


\title{Emission Line Properties of Seyfert Galaxies \\
    In the 12 Micron Sample}


\author{Matthew A. Malkan\altaffilmark{1}, Lisbeth D. Jensen\altaffilmark{1}, David R. Rodriguez\altaffilmark{1}, Luigi Spinoglio\altaffilmark{2}, and Brian Rush\altaffilmark{1,3}}
\email{malkan@astro.ucla.edu}


\altaffiltext{1}{Physics and Astronomy Department, University of California, Los Angeles, CA 90095}
\altaffiltext{2}{Istituto di Fisica dello Spazio Interplanetario, INAF, Via Fosso del Cavaliere 100, I-00133 Roma, Italy}
\altaffiltext{3}{Jet Propulsion Laboratory, California Institute of Technology, Pasadena, CA 91109}


\begin{abstract}
We present optical and ultraviolet spectroscopic measurements of the emission lines of 81 Seyfert 1 and 104 Seyfert 2 galaxies which comprise nearly all of the IRAS 12$\micron$ AGN sample. We have analyzed the emission-line luminosity functions, reddening, and other diagnostics. For example, the narrow-line regions (NLR) of Seyfert 1 and 2 galaxies do not significantly differ from each other in most of these diagnostics. Combining the H$\alpha$/H$\beta$ ratio with a new reddening indicator--the [SII]6720/[OII]3727 ratio, 
we find the average $E(B-V)$ is $0.49\pm0.35$ for Seyfert 1's and $0.52\pm0.26$ for Seyfert 2's. 
The NLR of Sy 1 galaxies has only insignificantly higher ionization level than in the Sy 2's. For the broad-line region (BLR), we find that the \ion{C}{4} equivalent width correlates more strongly with [\ion{O}{3}]/H$\beta$ than with UV luminosity. Our bright sample of local active galaxies includes 22 Seyfert nuclei with extremely weak broad wings in H$\alpha$, known as Seyfert 1.9's and 1.8's, depending on whether or not broad H$\beta$ wings are detected.  Aside from these weak broad lines, our low-luminosity Seyferts are more similar to the Sy2's than to the Sy 1's. In a a BPT diagram we find that Sy 1.8's and Sy 1.9's overlap the region occupied by the Sy 2 galaxies. We compare our results on optical emission lines with those obtained by previous investigators using AGN subsamples from the Sloan Digital Sky Survey.  The luminosity functions of forbidden emission lines [OII]$\lambda$3727\AA, [OIII]$\lambda$5007\AA, and [SII]$\lambda$6720\AA\;in Seyfert 1's and 2's are indistinguishable.  They all show strong downward curvature.  Unlike the LF's of Seyfert galaxies measured by the Sloan Digital Sky Survey,
ours are nearly flat at low luminosities.  The larger number of faint Sloan ``AGN" is attributable to their inclusion of weakly emitting 
LINERs and H II+AGN ``composite" nuclei, which do not meet our spectral classification criteria for Seyferts.

In an Appendix,  we have investigated which emission line luminosities can provide 
the most reliable measures of the total non-stellar luminosity, estimated from our extensive multi-wavelength database. The hard X-ray or near-ultraviolet continuum luminosity can be crudely predicted from either the [\ion{O}{3}]$\lambda$5007\AA\; luminosity, or the combination of [\ion{O}{3}]+H$\beta$, or [\ion{N}{2}]+H$\alpha$ lines,  with a scatter of $\pm\,4$ times for the Sy 1's and $\pm\,10$ times for the Sy 2's.
Although these uncertainties are large,
the latter two hybrid (NLR+BLR) indicators have the advantage of predicting the same HX luminosity independent of Seyfert type.
\end{abstract}


\section{Introduction}

Many previous studies have measured emission line ratios in samples of Seyfert nuclei, but the existence of strong selection effects (such as searches based on host galaxy properties, UV-excess or X-ray flux) raise questions about whether these results would hold for a representative sample of the \textit{full} unbiased Seyfert population.  A common limitation of most surveys is lack of complete data for the less luminous Seyfert nuclei. Another limitation of surveys at optical, ultra-violet, or soft X-ray wavelengths is their inability to find more reddened AGN. These limitations can be overcome with a nonstellar-flux-limited all-Sky survey at long wavelengths, which includes most of the bright Seyfert galaxies in the local universe. 
We pursue this approach in this paper.  One of our motivations is that a better observational understanding of the nearest and best observed Seyfert galaxies will help us to understand the population of AGN at high redshifts.  High-redshift studies
are now seeking to measure the cosmic evolution of AGN, but with much less complete data than we have locally. 
By emphasizing quality of the data over raw quantity, we hope to use extensive observations of local AGN to help calibrate the emission-line diagnostics in high-redshift samples.

\subsection{The Extended 12$\micron$ and CfA AGN Samples}

The extended 12$\micron$ galaxy catalogue of \citet{sm89} and \citet{rms93}, is a 12$\micron$ flux-limited sample containing 893 galaxies selected from the IRAS Faint Source Catalog, Version 2. 
The galaxies in this catalogue have galactic latitude of  $\left|b\right| \geq 25^{o}$ to decrease the extinction and avoid stellar contamination from the plane of our Galaxy. 
The 12$\micron$ galaxy sample contains 9 percent Sy 1's and quasars, and 11 percent  Sy 2's. These percentages are of course far higher than the percentages of Seyferts among ordinary optically-selected galaxies such as in the Sloan Digital Sky Survey (SDSS). This is because, by design, the 
12$\micron$ galaxies were selected at a wavelength where the continuum emission from warm dust in the Seyfert nucleus is especially bright relative 
to the normal emission from the underlying host galaxy.
The 12$\micron$ flux is a constant fraction of $\sim1/5$ of the bolometric flux in Seyfert 1 and 2 galaxies, three times more than in normal spiral galaxies 
\citep{sm89}.
The 12$\micron$ Seyferts are excellent representatives of the entire class, since they span nearly 6 orders of magnitude in luminosity. 
With a $\log(N) - \log(S)$ test \citet{sm89} showed that the sample is complete down to 0.30 Jy, and has a level of incompleteness of $\sim40$\% at 0.22 Jy, the chosen flux limit.  

The extended 12$\micron$ sample includes the brightest nearby Seyferts in the local universe.  It has been subjected to extensive observational follow-up across the entire electromagnetic spectrum, and thus has the most complete multi-wavelength dataset available for any AGN sample. The redshifts range from $z= -0.0001$ to $z= +0.1884$, with the majority at $z \leq 0.05$. The redshifts are obtained from the the NASA/IPAC Extragalactic Database (NED)\footnote{The NASA/IPAC Extragalactic Database (NED) is operated by the Jet Propulsion Laboratory, California Institute of Technology, under contract with the National Aeronautics and Space Administration.}. Throughout we adopt $H_{0} = 72$ km/s/Mpc when computing distances to the Seyferts. \footnote{The one exception is NGC 3031/M81, for which we use the measured Cepheid distance of 3.63 Mpc  \citep{f94}.}.

To further control possible selection bias, we supplement the data with Seyfert galaxies from the Center for Astrophysics (\textit{CfA}) galaxy sample. The \textit{CfA} galaxies are a host-galaxy flux-limited, spectroscopically selected sample defined by \citet{huchra83}. They come from 2399 galaxies with $m \leq 14.5$, with cuts in $b$ and $\delta$ to avoid contaminations from the galactic plane \citep{hb92}. \citet{ts92} provide IRAS fluxes for 1544 galaxies in the \textit{CfA} sample that are detected in the IRAS Faint Source Catalog. The overlap between  the \textit{CfA} and the 12$\micron$ sample is 47  Seyfert galaxies ($\sim 25\%$), as described in \citet{rms93}. 

Our data are compiled from multiple literature sources, the \textit{CfA} sample, and our own  previously unpublished data. The full sample, containing 185 Seyfert galaxies, 81 Seyfert 1's and 104 Seyfert 2 galaxies, is listed in Tables ~\ref{tbl1} and ~\ref{tbl2}. It lists spectral classifications, redshift, and  which galaxies are common to both the 12$\micron$ and \textit{CfA} samples. 

For simplicity we have adopted the spectral type classification given in NED's ``Basic Data". These are based on spectra obtained from a wide variety of publications. The ability to discern faint Balmer-line wings--required for an Sy 1 classification--depends on the quality the spectrum. The classification for weak Sy 1's and also composite Sy 1+HII galaxies is therefore sometimes ambiguous. But we only have one or two active galaxies for which our data indicate a different classification from NED. These special cases are mentioned below.

\subsection{Spectroscopic Measurements} \label{Data} 
We have averaged over multiple measurements and conservatively assign an uncertainty of 30\% in the line fluxes, 
although ratios of nearby lines are usually more accurate. 
Our combined optical and ultraviolet emission lines and their corresponding rest wavelengths can be found in Table~\ref{tbl3}. 
Optical line fluxes along with literature references are in Table~\ref{tbl4}, and the UV data are summarized in Table~\ref{tbl5}, 
also with literature references. 

We have supplemented our observational results with data from Sloan Digital Sky Survey Data Release 7 from ``The MPA-JHU DR7 release of spectrum measurements"\footnote{Obtained from  http://wwwmpa.mpa-garching.mpg.de/SDSS/DR7/. Raw data from http://wwwmpa.mpa-garching.mpg.de/SDSS/DR7/raw\_data.html.}. This data base contain 927,552 AGN galaxies. We excludeed two-thirds of those which lacked S/N emission-line ratios of at least $ > 10$. We have also used the SDSS DR6 SkyServer Explore Tool\footnote{The SDSS DR6 SkyServer Explore Tool can be found at \url{http://cas.sdss.org/astro/en/tools/explore/}.} to provide line fluxes for 5 Sy1's and 11 Sy 2's, including four objects (IRAS 13354+3924, IRAS 16146+3549, NGC 833, and UGC 6100) for which no prior data existed. We calculate the line-flux ratios of [\ion{O}{3}]/H$\beta$, [\ion{O}{3}]/[\ion{O}{2}], [\ion{O}{2}]/[\ion{O}{1}], [\ion{N}{2}]/H$\alpha$, [\ion{O}{2}]/[\ion{S}{2}], [\ion{O}{2}]/[\ion{N}{2}], [\ion{O}{3}]/[\ion{S}{2}], and [\ion{N}{2}]/[\ion{S}{2}], and find that the average difference between our sample and that of SDSS for the ratios is $-0.03 \pm 0.25$, consistent with no systematic difference.

This paper is organized as follows. In \S~\ref{LumFun} we describe our emission line luminosity functions. In \S~\ref{NLR} we use emission lines to diagnose  properties in the narrow line region, while in \S~\ref{BLR} we consider the broad line region. We summarize our results in \S~\ref{Conclude}. In Appendix A we investigate, which, if any, emission line luminosities can provide the most reliable estimates of the total non-stellar luminosity,

\section{Line Luminosity Functions} \label{LumFun}
We construct emission-line luminosity functions (LFs) for each Sy 1 and Sy 2 galaxy by taking $V_{max}$ values from the 12$\micron$ flux sample and binning them according to their individual emission-line luminosities. Since our sample is defined and limited by the 12$\micron$ flux, we apply correction factors where necessary to account for incompleteness previously determined by \citet{rms93}. 

Emission-line LFs are derived for [\ion{O}{1}], [\ion{O}{2}], [\ion{O}{3}], [\ion{N}{2}], [\ion{S}{2}], H$\alpha$, and H$\beta$ lines. A double power-law in Logarithmic space is fitted to the luminosity functions:
\begin{equation}\label{eqn1}
\Phi (L) = \frac{ \Phi_{\star} }{  \left( \frac{L}{L_{\star}} \right)^{\alpha} + \left( \frac{L}{L_{\star}} \right)^{\alpha + \beta} }
\end{equation}
\noindent where $L_{\star}$ is the emission-line luminosities of the characteristic break in the LF (in erg/s),  and $\Phi_{\star}$ is twice the number density at that break (in Mpc$^{-3}$). 
The points always require a bend in the LF, i.e. a steeper slope at high luminosities. However, the strength of this bend is not well determined due to the small of high- and low-luminosity galaxies in our sample. We therefore assumed a low-luminosity slope of $\alpha= -0.1$ and a slope steepening break of $\beta=+1.5$. These broken power laws, based on the shape of the 12$\micron$ LF, match all the emission line  LFs adequately.

\begin{figure*}
\centering
\includegraphics[angle=0,scale=0.46]{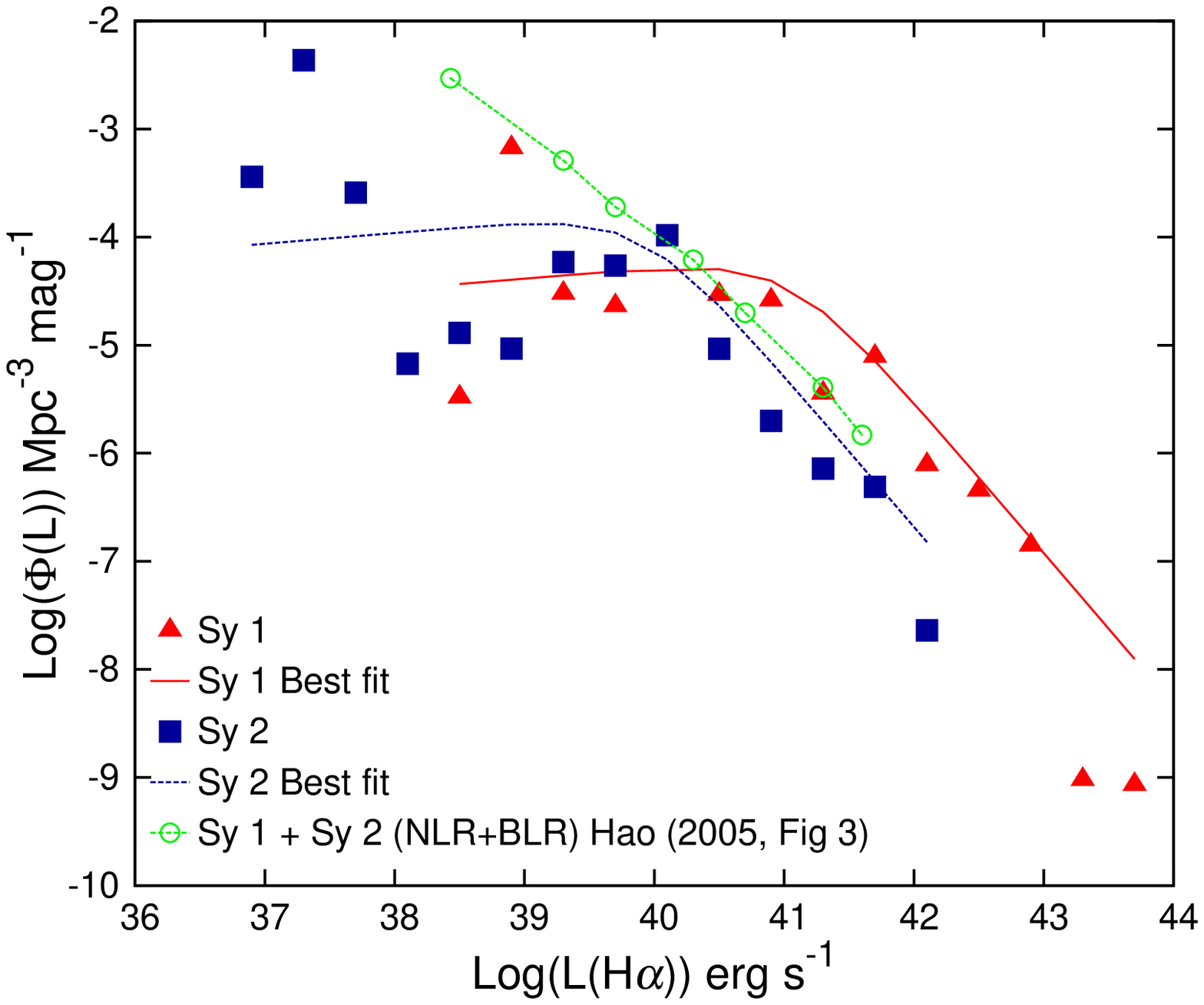}
\includegraphics[angle=0,scale=0.46]{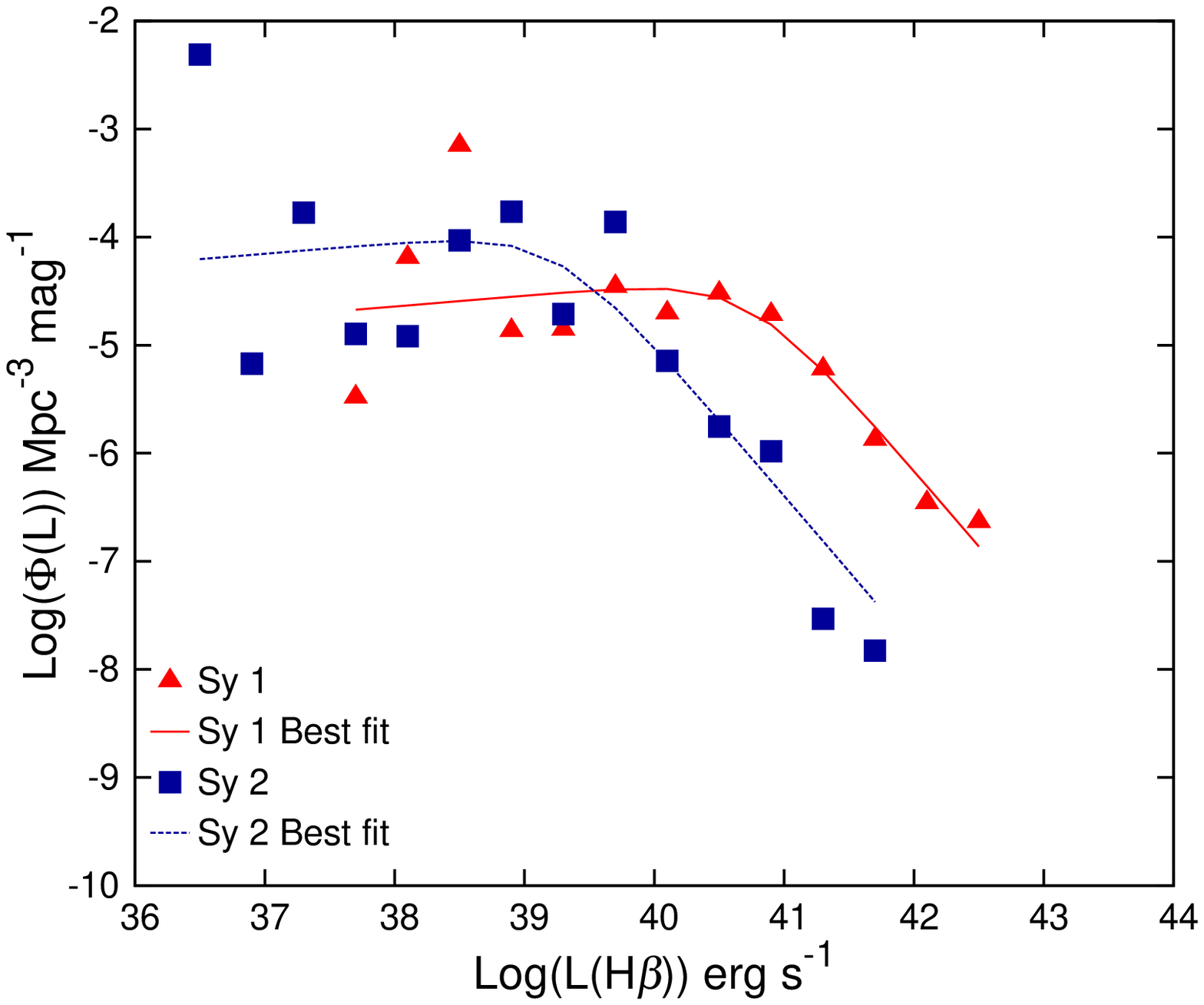}\\
\includegraphics[angle=0,scale=0.46]{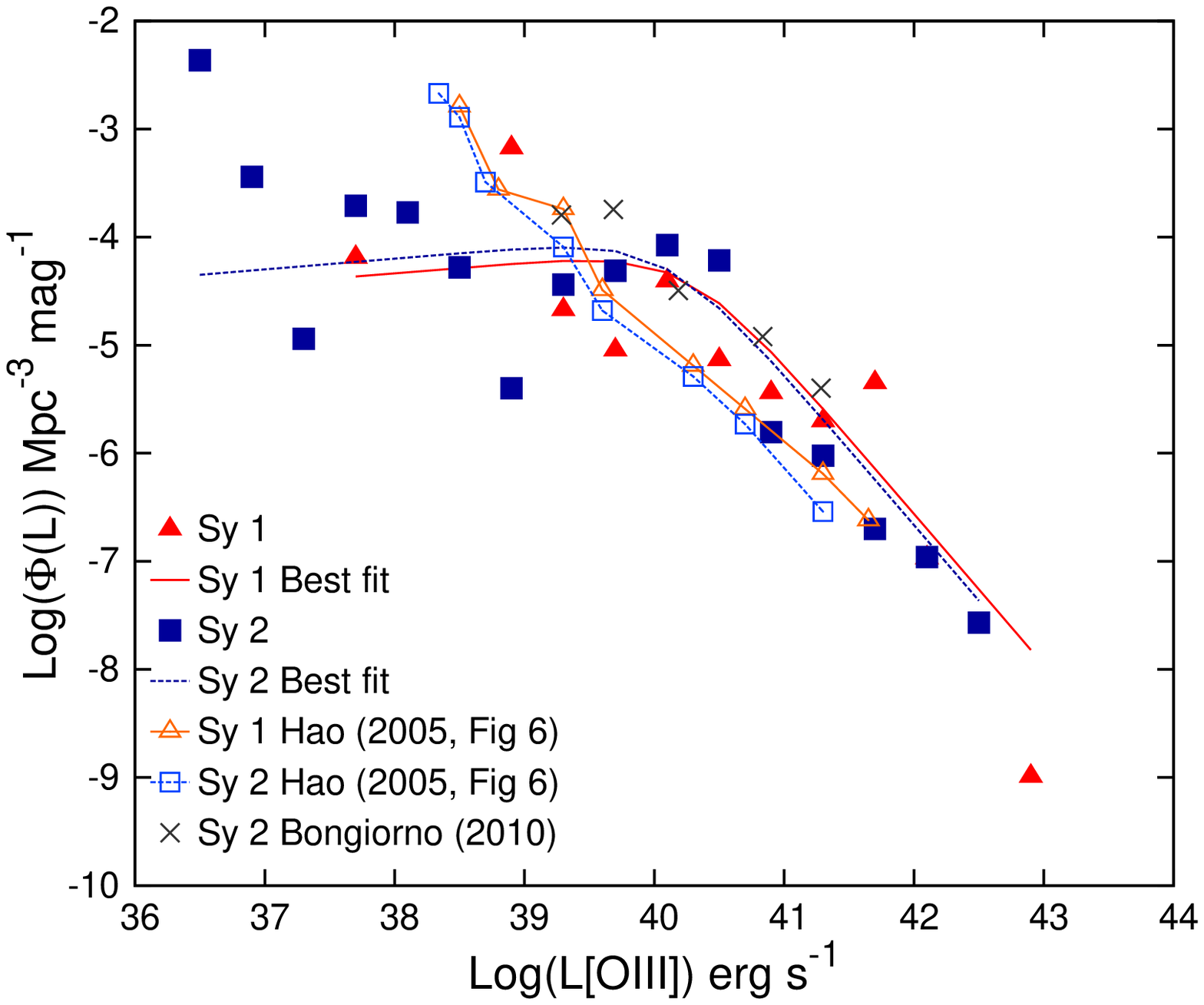}
\includegraphics[angle=0,scale=0.46]{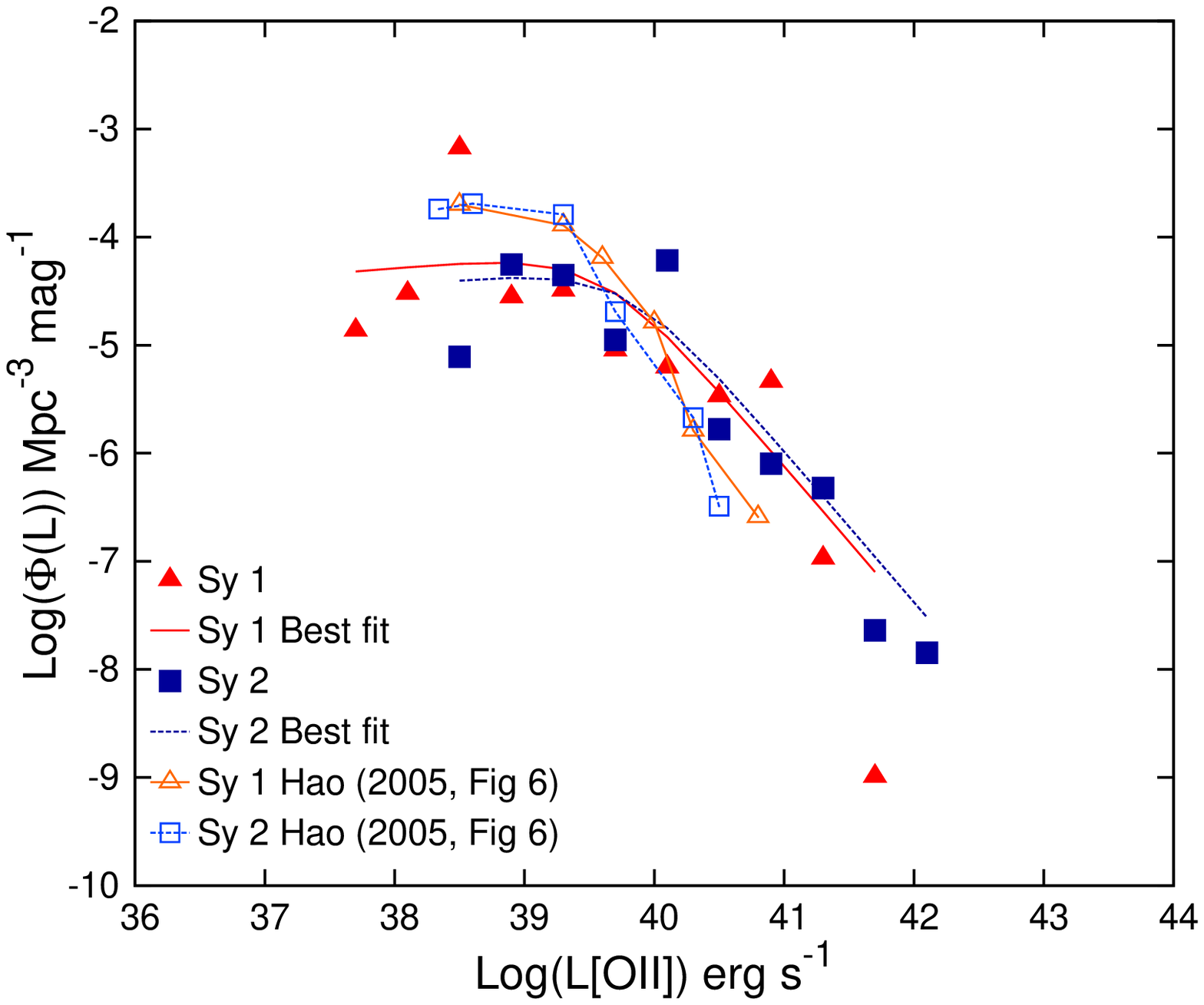}
\caption{Emission Line Luminosity Functions for H$\alpha$, H$\beta$, [\ion{O}{3}] $\lambda5007$, and [\ion{O}{2}], where $\Phi$ has units of Mpc$^{-3}$ mag$^{-1}$ and L has units of ergs s$^{-1}$. The solid line is the fit to the Seyfert 1 LF's, while the blue line is the fit to the Seyfert 2's. In the graphs of LF for H$\alpha$, [\ion{O}{3}], and [\ion{O}{2}] we have overplotted the Luminosity Functions of SDSS Seyfert galaxies from \cite{H05}. In the panel for [\ion{O}{3}] we have also added LF data from \cite{B10} (zCOSMOS). Although data from \cite{H05} covers our $z$ range ($0 <  z < 0.15$), the zCOSMOS data ($0.15 < z < 0.35$) agrees better with our LF's. \label{Fig1}}
\end{figure*}

The best-fit parameters of the LFs are given in Table \ref{tbl6} and displayed in Figure \ref{Fig1}. We plot the LF for the narrow lines [\ion{O}{2}] and [\ion{O}{3}], and the H$\alpha$ and H$\beta$ lines (with the broad and narrow components combined) in Figure \ref{Fig1}.

The Balmer-line luminosity function of the Sy 1s extends to higher luminosities than that of the Sy 2s. Because the Sy 1 permitted lines are brightened by the contribution 
from their broad line region ($BLR$), they are more numerous at the high line luminosities. However, for the [\ion{O}{3}] and [\ion{O}{2}] LFs, there is only a small difference between the Sy 1 and Sy 2 galaxies. Thus at a given narrow line luminosity, the two types of Seyferts have similar space densities. There is also little difference in the 12$\micron$ Sy 1 and Sy 2 continuum LFs \citep{rms93,toba14}, at luminosities below that of quasars. So if the narrow line luminosity is emitted approximately isotropically, the similarity of these Sy1 and Sy2 space densities suggests that the 12$\micron$ continuum emission is also relatively isotropic. 

Figure \ref{Fig1}a, \ref{Fig1}c, and \ref{Fig1}d compares the Sy 1 and Sy 2 emission-line Luminosity Functions for our 12$\micron$ selected sample with the optically selected AGN sample from the SDSS \citep{H05, S05}. The optical and IR selection methods show good agreement in the derived LFs around the `knee', i.e. the line-luminosities around $10^{39-41}\;\mathrm{erg/s}$, for both Seyfert types. However, compared with the 12$\micron$ selection, the SDSS finds relatively \textit{more} low-luminosity AGN and relatively \textit{less} high-luminosity AGN. Thus the SDSS LFs are steeper at low luminosities than our 12$\micron$ sample. The SDSS LFs thus show much weaker breaks to high luminosities. We attribute this difference to the inclusion in SDSS of more ``composite" Sy 2's. These low-luminosity AGN have substantial line emission contributed by star formation, which tends to prevent them for being classified in NED as Seyfert Galaxies.  On the other hand, the 12$\micron$ selection is particularly efficient for finding luminous AGN, which tend to be at high redshifts.
Figure~\ref{Fig1} indicates that it is the optical selection of SDSS that becomes significantly incomplete at high luminosities (emission-lines $\gtrsim 10^{41}$). We note that the [\ion{O}{3}] LF found by \cite{B10} from zCOSMOS, at slightly higher redshifts ($0.15 < z < 0.3$) agrees closely with ours, and is similarly flatter than that of \cite{H05}.

We have also compared our Seyfert galaxy LF's to the those of the local normal galaxies in H$\alpha$ and [\ion{O}{3}] (not shown), using data from \cite{g95,g96}. For most line luminosities, the space density of the galaxies without Seyfert nuclei exceeds that of the Seyfert by up to two orders of magnitude. However, the H$\alpha$ and [\ion{O}{3}] LF's of non-AGN cut off exponentially above L$_{\mathrm{line}} > 10^{41}$ erg/sec. The result is that with H$\alpha$ or [\ion{O}{3]} luminosities of $10^{41}$ erg/sec or above, most of the galaxies are Seyferts.

\section{Seyfert 1 and 2 Narrow Line comparison} \label{NLR}
The main observational difference between the two types of Seyfert galaxies is the presence or absence of broad permitted lines with widths of $10^{3}$ km/s or higher. The Seyfert 1 galaxies are further divided into Seyfert 1.2, 1.5, 1.8, and 1.9's, based on the increasing relative strength of the Narrow to Broad line components \citep{o81}.

In this Section we investigate differences and similarities in the emission line properties of Seyfert 1 and 2 galaxies in our sample. Although we consider a galaxy with {\it any} broad lines to be a Sy 1, we keep in mind the possibility that Seyferts with only very weak wings on H$\beta$ (Sy 1.8), or only on H$\alpha$ (Sy 1.9), may turn out to be more similar to Seyfert 2 galaxies in most observational respects.

\subsection{Nuclear Reddening} \label{Red}

Various physical models have been proposed to connect Sy 1 and Sy 2 galaxies.
The ``unification model" assumes that the sole difference is the viewing orientation of the Seyfert nucleus with respect to our line of sight \citep{ar93}. Thus there would be no difference between the Sy 1 and the Sy 2 nuclei if these galaxies were viewed from the same direction. 
There are, however, alternate hypotheses. For example, in the Galactic Dust Model \citep{mgt98}, Sy 2's have intrinsically larger dust covering fractions due to the presence of galactic dust lanes. This dust obstructs our view of the inner nuclear regions where the engine and broad line region (BLR) are located. One consequence is that the emission lines in Sy 2's should on average have a greater degree of reddening than Sy 1's.

\begin{figure*}
\centering
\includegraphics[scale=1.0,width=6in]{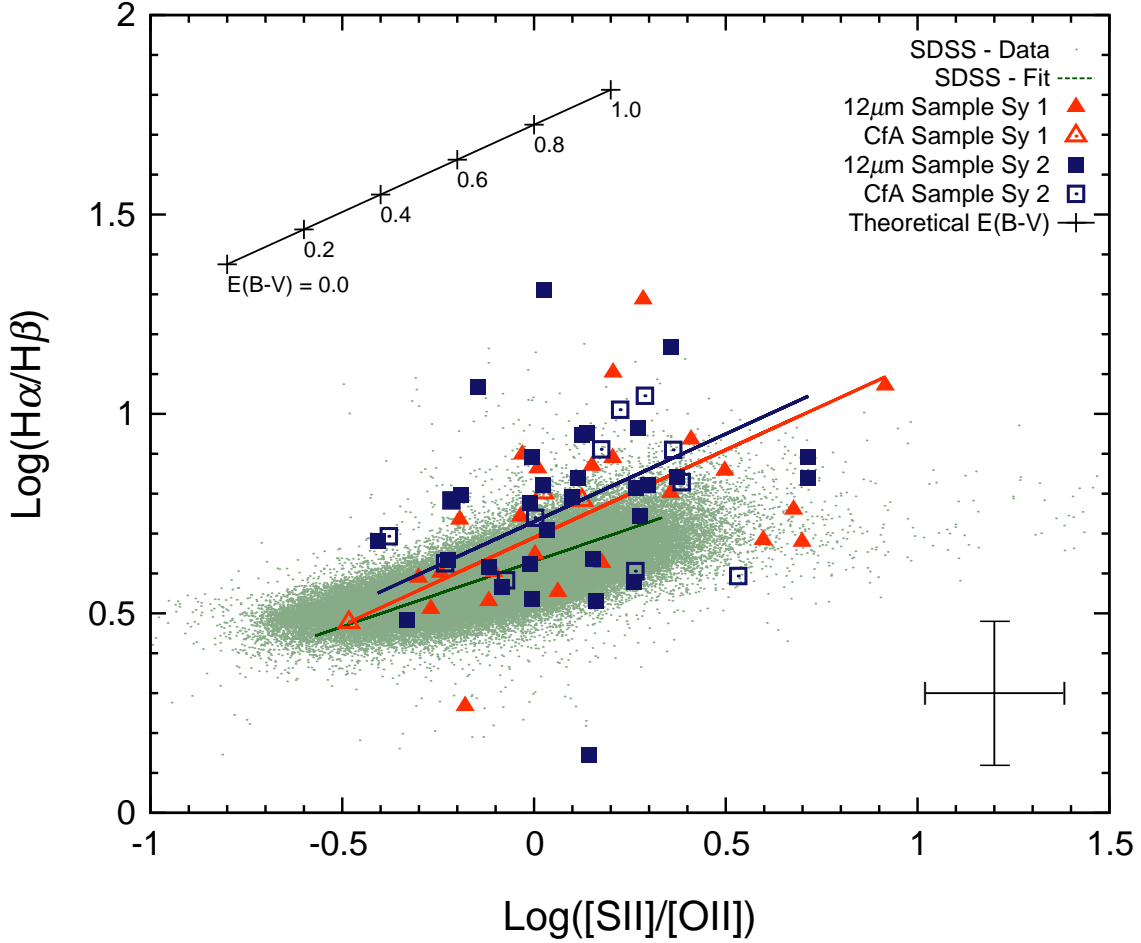}
\caption{Nuclear reddening-sensitive emission-line ratios log(H$\alpha$/H$\beta$) vs. log([\ion{S}{2}]/[\ion{O}{2}]). Open symbols represent galaxies added data from the \textit{CfA} sample. The best fit slopes for the Sy 1'sand Sy 2's are identical; Sy 1: $0.44\pm0.12$ (red line), Sy 2: $0.44\pm0.22$ (blue line). The best fit to the SDSS data is $y=(0.33\pm 0.004)x - (0.64\pm0.08)$. The theoretical reddening vector, derived from \cite{cardelli}, is arbitrarily offset for clarity. Our Sy 1 and Sy 2 galaxies are plotted with red triangles and blue squares, respectively, with typical uncertainty shown by the black error bar. The large number of (135,116) of SDSS DR7 galaxies with strong emission-lines (of any origin--AGN or starburst) are plotted as a cloud of light green/gray dots.\label{Fig2}}
\end{figure*}

To test this we  compare two optical emission line-ratios in our data that are widely separated in wavelength--the Balmer decrement (H$\alpha$/H$\beta$) and the narrow line region (NLR) ratio  [\ion{O}{2}]/[\ion{S}{2}]. The positive correlation between these two reddening-sensitive line ratios is shown in Figure \ref{Fig2}. We note that both the Sy1's and the Sy2's appear to lie along the same correlation.  Furthermore, this trend looks similar to the one defined by emission-line galaxies measured in the SDSS, which are shown by the cloud of small grey/green points. There are 135,116 galaxies of any origin (AGN or starburst) with spectra from DR7, which have all four emission lines detected at greater than the 10-sigma level.

To quantify this trend, we compute proper least-squares fits for these groups of galaxies separately. For this and all subsequent correlation analyses, we used a proper least-squares (LSQ) fit (to account for the comparable errors in both line ratios). We use a \texttt{FORTRAN} fitting program (\textit{Linear regression with measurement errors and scatter}\footnote{Obtained from the website for statistical packages: http://www2.astro.psu.edu/statcodes/sc\_correlregr.html.}) written and described by \cite{ak96}, to compute the orthogonal proper LSQ fits. The slope of the fit for the 12$\micron$ sample is $0.44\pm 0.14$ for the Sy 1 Galaxies and $0.44 \pm 0.22$ for the Sy 2's. The large sample of SDSS emission line galaxies also shows a similar line-ratio correlation, with a slope of $0.33 \pm 0.004$ and intercept of $0.64 \pm 0.08$. For comparison we calculate the predicted slope in this line ratio correlation that should be produced solely by reddening of two fixed (constant) intrinsic line ratios. Adopting the standard reddening law of \citet{cardelli} for $R=3.1$, the predicted reddening slope should be 0.44. We placed the straight-line reddening vector in the upper left in Figure \ref{Fig2} with tick marks showing increasing amounts of $E(B-V)$.

The quantitative fits confirm the visual impression of the figure:
all three of the groups of galaxies (Sy1s, Sy2s, and SDSS emission line galaxies)
show a consistent correlation.  The positive slopes of each correlation are all consistent
with each other, and with the prediction from a standard reddening law. 
The Seyfert galaxies have more widely ranging line ratios than those normally seen in the SDSS spiral galaxies, although the distributions mostly overlap. We also compared the $H\alpha/H\beta$ and [\ion{S}{2}]/[\ion{O}{2}] ratios with two other emission-line ratios that should be sensitive to reddening. However, these other line ratios--[\ion{N}{2}]/[\ion{O}{2}] and [\ion{O}{3}]/[\ion{Ne}{3}]--did not correlate so well the others, suggesting that they are influenced by other factors beyond mere reddening. Thus, their intrinsic value can not be considered constant, and we do not consider them valid reddening indicators. The results of all the correlations are presented in Table \ref{tbl8}.  
 
There is reasonable consistency with the position of the lower-left  (bluest) extent
of the line ratios in all three galaxy groups. This limit should correspond to galaxies with
essentially unreddened emission-line regions. We therefore interpret this lower left
boundary as indicating that both types of Seyferts, as well as normal star-forming galaxies, have roughly the same intrinsic (unreddened) line ratios. Specifically, the line ratios in both Sy1s and Sy2s have unreddened H$\alpha$/H$\beta$ ratios which appear consistent with
\textit{Case B'} (log H$\alpha$/H$\beta$ = 0.5), \citep{gf84}, also recommended by \cite{m83}).
The normal spiral galaxies from SDSS have slightly lower Balmer decrements--they are
consistent with the normal Case B value of log H$\alpha$/H$\beta$ = 0.45, although this
small difference is only marginally significant.

\begin{figure*}
\centering
\includegraphics[scale=1.0,width=6in]{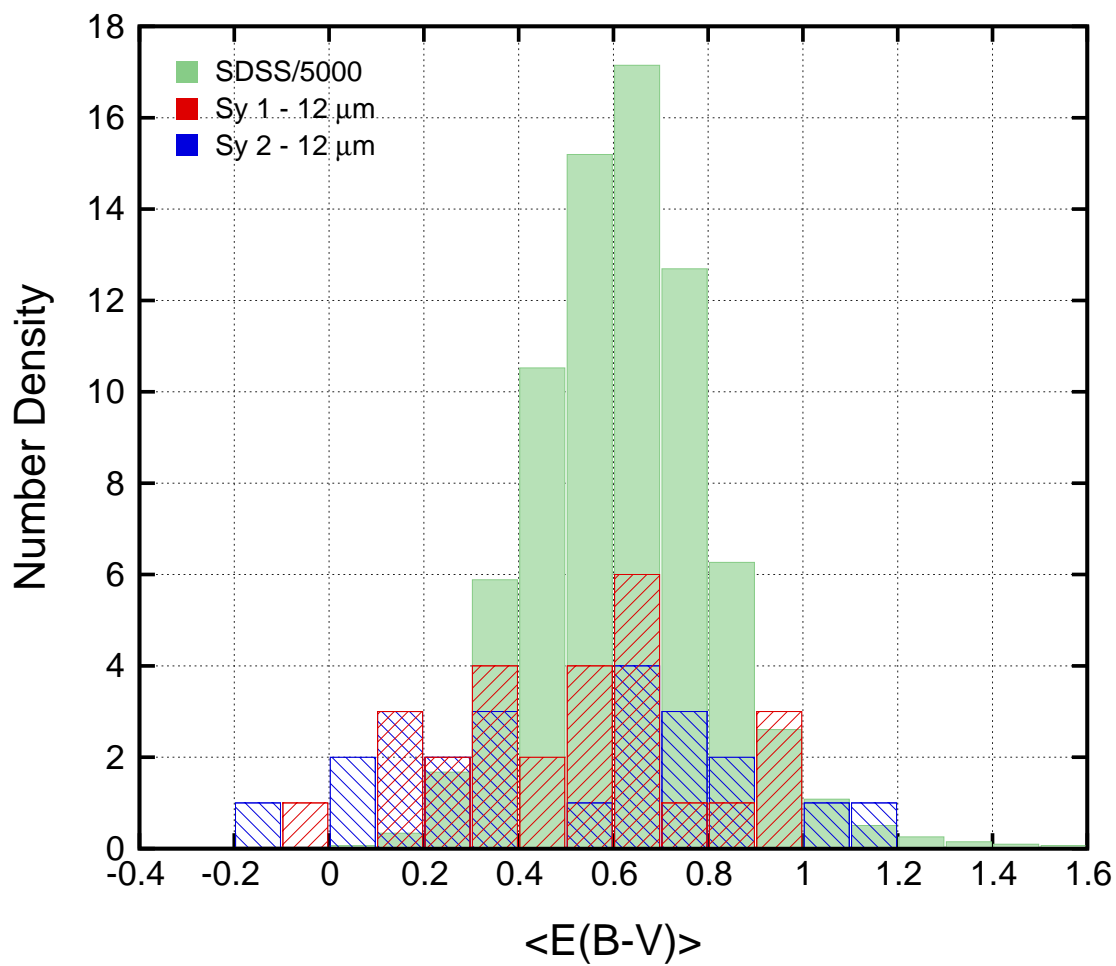}
\caption{Histogram of the average E(B-V) using H$\alpha$/H$\beta$ and [\ion{S}{2}]/[\ion{O}{2}] line ratios for the $12\;\mu$m Seyferts and the SDSS galaxies. The average reddening is essentially the same in both Seyfert types: $<E(B-V)> = 0.49 \pm 0.35$ for the Sy1's, and  $<E(B-V)> = 0.52 \pm 0.26$ for the Sy2's. The average reddening for the SDSS galaxies is $<E(B-V)> = 0.63 \pm 0.21.$ \label{Fig3}}
\end{figure*}

Supported by the consistency of the (H$\alpha$/H$\beta$) correlation with [\ion{S}{2}]/[\ion{O}{2}], we will now assume that the primary determinant of where each galaxy lies in Figure \ref{Fig2} is its amount of internal dust reddening.  We therefore estimated  $E(B-V)$ values in each Seyfert galaxy assuming the standard reddening law and intrinsic line ratios found above. The average of the two line ratio estimators gives the $E(B-V)$ values for individual Seyfert galaxies plotted in Figure~\ref{Fig3}. 
This histogram shows that the average reddening is the same in both Seyfert types:
$<E(B-V)> = 0.49 \pm 0.35$ for the Sy1's, and  $<E(B-V)> = 0.52 \pm 0.26$ for the Sy2's.
A Kolmogorov-Smirnov (K-S) test yields a probability of $p$(0.31) that the two groups are drawn from the same distribution. In both Seyfert classes, $E(B-V)$ ranges uniformly from 0.0 to 1.0 mag. Furthermore, there is no trend for reddening to vary systematically with the luminosity of the Seyfert galaxy, over more than two orders of magnitude in 12 $\mu$m luminosity. For comparison, the average $E(B-V)$ inferred from normal SDSS spirals is 0.63 mag. As shown in Figure \ref{Fig2}, the Seyfert reddenings strongly overlap with these, except for the larger scatter which may be due to observational uncertainties.

Other studies have been done on reddening in Seyfert galaxies. \citet{rl05} report substantial reddening in the narrow-line regions of 11 Seyfert 2 galaxies\footnote{These include Seyfert 1.8 and 1.9 galaxies.} and \citet{ti89} also finds that Sy 2's are more reddened than the Sy 1's for their sample of 24 Seyferts. However, \citet{m83} found only a marginal tendency for Seyfert 1's to have smaller reddening in their forbidden line region, consistent with our new findings. Our finding of no significant difference between the reddening of the NLRs in Seyfert 1 and Seyfert 2 galaxies is consistent with the expectation of the strong unification hypothesis.

We note that some narrow line regions appear to be optically thick at optical wavelengths, with $E(B-V) > 0.8$ mag, which implies $A_V > 2.5$. If this reddening is also applied to the non-stellar continuum, then the $UV$ would be more then 99\% extinguished -- i.e. virtually obliterated. It appears that some minority of the Seyfert nuclei we uncover in the 12 $\mu$m sample--unlike many other samples--are so reddened that some special explanation is required for their observed UV continuum emission. The two possibilities are:
\begin{enumerate}
\item The UV continuum is purely from the stellar photospheres in the host galaxy. If the UV continuum is particularly bright, this would be coming from a young population of hot stars.
\item The observed UV continuum does come from the non-stellar engine (thought to be an accretion flow around the central massive black hole), but suffers from less extinction than the NLR. For example, we might have indirect unobscured views to the central engine, which we see through scattered light \citep{ar93}.
\end{enumerate}
In most cases neither of these possibilities can be ruled out.

\subsection{Degree of Gas Ionization}

The blue portion of the optical spectrum contains several emission line ratios which are sensitive to the level of ionization in the gas. Table \ref{tbl3} summarizes the ionization potential of our emission lines. We select line ratios that differ in their ionization potential by 20 eV or more. These diagnostic line ratios are observable by CCD spectrographs up to redshifts of $z \sim 0.9$.

As summarized in Table \ref{tbl7}, our averages for log([\ion{Ne}{5}]/[\ion{Ne}{3}]) are $-0.07 \pm 0.24$ and $-0.16 \pm 0.40$ (individual scatter) for 32 Sy 1's and 26 Sy 2's, respectively. For log([\ion{O}{3}]/[\ion{O}{2}]) the averages are $0.59 \pm 0.51$ and $0.43 \pm 0.52$ for 50 Sy 1's and 51 Sy 2's, respectively. For both line ratios, the K-S test shows that the small differences between the Sy1's and Sy 2's are not significant at the 95\% level
\footnote{This finding contradicts \cite{S98}. A possible explanation is the \cite{S98}'s selection of 52 Sy 1 galaxies from the literature may have missed some low-luminosity Sy 1's. They are included in our complete sample, and tend to have less highly ionized narrow lines. The only possible significant ionization difference seen in our sample is that log([\ion{Ne}{3}]/[\ion{O}{3}]) is $-0.17\pm0.40$ in the Sy 1's and $-0.48\pm0.30$ in the Sy 2's. However, many of the measurements of ``[\ion{Ne}{3}]" in the Sy 1's were obtained with low-resolution spectroscopy, with moderate SNRs. We therefore suspect that some of the line flux attributed to [\ion{Ne}{3}]$\lambda 3869$ in some Sy 1's in our study, and in Schmitt's, may instead be contaminated from the weak broad line emission line \ion{He}{1}$\lambda 3889$. }. 

\begin{figure*}
\centering
\includegraphics[scale=1.0,width=6in]{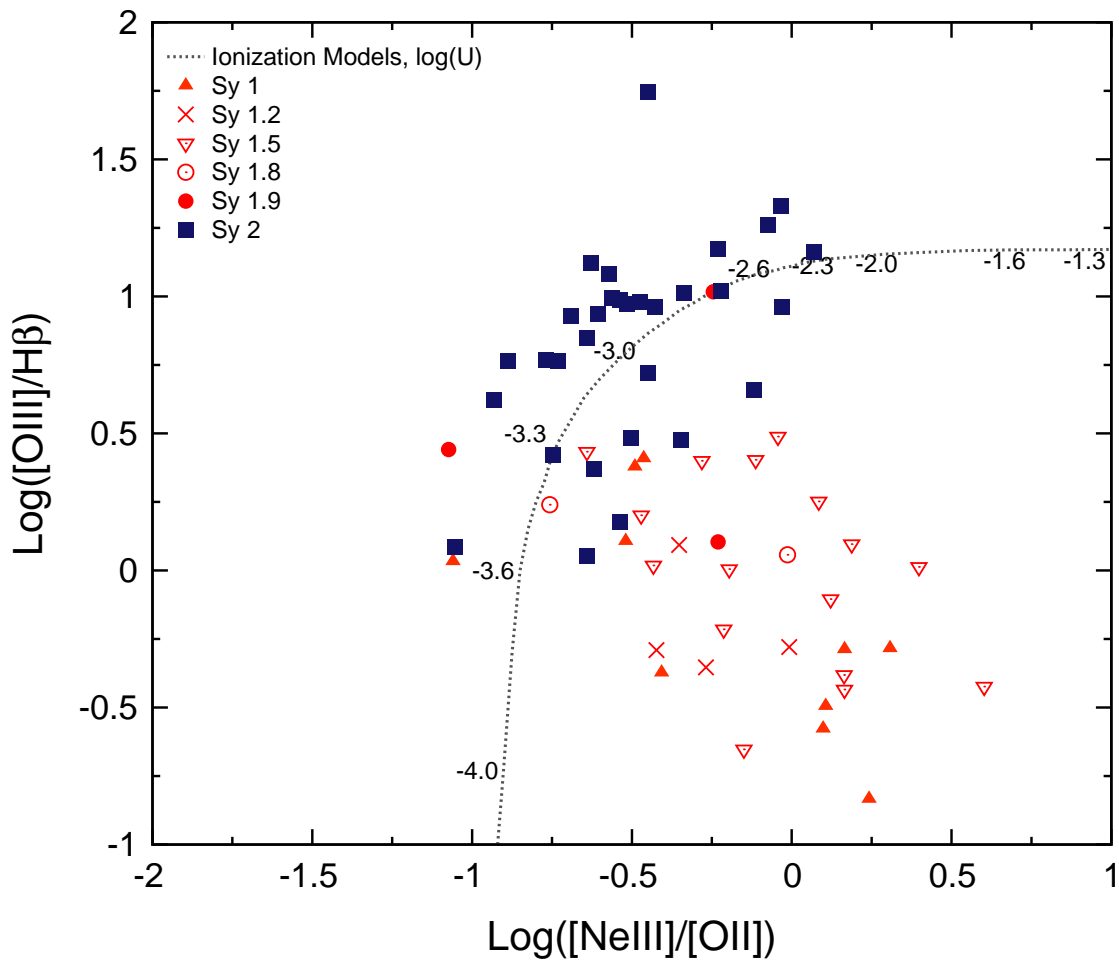}
\caption{Logarithmic line ratios of[\ion{O}{3}]/H$\beta$ vs. [\ion{Ne}{3}]/[\ion{O}{2}]. The dashed line is a sequence of models for the narrow line region from \citet{gds04} with the numbers along the line indicating the ionization parameter $U=S_{\star}/(nc)$, where $S_{\star}$ is the flux of ionizing photons and $n$ is the number density of hydrogen atoms. Since this model only includes the NLR, most of the Sy 1 galaxies, especially Seyfert 1, 1.2, and 1.5's, falls in the lower right section of the graph, because of their additional $H\beta$ contribution from the BLR.\label{Fig4}}
\end{figure*}

In Figure \ref{Fig4} we plot log[\ion{O}{3}]/H$\beta$ versus log[\ion{Ne}{3}]/[\ion{O}{2}]. The dotted line is a model for the NLR from \citet{gds04}. The numbers along the line indicate the ionization parameter $U=S_{\star}/(nc)$, where $S_{\star}$ is the flux of ionizing photons and $n$ is the number density of hydrogen atoms. Their model is of a dusty, radiation pressure-dominated region surrounding a photo-evaporating molecular cloud, which in turn is surrounded by a coronal halo where the dust has been largely destroyed \citep{dg02}. In this model only the NLR is included. The particular model that matches our data is un-reddened and un-depleted. \citet{gds04} adopt a power-law ionizing continuum, $F_{\nu} \propto \nu^{\alpha}$ with the best slope of $\alpha = -1.4$, and a number density of $10^{3}$ cm$^{-3}$. The chemical abundances are solar. As can be seen in Figure \ref{Fig4}, the Sy 1's are generally in the lower/right of the dotted line in the graph, while the Sy 2's are to the left and above. Since H$\beta$ is a broad line and the BLR is not included in this particular model, we expect the Sy 1's to have lower log[\ion{O}{3}]/H$\beta$ values. The NLR ratios indicates an ionization parameter of $\log U = -2.5$ for the average of the Sy 1's, and $-2.8$ for the average of the Sy 2's. 

\subsection{[\ion{S}{2}]/[\ion{N}{2}] Ratio} \label{S2N2}
We found that the log([\ion{S}{2}]/[\ion{N}{2}]) line ratio has almost the same value for Sy 1's and Sy 2's, $-0.23 \pm 0.24$ and $-0.28 \pm 0.20$ respectively, with a K-S probability of $p(0.97)$. For the previously selected subset of Seyfert 2 galaxies we took from DR7, we find an average value of $\log([SII]/NII])$ of $-0.19\pm 0.1$. This is also reasonably consistent with our values, allowing for observational uncertainties (Figure \ref{Fig5}). An average value of $-0.23 \pm 0.18$ for this ratio can be used regardless of whether the galaxy is a type 1 or type 2 Seyfert. These lines have about the same ionization potential (10.4 and 14.5 eV), but their critical densities are different ($\sim10^3$ cm$^{-3}$ for [\ion{S}{2}] and  $\sim10^5$ cm$^{-3}$ for [\ion{N}{2}]). The fact that this ratio is the same for both Seyfert types implies that the density structure of the narrow line region is the same at least up to $10^5$ cm$^{-3}$. This is consistent with the result found by \citet{n01}. They find that the ratio [\ion{S}{2}]/[\ion{N}{2}], among other low-ionization line ratios, shows no difference between Sy 1's and Sy 2's. 
Our [\ion{S}{2}]/[\ion{N}{2}] ratio is the same as that found in Sy 1 galaxies in \citep{SL13}. They obtained log([\ion{S}{2}]/[\ion{N}{2}]) = -0.25, independent of bolometric luminosity, for the mean stellar mass in their sample $<\log M_{*}\sim 10.8>$. We caution that, in the least-massive galaxies with $<\log M_{*}\sim 10.4>$, \citep{SL13} found slightly higher ratios of log([\ion{S}{2}]/[\ion{N}{2}]) $\gtrsim -0.20$. Since we do not generally know $M_{*}$ for all Seyfert host galaxies, this introduces a small uncertainty that might contribute to the scatter we observe.
Nonetheless, the near constancy of the [\ion{S}{2}]/[\ion{N}{2}] can be useful in de-blending [\ion{N}{2}] from broad H$\alpha$ in Sy 1's. Some studies use other forbidden line fluxes to remove the [\ion{N}{2}] which is blended with H$\alpha$ \citep{L82}.

\subsection{Warm Dust}
\begin{figure*}
\centering
\includegraphics[scale=1.0,width=6in]{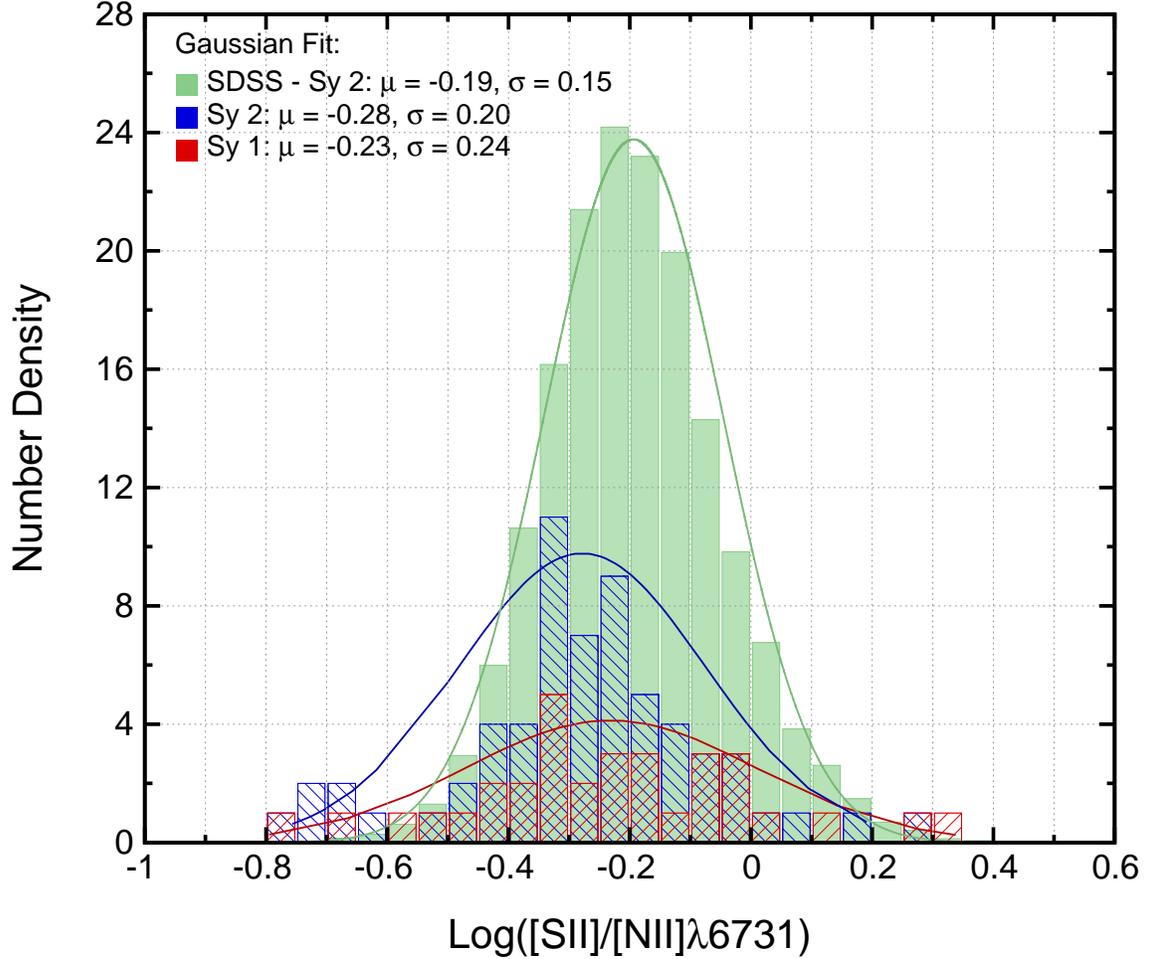}
\caption{Histogram of [\ion{S}{2}]/[\ion{N}{2}] line raions for our 12 $\mu$m Seyfert 1 and Seyfert 2 galaxies (red and blue hatching, respectively). These distributions are quite similar to the line ratios observed in our SDSS DR7 sample of 16,708 Sy 2 galaxies, shown in light green. Each of these distributions can be fitted approximately by a Gaussian. The mean and standard deviation values are given. All of the distributions overlap substaintially, justifying our adoption of a fairly ``universal" of $\log$([\ion{S}{2}]/[\ion{N}{2}]])$=-0.23$. The histogram widths are likely to be dominated by observational uncertainties.\label{Fig6}}
\end{figure*}

\begin{figure*}
\includegraphics[angle=0,scale=1.0]{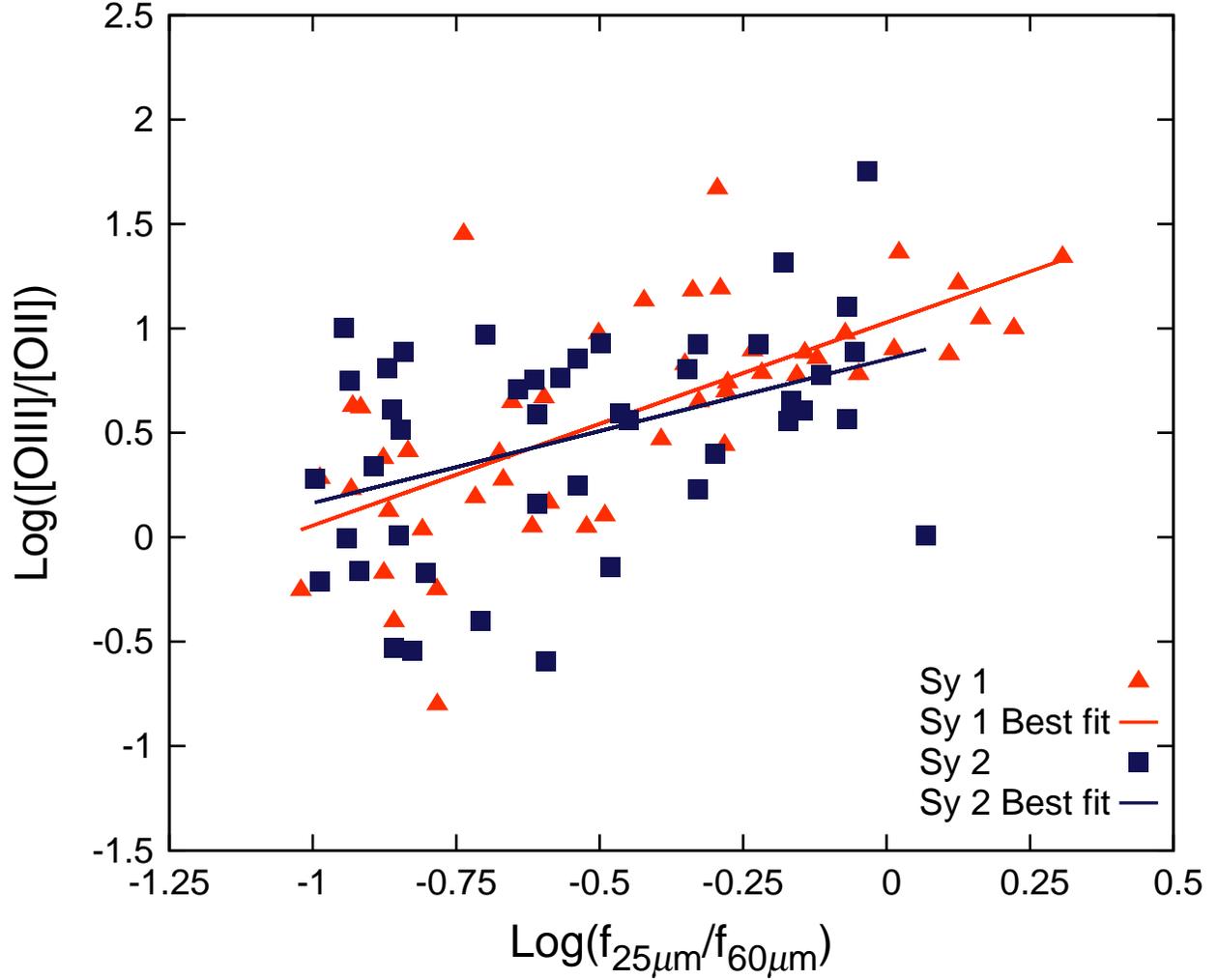}
\caption{Logarithmic line-ratios [\ion{O}{3}]/[\ion{O}{2}] vs. $f_{25\micron}/f_{60\micron}$, which both measure the relative Seyfert nucleus power, compared with the lower-ionization gas and the cooler dust in HII regions. Although this positive correlation shows large scatter, it is significant, with a Sy 1 regression fit of log([\ion{O}{3}]/[\ion{O}{2}]) = (0.97 $\pm$ 0.12)log($f_{25\micron}/f_{60\micron}$) + ($1.03\pm0.63$) and log([\ion{O}{3}]/[\ion{O}{2}]) = (0.69 $\pm$ 0.16)log($f_{25\micron}/f_{60\micron}$) + ($0.85\pm0.76$) for Sy 2 types. A Kendall's Tau significant  test confirms that the gas ionization/dust temperature correlation holds for both Sy 1's and Sy 2's, at a CL $ > 97.5\%$.\label{Fig5}}
\end{figure*}

Our IRAS data give the ratio $f_{25\micron}/f_{60\micron}$ integrated over each entire galaxy, which increases when the
proportion of warm dust (heated by the AGN) increases. The [\ion{O}{3}]/[\ion{O}{2}] ratio 
was previously used as a measure of the the relative strength of the Seyfert nucleus with respect to the \ion{H}{2} regions in the host galaxy. We plot [\ion{O}{3}]/[\ion{O}{2}] vs. $f_{25\micron}/f_{60\micron}$ (Figure \ref{Fig6}),
and find a positive correlation for both Seyfert types. A Kendall's Tau test reveals that this gas ionization/dust temperature correlation holds for both Sy 1's and Sy 2's separately, with confidence levels of CL = 99.9\% and CL = 97.5\% respectively. The individual regression fits for the Sy galaxies are log([\ion{O}{3}]/[\ion{O}{2}]) = (0.97 $\pm$ 0.12)log($f_{25\micron}/f_{60\micron}$) + ($1.03\pm0.63$) for the Sy 1 and log([\ion{O}{3}]/[\ion{O}{2}]) = (0.69 $\pm$ 0.16)log($f_{25\micron}/f_{60\micron}$) + ($0.85\pm0.76$) for the Sy 2 types.

\subsection{Correlation of [OI] and [OIII]}
\begin{figure*}
\includegraphics[angle=0,scale=1.0]{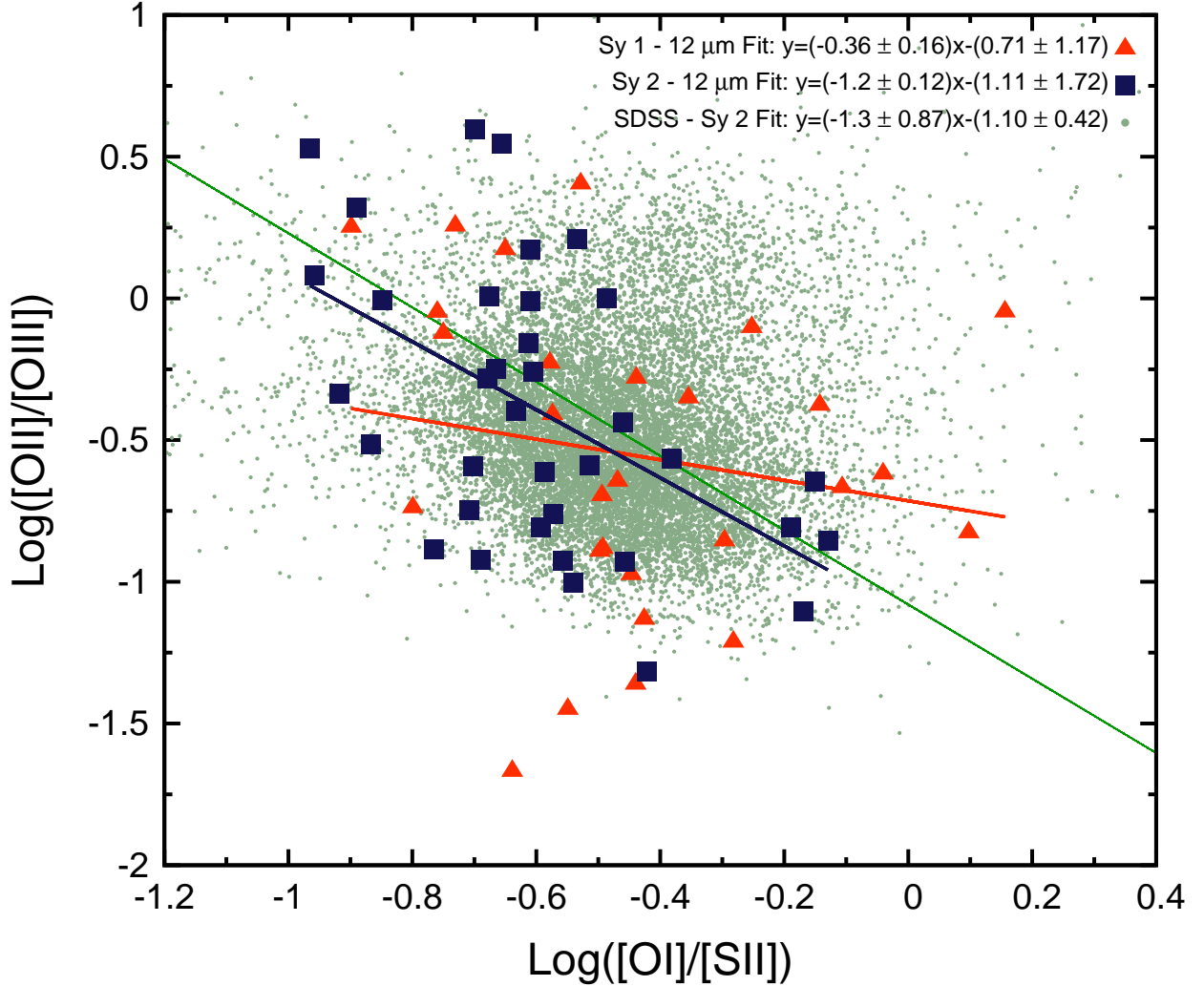}
\caption{Correlation of log([\ion{O}{2}]/[\ion{O}{3}]) vs. log([\ion{O}{1}]/[\ion{S}{2}]). We again show our Sy 1 and Sy 2 galaxies as red triangles and blue squares, respectively. The light green/grey points is a subsample of 13,688 Seyfert 2 galaxies from the DR7 of the SDSS. The inverse correlation is the opposite of an ionization effect, and indicates that [\ion{O}{1}] follows [\ion{O}{3}] more closely than [\ion{S}{2}]. This implies that the [\ion{O}{1}] emission comes predominantly from the Seyfert nucleus, not from HII regions. The Kendall's Tau significant test reveals that it is significant for the Sy 2 galaxies but not for the Sy 1's, CL $ = 99.9\% $ and CL $ < 90\%$ respectively. The slope of the Sy 2 fit is $-1.2\pm0.12$ and the slope for the SDSS fit is $-1.3\pm0.87$.\label{Fig7}}
\end{figure*}

The ionization potential of \ion{O}{1} is 13.6 eV and the critical density of [\ion{O}{1}]$\lambda$6300\AA\; is $1.8\times 10^{6}$ cm$^{-3}$ \citep{do86}. The [\ion{O}{1}] line is formed beyond the classical ionization front, in a partially ionized region heated by X-rays from the AGN \citep{vo87,sm92,gds04}. We therefore test the possibility that [\ion{O}{1}]  correlates better with the high-ionization  fine structure emission lines of the Seyfert NLR, than with the low-ionization emission lines from H~II regions. In Figure~\ref{Fig7} we plot the narrow line ratios log[\ion{O}{2}]/[\ion{O}{3}] vs. log[\ion{O}{1}]/[\ion{S}{2}]. As we found in other NLR plots, the vertical axis is inversely proportional to the average ionization level of the gas, and therefore increases in the relatively ``weak" Seyferts, which have larger contributions to their [\ion{O}{2}] line emission from H~II regions in their host galaxies. If the main difference along the horizontal axis is also degree of gas ionization, 
then we would expect a positive correlation in this graph.  Instead, we find the opposite--an {\it inverse} correlation. The orthogonal regression fit for the Sy 2 galaxies is: log([\ion{O}{2}]/[\ion{O}{3}]) = ($-1.2\pm0.12$) log([\ion{O}{1}]/[\ion{S}{2})] - ($1.11\pm1.72$). 
Although the spread is large, we find that a strong correlation exists (Kendall's Tau $ > 99\%$). 
The Sy 1 sample also gives a fit with a negative slope: log([\ion{O}{2}]/[\ion{O}{3}]) = ($-0.36\pm0.16$)log([\ion{O}{1}]/[\ion{S}{2}]) - ($0.71\pm1.17$). 
Here the Kendall's Tau is $ < 90\%$ and the null hypotheses can not be rejected, i.e. the Sy 1 correlation is not significant. But for Seyferts, overall, the [\ion{O}{1}] line tracks [\ion{O}{3}] more closely than the [\ion{O}{2}] line. 

Since this conclusion comes from the small dataset of our 12$\micron$ Seyferts, we sought confirmation of  this same inverse correlation in the much larger database of the SDSS DR7. From this database of $>900,000$ galaxies, we restricted our consideration to 239,795 galaxies in which the [NII], H$\alpha$, and [OIII] emission lines were detected at the 10-$\sigma$ level or better, while the [OII] line was detected at the 20-$\sigma$ level. Using TOPCAT\footnote{Available at http://www.star.bris.ac.uk/~mbt/topcat/.} to plot these line ratios in the standard BPT diagram, we used the Select Tool feature to construct a subset of 15,190 ``pure" Seyfert galaxies, whose emission spectra are dominated by the NLR. These DR7 Seyferts are plotted in the figure with small grey/green dots, which were fitted by the solid gray/green line with the proper LSQ. This fit is indistinguishable from what we obtained fitting our much smaller Sy2 sample, but with a larger scatter (log([\ion{O}{2}]/[\ion{O}{3}]) = ($-1.3\pm0.87$)log([\ion{O}{1}]/[\ion{S}{2}]) - ($1.10\pm0.42$)).  Thus SDSS confirms the trend we found that [\ion{O}{1}]$\lambda$6300\AA\; tracks the high-ionization gas in Seyferts more closely than the low-ionization gas, because both are produced primarily by the AGN, not HII regions. \footnote{This finding holds for Seyfert nuclei, \textit{not for LINERs} (see also \cite{N09}).}

\subsection{BPT Diagram} \label{BPT}

The BPT diagrams \citep{bpt81} were developed to identify different photoionization mechanisms in galaxies, using ratios of lines at similar wavelengths, to minimize the effects of reddening.
In general, galaxies dominated by stellar photoionization in  \ion{H}{2} regions have relatively stronger emission lines from less ionized gas, like [\ion{O}{2}] and [\ion{N}{2}]. Active galaxies, in contrast, have a power-law ionizing continuum, which tends to produce more lines from highly ionized gas. In addition, X-rays from the AGN can penetrate into neutral or partially ionized zones to produce low-ionization lines that would typically not be produced in \ion{H}{2} regions.  Thus AGN emission line spectra show a wider {\it range} of ionization.

The most common BPT diagram uses the ratio of [\ion{O}{3}]/H$\beta$ to [\ion{N}{2}]/H$\alpha$, though sometimes [\ion{S}{2}] or [\ion{O}{1}] is used in place of [\ion{N}{2}]. These diagrams are designed for narrow-line objects, and therefore the Sy 1 galaxies are generally not displayed because of the presence of broad permitted lines. However, we decided to include the weak broad-line objects to determine where they would be located in the BPT diagram, for situations in which the BLR and NLR are not separated. 

Our BPT diagram is shown in Figure~\ref{Fig8}. The gray/green dots represent all 239,795 SDSS DR7 galaxies with highly significant detection in all four emission-lines. We include the heavy dashed-dot line from \citet{k03} defined by: $\log([OIII]/H\beta) > 0.61/[\log([NII]/H\alpha) - 0.05] +1.3$, which separates \ion{H}{2} regions from active galaxies. The light dotted line  from \citet{k01} defined by: $\log([OIII]/H\beta) > 0.61/[\log([NII]/H\alpha) - 0.47)] +1.19$, also excludes ``composite" galaxies which include too much emission form \ion{H}{2} regions to be classified as ``pure" AGN.

The Sy 2 Galaxies cluster in the upper right of the BPT diagram, around ($x,y$) = (0.05, 0.85). However, a significant minority of the Sy 2 galaxies fall close to the Kauffman boundary: 17 out of a total of  82 Sy 2's (21\%), would be classified as ``composite" (Seyfet 2 + HII) galaxies. This contamination of narrow emission line fluxes from the host spiral galaxy is especially serious for less luminous AGN, when they are observed with relatively poor spatial resolution \citep{T16}.

\subsection{BPT Classification of Broad Line AGN}
The BLR contamination should in principle be removed before using the narrow line for BPT classification. However, in some studies the spectra do not or cannot have the broad Balmer components removed. This is especially true if the BLR component is very faint compared to the NLR.  We therefore plot the total line-flux ratios (broad + narrow components) for Sy 1.5, 1.8, 1.8 and Sy 2 galaxies in Figure~\ref{Fig8}.
In addition to the AGN/HII boundary line from \cite{k03} we also plot the boundary line defined by \cite{k01}. The Sy 1.9 galaxies occupy the same region as the Sy 2's (except for NGC 7314 which has a very weak nucleus and is located below the AGN/HII boundary line). This is because their broad Balmer line components are so weak. Thus the BLR hardly alters the ratio of forbidden lines to the permitted lines, away from the NLR values. This agrees with the findings of \cite{S05}, who showed that his ``Sy 1.x galaxies" having broad H$\alpha$ wings (what we call Sy 1.8 and Sy 1.9's) are indistinguishable from the Sy 2's in the BPT diagram. Indeed, Simpson's NLR mixing line - also plotted in Figure~\ref{Fig8} - goes through most of our Sy 1.8's and Sy 1.9's, with the latter lying further away from the AGN/HII boundary. Our few Sy 1.8's are all at an intermediate location in the BPT diagram. Their combination of NLR and BLR compontents make them appear like \textit{composite AGN} - mixtures of HII and Seyfert 2 line emissions.
The distribution of our own 12 $\mu$m Seyferts in the BPT diagram is somewhat similar to that of the hard X-ray selected Seyferts from the BAT survey \citep{O17}. However, our sample includes more low-luminosity Sy 2's. Their line emission is more dominated by HII regions. The relatively weaker NLR emission corresponds to lower hard X-ray luminosity (see Appendix \ref{appxA}). Thus they are likely to be missed by the BAT survey. Another difference is that a substantial fraction of the BAT Seyfert 1's fall left of the NLR area in the BPT diagram. Their [\ion{N}{2}]/H$\alpha$ ratios are anomalously low. This could have resulted from mistakenly attributing some of their broad H$\alpha$ emission to the NLR component.\footnote{The BPT diagram is supposed to include only the narrow emission lines. When their broad emission lines are not removed, their contamination would push Sy galaxies to the left downward as in Figure \ref{Fig8}. Indeed, the BLR-dominated Seyferts nearly all fall below the AGN/HII separation line. For all the 32 Seyfert galaxies classified as any type of broad line AGN (1, 1.2, 1.5, 1.8 and/or 1.9) we find that according to the definition of the BPT diagram 59\% and 75\% of them lie below the Kauffman and Kewley lines, respectively.}

\citep{SL13} used SDSS spectra of 3175 Sy 1 galaxies to separate out the narrow line components of H$\alpha$ and H$\beta$. This allowed them to place the pure NLR line ratios on the BPT diagram. Since we did not make this BLR/NLR decomposition, we can only compare the their results to our own Sy 2's and Sy 1.9's, since they have negligible contribution from the broad Balmer lines. \citep{SL13} found that the BPT ratios classify 5\% of the NLRs in their Sy 1 sample as ``Star-Forming galaxies" (i.e. below the Kauffman line) and 15\%  of them as HII/AGN ( i.e. ``Composites", between the Kauffman and Kewley lines ). Our narrow-line Seyferts show exactly the same distribution: three out of 65 are classified as Star-Forming, while 12 out of 65 are classified as Composites. Thus the BPT distribution of our NLR AGN sample is indistinguishable from SDSS Seyfert 1's.

\subsubsection{Quantative Decomposition of AGN Components in the BPT Diagram}
\begin{figure*}
\begin{center}
\includegraphics[angle=0,scale=.75]{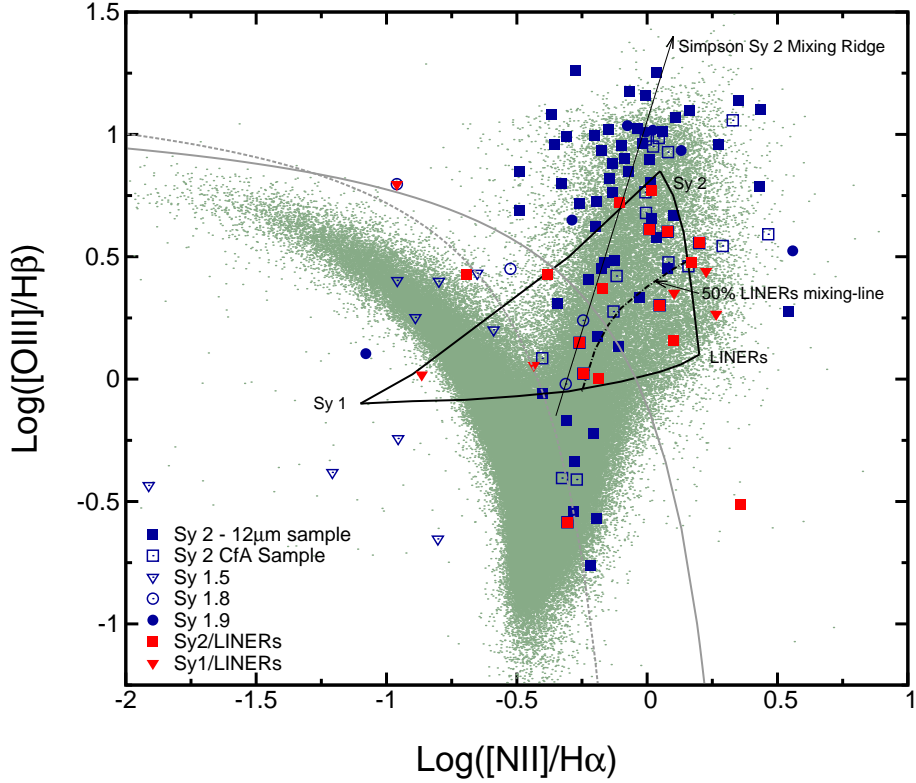}
\end{center}
\caption{BPT line-ratio diagram of our Sy 1.5, 1.8, 1.9, and Sy 2 galaxies, shown by the same color symbols as before. The green/grey dots represent the data form emission-line galaxies from SDSS DR7. We include the boundaries separating AGN and SFG/HII regions from \cite{k03}) (dashed light-grey) and \cite{k01} (solid light-grey). Most of the Sy 2's fall in the upper right corner as expected, as do Sy 1.9 's. Because of their BLR contimination of the H line-fluxes, the Sy 1.8's fall in the SFR/AGN ``composite' region of the BPT diagram, while the Sy 1.5's scatter below the HII/AGN boundary. The mixture of the pure Sy 2 line emmision and HII regions seen in our 12 $\mu$m Seyferts are well matched by the Sy 2 ``ridge line" from \cite{S05}. The three vertices of the black triangle show our adopted line rations for pure ``NLR" (Sy 2, upper right), ``BLR" (Sy 1, lower left), and "LINERs" (lower right). The curved line inside the triangle shows a 50\% mix of LINER emissions and Seyfert lines. This curve corresponds closely to the boundary between Seyfert 2 (above) and LINER (below) galaxies.\label{Fig8}}
\end{figure*}

To quantitatively interpret the location of AGNs in this BPT diagram, we make the simplifying assumption that each galaxy has observed emission lines which are the sums of three ``pure" Seyfert 1, 2, and LINER components. To accomplish this crude separation we have adopted the values for log([\ion{N}{2}]/H$\alpha$) of -1.1, 0.05, 0.2 and log([\ion{O}{3}]/H$\beta$) of -0.1, 0.85, 0.1, for the three respective components. 
The solid lines in Figure \ref{Fig8} connect each of these three components, showing the `~mixing curves" obtained by combining varying proportions of two of our ``pure" components.  Along the perimeter we have omitted the third emission component; for the interior points we compute the contribution from all three types.

To determine the relative contribution from each particular type, we take the distance, $d_i$ in the BPT diagram of each Seyfert to the three points defined as pure Seyfert 1, 2, and LINER.  The contribution is defined as 
\begin{equation}\label{eq3}
C_i = \frac{1/d_i}{\displaystyle\sum_i 1/d_i}
\end{equation}
where $i$ refers to Seyfert 1, 2, and LINER emission components, and 
\begin{eqnarray}\label{eq4}
d_{i}^{2} &=  \left(\log\frac{[OIII]}{H\beta} - \left(\log\frac{[OIII]}{H\beta}\right)_i \right)^2\nonumber \\
\nonumber \\
&+ \left( \log\frac{[NII]}{H\alpha} - \left(\log\frac{[NII]}{H\alpha}\right)_i \right)^2.
\end{eqnarray}

\begin{figure*}
\centering
\includegraphics[angle=0,scale=1.0,width=6in]{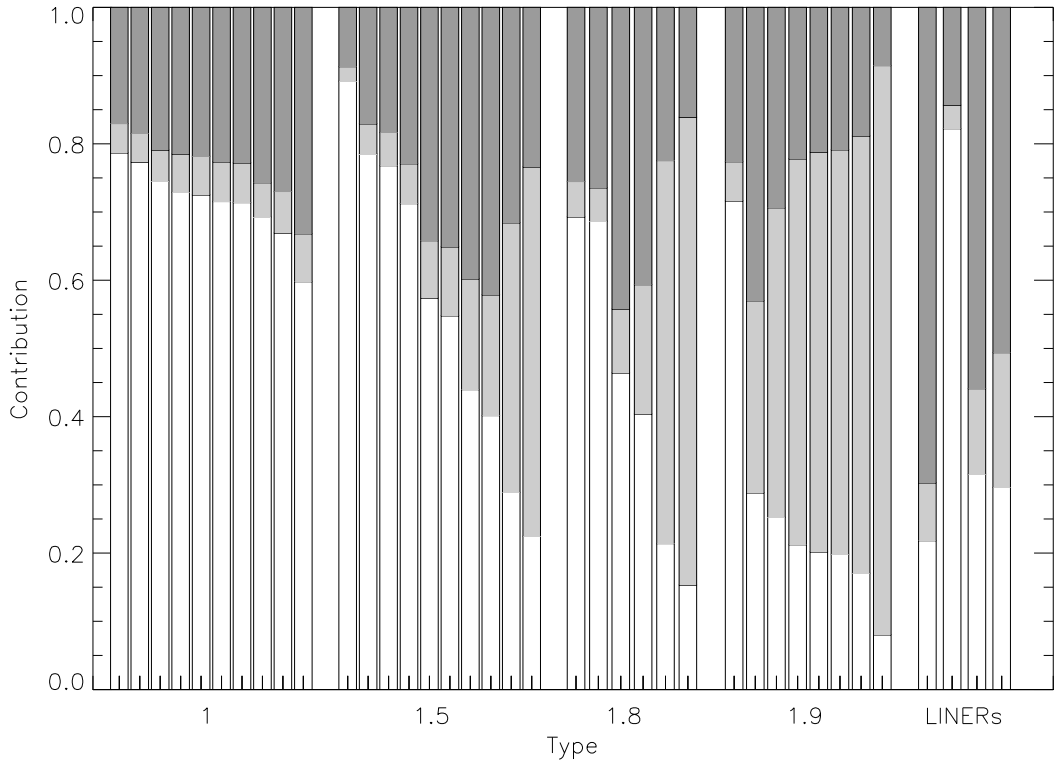}
\caption{Emisson line components of 12 $\mu$m of Seyfert 1 galaxies based on their locations in the BPT diagram. White corresponds to the Seyfert 1 contribution, the light gray corresponds to the Seyfert 2 contribution, and the dark gray corresponds to LINER. AGN components classified as Sy 1.8 and 1.9 are dominated by the narrow line component as in the Sy 2's, while the Sy 1.5's are a mix of the Sy 1 and the Sy 2 line components. Spectroscopically classified LINERs have 50\% or more of the emission from the LINER component.\label{Fig9}}
\end{figure*}

Our three-component decompositions of the relative AGN contribution are shown graphically in Figure \ref{Fig8}. The resulting components for individual galaxies are shown in Figure \ref{Fig9}.

The average BPT contributions of the ``BLR" and ``NLR" components change from 72\%/6\% in pure Sy 1's to 56\%/17\% in Sy 1.5's, to 43\%/12\% in Sy 1.8's, and to 20\%/58\% in Sy 1.9's. In this highly simplified BPT decomposition, all types of Seyfert 1 nuclei show an average contribution of 24\% from a ``LINER" component, regardless of the BLR/NLR ratios. 

As Figure \ref{Fig8} indicates, actual AGN show a continuously ranging mixture of NLR and LINER components. There is no clear-cut separation between the two components. However, an emission line galaxy cannot be reliably classified as predominantly a LINER unless it shows a $>50$\% contribution from the LINER component. The dot-dashed ``mixing curve" in Figure \ref{Fig8} shows the locus along which the LINER component, $C_{LINER}$, equals 50\%. As we expected, the only galaxies lying to of the right of the boundary are classified as LINERs (the rightmost filled circle is NGC 2639, classified in NED basic data as a Sy 1.9, but based on its Activity Type it is considered a LINER). Interestingly, this line corresponds to the transition between LINERs and Seyfert galaxies in \cite{ghk06}. That is, our $C_{LINERs}=50\%$ line accurately divides most of the \cite{ghk06} Seyferts from the LINERs--the blue and the red datapoints, respectively in their Figure 1. As expected from our use of NED classification, hardly any of our Seyfert galaxies turn out to be LINER-dominated. The same result applies to the BAT-selected survey \citep{O17}.

\section{Broad Line Region, Eigenvector 1, and the Baldwin Effect Relationship} \label{BLR}

The luminosities of broad emission lines generally increase linearly with the non-stellar continuum luminosity, but there are some exceptions. \citet{b77} found a negative correlation between the equivalent width of \ion{C}{4}$\lambda$1549\AA , and UV continuum luminosity ($L_{UV}= \nu L_{\nu}\;\lambda$1449\AA) in quasars, commonly referred to as the \textit{Baldwin effect}.  This less-than-linear increase in \ion{C}{4} with underlying luminosity has been confirmed in many different quasar and Sy 1 samples \citep{j16}. One interpretation is that the wavelength peak of accretion disk luminosity shifts from the UV in quasars to the EUV in the less luminous Seyferts, because the latter have smaller black holes with hotter accretion disks \citep{zm93}. 

Our sample includes 23 Sy 1 galaxies with measured \ion{C}{4} emission lines (Table \ref{tbl5}). We exclude the galaxies \object{MKN 231} and \object{NGC 2841} galaxies because of their anomalously weak \ion{C}{4}. Figure \ref{Fig10}a shows a plot of log(\ion{C}{4} EW) versus log($L_{UV}/10^{40}$) (color coding is described in the following section).
The solid line shows our proper least-squares fit to this Baldwin relation
\begin{eqnarray}\label{eqn5}
\log(C IV EW)= 
(-0.12\pm0.05)\log\left(\frac{L_{UV}}{10^{40}}\right) + (2.4\pm0.16).
\end{eqnarray}

The amplitude of this slope is much smaller than what \citet{b77} found for his sample of 20 quasars ($-0.63$). These is so much scatter in our sample that the anti-correlation is only weakly significant.

\subsection{Improving the Baldwin Effect with Eigenvector 1}

Many studies have attempted to connect the Baldwin effect with some other parameter, 
which could be more astrophysically fundamental than L$_{UV}$ \citep{zm93}. The most significant way in which the emission-line regions of quasars  differ from each other is in the strength of the so called `Eigenvector 1' \citep{bg92}. This parameter is associated with stronger [\ion{O}{3}] from the NLR, which is in turn strongly anti-correlated with the permitted \ion{Fe}{2} emission lines from the BLR. \cite{bg92} also reported that EW of [\ion{O}{3}] and the luminosity of [\ion{O}{3}] are both correlated with Eigenvector 1. They suggested that a weaker Eigenvector 1 results when the Eddington ratio (L/L$_{\mathrm{Edd}}$) is larger. Our sample does not contain reliable \ion{Fe}{2} measurements. We therefore use the [\ion{O}{3}]/H$\beta$ ratio instead to indicate the strength of Eigenvector 1.

\begin{figure*}
\centering
\includegraphics[width=3in]{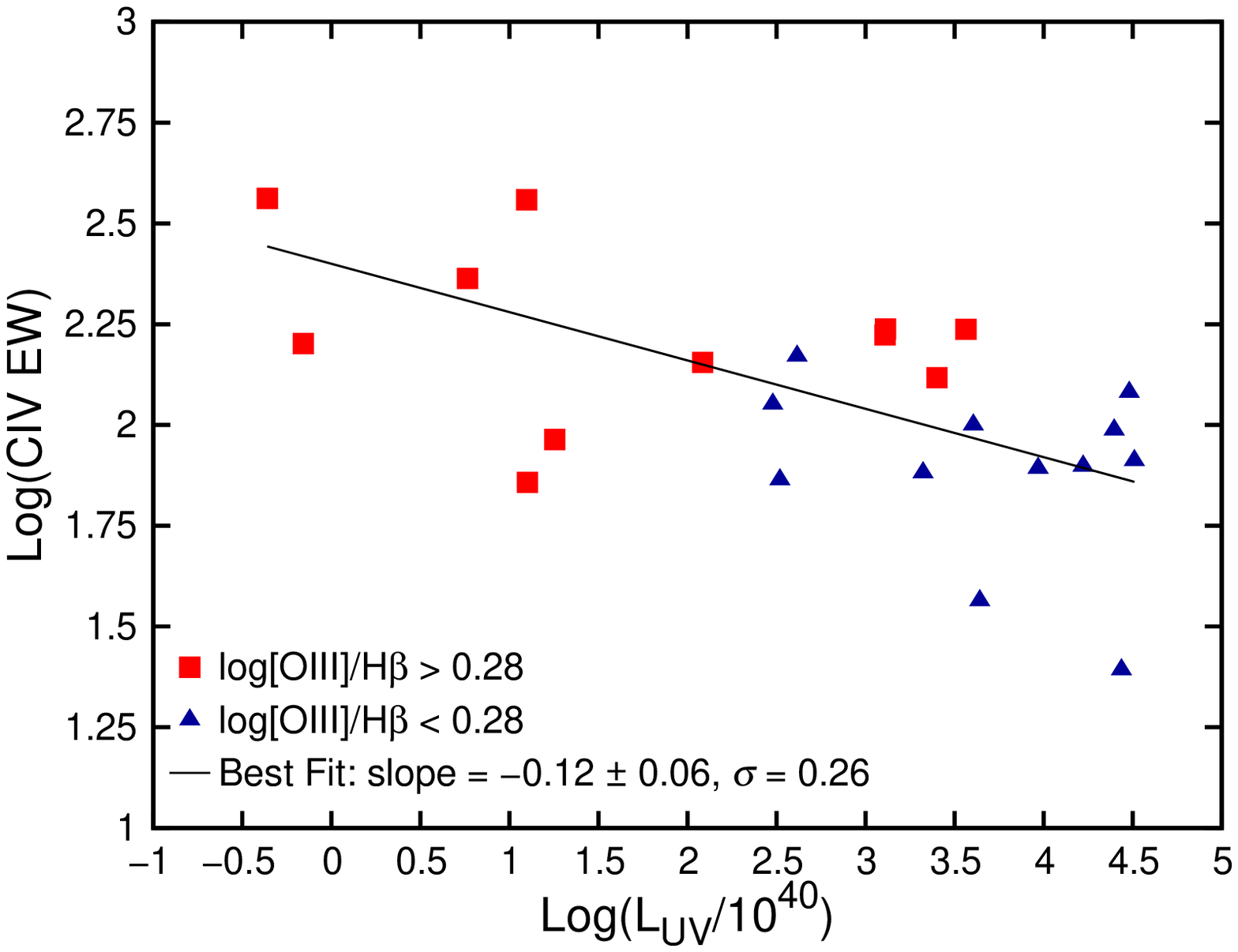}
\includegraphics[width=3in]{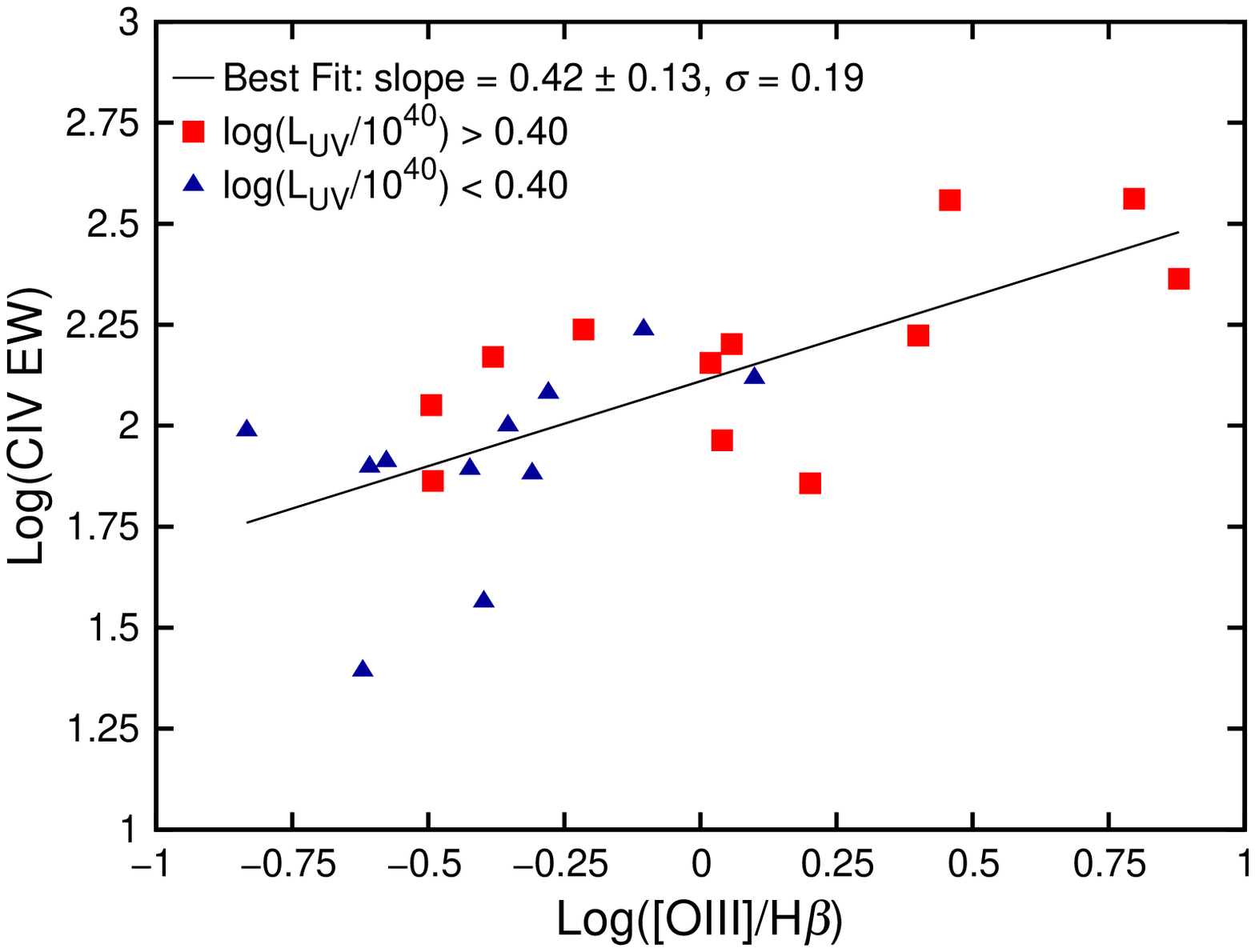}\\
\includegraphics[width=4in]{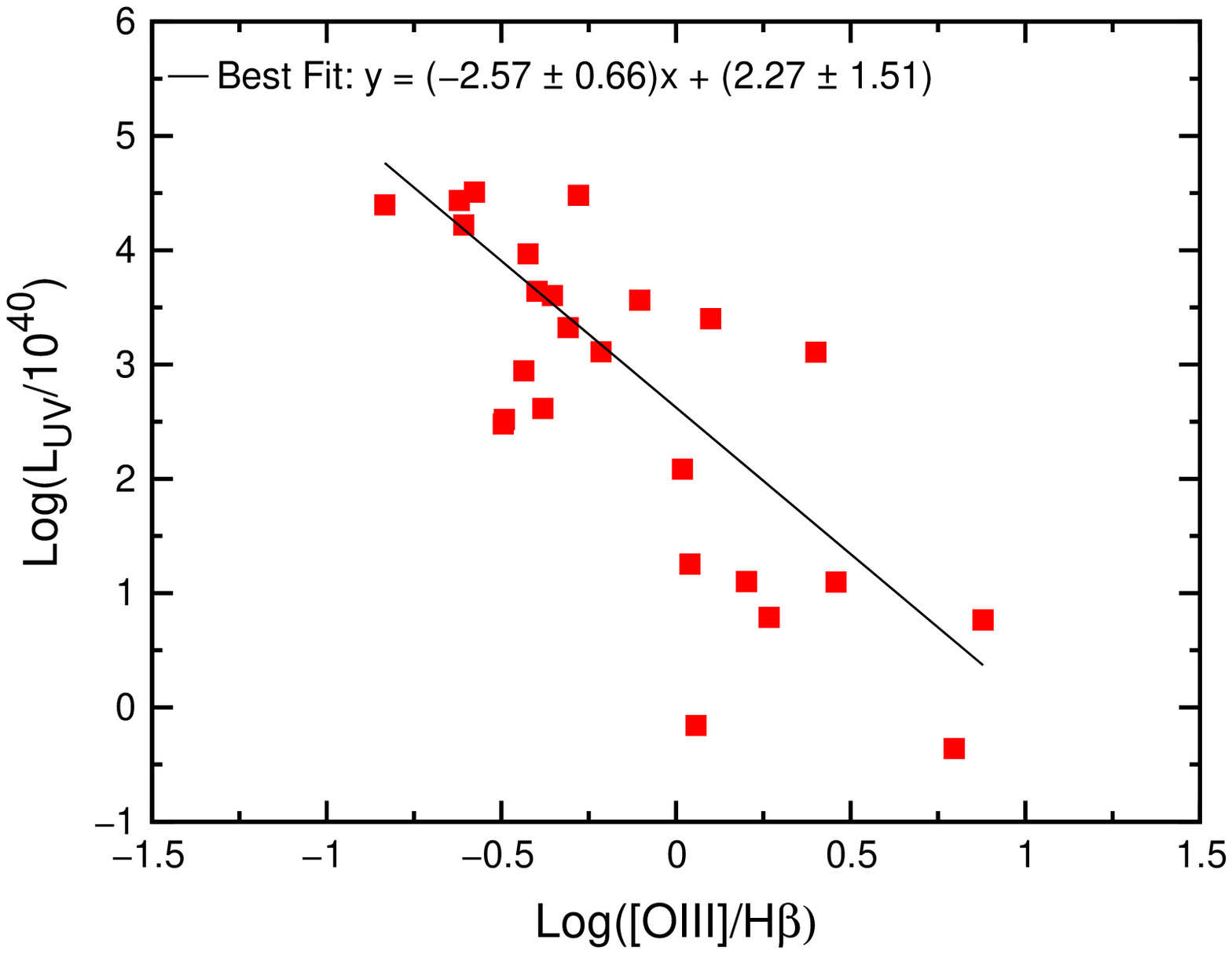}
\caption{\textit{Top Panels}: The classical Baldwin effect is displayed in the left panel and our ``substitute Baldwin relation" is shown in the right panel. The symbols for each 12 $\mu$m Seyfert 1 are coded by the strength of [\ion{O}{3}]/H$\beta$ ratio (left) and $L_{UV}/10^{40}$ (right). The strong segregation between the red and blue points in the left uppper panel indicates that [\ion{O}{3}]/H$\beta$ is a good predictor of EW(\ion{C}{4}). The classical Baldwin effect is weak, compared to log(\ion{C}{4} EW) vs. log([\ion{O}{3}]/H$\beta$) (right). The EW(CIV) is much better predicted from log([\ion{O}{3}]/H$\beta$). The $L_{UV}/10^{40}$ parameter hardly improves this. \textit{Bottom Panel}: The strong anticorrelation between $\log L_{UV}/10^{40}$ and log([\ion{O}{3}]/H$\beta$) suggests that the classical Baldwin relation is a secondary effect, resulting from this correlation.\label{Fig10}}
\end{figure*}

We coded the Seyfert galaxy points in Figure \ref{Fig10}a by their [\ion{O}{3}]/H$\beta$ ratio. Those with log([\ion{O}{3}]/H$\beta$) $> 0.28$ are shown as red squares; Seyferts with weaker [\ion{O}{3}]/H$\beta$ are shown as blue diamonds. The strong segregation between the red and blue points shows that the [\ion{O}{3}]/H$\beta$ is indeed a good a predictor of EW(\ion{C}{4}). To find out how good, we plot log(\ion{C}{4} EW) versus log([\ion{O}{3}]/H$\beta$) in Figure \ref{Fig10}b. It shows a significant positive correlation:
\begin{eqnarray}\label{eqn7}
\log(C IV EW)=
(0.42\pm0.1)\log\left(\frac{[OIII]}{H\beta}\right) + (2.11\pm0.27).
\end{eqnarray}
with a $\chi^2/dof = 1.11$ for 22 degrees of freedom. This new ``substitute Baldwin relation" works much better than the original ($\chi^2/dof = 1.48$, 21 degrees of freedom). We can also see this in Figure \ref{Fig10}b by the color coding. Seyferts with log(L$_{UV}$) $> 43.15$ as blue diamonds, while the less luminous galaxies are shown as red squares. The large overlap of red and blue points in this graph demonstrates that L$_{UV}$ is not so good predictor of EW(\ion{C}{4}) as is [\ion{O}{3}]/H$\beta$. 
Our finding of a positive correlation between
EW (CIV) and Eigenvector 1 in Seyfert 1 nuclei
is fully consistent with Baskin and Laor's
finding of a similar correlation in the PG Quasars \citep{bl04}.

Could both L$_{UV}$ and [\ion{O}{3}]/H$\beta$ be combined to make an even better Baldwin relation?
We made a bivariate least-squares fit:
\begin{eqnarray}\label{eqn6}
\log(C IV EW) = (2.16\pm0.1)
+ (-0.02\pm0.04)\log\left(\frac{L_{UV}}{10^{40}}\right)
+ (0.37\pm0.12)\log\left(\frac{[OIII]}{H\beta}\right)\nonumber \\ 
\end{eqnarray}
The  coefficient of the UV luminosity is only slightly negative, not significantly different 
from zero. In contrast, the coefficient of log([\ion{O}{3}]/H$\beta$) differs from zero by $3\sigma$, indicating that the Baldwin effect may be tightened by either adding this optical line ratio or replacing L$_{UV}$ altogether. We use the Bayesian Information Criterion difference ($\Delta BIC$) to compare the bivariate fit in Eqn. \ref{eqn6}. The parameters from the original Baldwin effect (Eqn. 4) give a $BIC=96.93$ and the new ``substitute Baldwin relation" gives $BIC= 89.11$. The inclusion of the [\ion{O}{3}]/H$\beta$ in the Baldwin relation give $\chi^2/dof = 1.15$ and $BIC= 85.79$. The $\Delta BIC$ between $L_{UV}$ and the bivariate fit is $>10$.  Thus the original Baldwin relation, using $L_{UV}$ is significantly inferior. The $\Delta BIC$ obtained when adding $L_{UV}$ to the regression (Eqn. \ref{eqn6}) is 3.32, which is considered only marginal evidence in favour of adding $L_{UV}$. Evidently, Eigenvector 1 may be a more fundamental driver of the Baldwin effect. Perhaps L$_{UV}$ is only a secondary parameter--it might correlate inversely with EW(CIV) merely because it correlates inversely with log([\ion{O}{3}]/H$\beta$). We graph this relation in Figure \ref{Fig10}c, and indeed there is a strong anticorrelation: log(L$_{UV}$)$ = (-2.57 \pm 0.66$)log([\ion{O}{3}]/H$\beta) + $($2.57\pm 1.51$) (Kendall's Tau $>$ 99\% significance).

\begin{figure*}
\centering
\includegraphics[angle=0,scale=1.0,width=6in]{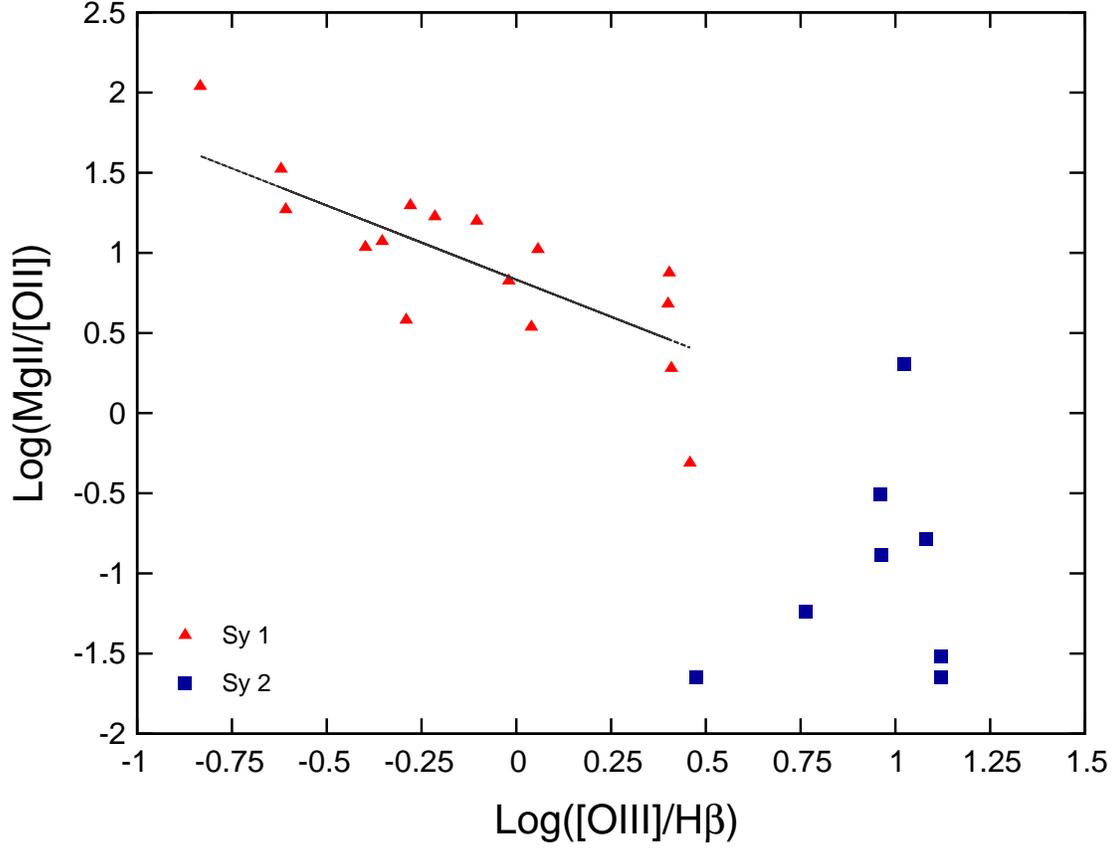}
\caption{Plot of \ion{Mg}{2}/[\ion{O}{2}] vs. [\ion{O}{3}]/H$\beta$. The best fitting line for the Sy 1 galaxies (red triangles) is log(\ion{Mg}{2}/[\ion{O}{2}])=$(-0.93\pm 0.12)$log([\ion{O}{3}]/$H\beta) + 0.84 \pm 0.32$. We use [\ion{O}{3}]/H$\beta$ as a proxy for weak \ion{Fe}{2} (Eigenvector 1), so an inverse correlation for the Sy 1's implies that \ion{Mg}{2} and \ion{Fe}{2} are correlated. The Seyfert 2 galaxies (blue squares) have no BLR and therefore have much lower \ion{Mg}{2}/[\ion{O}{2}], and higher [\ion{O}{2}]/H$\beta$ ratios. \label{Fig11}}
\end{figure*}

Our data also show a weak positive correlation between EW(\ion{C}{3}]1909) and [\ion{O}{3}]/H$\beta$ . However, there is so much scatter that, unlike with the EW(\ion{C}{4}), this does not provide a meaningful Baldwin relation. We also plot log(\ion{Mg}{2}/[\ion{O}{2}]) versus log([\ion{O}{3}]/H$\beta$) in Figure~\ref{Fig11}. The solid black line shows the fit to the broad-line objects only (Sy 1's are indicated by red traingles). The best fit slope is very significantly non-zero ($-0.93\pm0.12$), i.e. the anti-correlation is significant at the 99\% level. Since we use [\ion{O}{3}]/H$\beta$ as our proxy for ``Eigenvector 1", this anti-correlation shows that the relative strength of \ion{Mg}{2} decreases in strong ``Eigenvector 1" Seyfert 1 galaxies. Although we lack sufficient \ion{Fe}{2} data for our 12$\mu$m sample, this suggests the UV \ion{Mg}{2} and optical \ion{Fe}{2} emission line strengths should be closely correlated. The average log(\ion{Mg}{2}/[\ion{O}{2}]) ratio in the NLR alone (Seyfert 2's) is $-0.99 \pm 0.63$.

\section{Conclusions} \label{Conclude}
We have shown that the narrow-line regions of Seyfert 1's and Seyfert 2's galaxies have little systematic difference in many properties:

\begin{enumerate}
\item The luminosity functions of the narrow lines are the same for both types of Seyferts. Therefore the space densities of Seyfert 1 and Seyfert 2 galaxies are roughly equal.
(Only the H$\alpha$ and H$\beta$ luminosity functions differ in Sy 1's at higher luminosities.)

\item
Measured by two independent emission-line ratios (H$\alpha$/H$\beta$ and [\ion{S}{2}]/[\ion{O}{2}]), the Sy 2's are not more reddened than the Sy 1's. This indicates that the amount of dust present in the narrow line region in both types of Seyferts is similar. Nor do the Sy1's show a significantly higher ionization in their NLRs, also consistent with the  premise of simple geometric unification. We have demonstrated that a value of $-0.25 \pm 0.16$ can be used for log([\ion{S}{2}]/[\ion{N}{2}]), regardless of whether the galaxy is a type 1 or 2 Seyfert. This can be useful in determining the flux of [\ion{N}{2}] when it is blended in H$\alpha$.

\item We identify several ratios indicative of the ratio of the AGN luminosity to that of the host galaxy (the ``Seyfert dominance"). For example the dust in Seyfert galaxies is warmer for those objects more dominated by their AGN contribution than by their starburst or \ion{H}{2} region contributions. And we have found that [\ion{O}{1}]$\lambda6300$\AA\; correlates most closely with the high-ionization lines, both being powered primarily by the AGN. In the BPT diagram, we find that 15 \% of our Seyfert galaxies would be classified by \cite{k03} as ``HII/AGN Composites". A further 5 \% have such a weak Seyfert nuclei that their spectra are not distinguishable from those of normal star-forming spiral galaxies.

\item By separating the Seyfert galaxies into sub-types ($1 + 1.2 + 1.5$ and $1.8 + 1.9$) we showed that Seyfert 1.8 and 1.9 galaxies have similar ionization levels to Sy 2's. The Sy 1.8 and Sy 1.9's have very weak broad lines, so their narrow line ratios are similar to that of Sy 2's. The BPT diagram demonstrates that these objects lie in the same area as Sy 2's as well. 

\item We use the BPT diagram to make a simple decomposition of emission line ratios into three ``pure" components --Sy1, Sy2, and LINERs. As expected, the relative importance of the Sy 1 component falls steadily from the Sy 1.2's to the Sy 1.5's, the Sy 1.8's, and finally the Sy 1.9's. Although AGNs show a continuous range in NLR/LINER ratios, we find that the LINER component needs to be $>50$\% for an AGN to be specroscopically classified as a LINER.

\item In the broad-line region we studied the Baldwin effect and found that \ion{C}{4} equivalent width correlates more strongly with [\ion{O}{3}]/H$\beta$, rather than with UV luminosity. This may imply the Baldwin effect is more strongly dependent on the Eddington ratio, $L/L_{Edd}$. An additional implication was that broad \ion{Mg}{2} emission correlates with \ion{Fe}{2}. 

\end{enumerate}

\acknowledgments
This paper has benefited greatly from the extensive valuable comments of an anonymous referee, whom we thank. This research has made use of the NASA/IPAC Extragalactic Database (NED) which is operated by the Jet Propulsion Laboratory, California Institute of Technology, under contract with the National Aeronautics and Space Administration.

Funding for the Sloan Digital Sky Survey (SDSS) has been provided by the Alfred P. Sloan Foundation, the Participating Institutions, the National Aeronautics and Space Administration, the National Science Foundation, the U.S. Department of Energy, the Japanese Monbukagakusho, and the Max Planck Society. 

\begin{appendix}

\section{APPENDIX A.  Emission Line Proxies for the Nonstellar Luminosity}\label{appxA}
Many previous studies pursued the goal of estimating the total non-stellar power of an AGN by simply measuring one of the strongest emission lines in its spectrum.  Any such ``shortcut", if reliable over a wide range of Seyfert galaxies, would be of value in analyzing large samples, where good data at all wavelengths may not be available.  However, the danger of correlating any two luminosities against each other in astronomical samples is well known. When a wide range of luminosities is present, as in our Seyfert sample, an apparent correlation of one luminosity
with another luminosity is often found, even when there may be little physical connection between the two quantities.
Correlations of line luminosities with continuum luminosities in the 12$\micron$ Seyfert sample
should be expected, and do not necessarily prove how they are linked.
 
Keeping this caveat in mind, we now examine these line/continuum correlations to determine the intrinsic scatter, possible non-linearities and systematic differences depending on which quantities and which AGN types are included.  Our dataset--a representative sampling of a very wide range of local AGN properties--has advantages for finding and testing various ``scaling relations" between emission lines and the broadband AGN continuum. 
The various AGN-powered luminosities we are seeking to correlate span almost six orders of magnitude in our Seyfert galaxies. Our estimates of non-stellar luminosities from the X-ray and UV to the IR are given in Table \ref{tbl9}. 
It is important to base correlations on a complete sample, where
the observed fluxes for nearly all the AGN are detections--not upper limits.

The key to all proposed scaling relations is finding easily measured quantities which are dominated by the non-stellar (AGN) component. We start with the most obvious non-stellar continuum -- hard X-rays -- since they are thought to be produced almost entirely by the central engine of the AGN, except at the lowest luminosities. 
As \cite{BN11a} and others point out, it is very difficult for any galaxy lacking a Seyfert nucleus to produce (through normal stellar processes, including X-ray binaries) more than $10^{40}$ erg/sec in hard X-rays.
In Figures ~\ref{Fig12}a to ~\ref{Fig15}a (top left in these panels) we compare the luminosities of several of the strongest optical emission lines with the hard X-ray luminosities ($L_{HX}$) reported by \cite{BN11a, BN11b, P08}.

In the Sy 1's, $L_{HX}$ is reasonably well correlated with $L_{H\alpha}$. The plotted lines assumed a linear correlation, i.e., fixing the slope in log(line) versus log(continuum) to be 1.  There are no cases when a deviation from this linear slope gives a statistically superior fit. The result that $L_{HX} \sim 15 L_{H\alpha}$  is not surprising, since it has long been known that the broad Balmer emission line luminosities are closely correlated with the non-stellar continuum \citep{Y80, MS82}. As summarized in Table ~\ref{tbl10}, we find an individual scatter of a factor of 3 (solid blue squares in Figure \ref{Fig13}a).  This is far from perfect, but might be useful in cases where only a rough individual estimate, or average of a sample, is wanted.

A limitation of that relation is that a high-quality spectrum of the H$\alpha$ region may not be available, so it is sometimes unclear whether an AGN should be classified as Sy 1 or Sy 2.
Then this correlation runs into trouble, since Sy 2's lack any directly detectable broad H$\alpha$ emission. They produce more hard X-rays for a given (narrow) H$\alpha$ luminosity, and the scatter is so large it nearly destroys the HX/H$\alpha$ correlation in Sy 2's.

To include both Sy 1's and Sy 2's in one single AGN correlation, we
have followed previous studies, such as \cite{tom10}. 
We considered whether the strongest emission from the NLR -- the [\ion{O}{3}]$\lambda$5007\AA\;line --
could instead serve as a proxy to measure the non-stellar luminosity \citep{M94,D08, goto11}.
Indeed, the blue symbols in Figures ~\ref{Fig14}a and ~\ref{Fig15}a do show that the [\ion{O}{3}] luminosity can {\it roughly} predict the HX luminosity,
with an uncertainty of 4 times for Sy 1's and 10 times for Sy 2's.  For a given $\lambda$5007 line luminosity, the Sy 1's
tend to be 30\% brighter in hard X-rays.  That is a small difference compared with the large intrinsic scatter. 

Our $L_{\mathrm{HX}}/L_{\mathrm{[OIII]}}$ correlation for the Sy 1's has nearly the same normalization and scatter as \cite{hk05} found  for local AGN selected by \ion{O}{3}$\lambda5007$\AA\;($L_{\mathrm{HX}}/L_{\mathrm{[OIII]}} =10^{1.64}$ compared with our $L_{\mathrm{HX}}/L_{\mathrm{[OIII]}} =10^{1.59}$). This is also the same correlation found by \cite{X99} ($L_{\mathrm{HX}}/L_{\mathrm{[OIII]}}=10^{1.60}$), as well as \citep{SL12}, (their Equation 6). We find that the Sy 2's have relatively weaker hard X-rays and larger scatter in the $L_{\mathrm{[OIII]}}$ correlation. But this Sy 1/2 difference is larger in the \cite{H05} sample, which includes a substantial tail of Sy 2's with very weak hard X-rays ($log(L_{\mathrm{[OIII]}}) = 39-41$). \cite{H05} points out that the X-ray selected AGN will have relatively larger $L_{\mathrm{HX}}/L_{\mathrm{[OIII]}}$ ratios than our sample selected at longer wavelengths. Indeed this trend is seen in the luminous AGN sample found in Chandra Deep Field South Survey at $0.3 < z < 0.8$. The best fit correlation for AGN, found by \cite{N06} is shown by the dashed line in Figures ~\ref{Fig14}a and ~\ref{Fig15}a. In contrast, up to $L_{\mathrm{HX}}\sim 10^{44}$ we see no evidence of non-linearities in the correlation with any of the emission-line fluxes.

But we can improve on using $L_{\mathrm{OIII}}$ as a general measure of AGN power. We are motivated by the Stern and Laor studies (papers II and III) of 3175 SDSS spectra of Sy 1 galaxies. \citep{SL12} showed that the luminosity of the broad component of H$\alpha$ increase more rapid then the luminosities of any of the narrow lines. We propose to account for this simply by adding measures of NLR and BLR luminosities together. To include Seyferts of Type 1 and Type 2, we favor a ``hybrid" indicator of non-stellar luminosity, the sum of the [\ion{O}{3}]  and H$\beta$ lines - shown by
the open symbols in Figures ~\ref{Fig14}a and ~\ref{Fig15}a.  This empirical compromise  captures the AGN luminosity in Sy 2's emerging in the NLR, but also the BLR luminosity in Sy 1's. And as shown by Table \ref{tbl9}, one single relation, $L_{HX} \sim 25 L_{([OIII] + H\beta)} $, gives a rough estimate for any Seyfert galaxy, regardless of type.

The usefulness of this hybrid (NLR+BLR) AGN luminosity indicator lead us to also consider correlations with the blend of H$\alpha$+[\ion{N}{2}] (open symbols in Figures ~\ref{Fig13}a and ~\ref{Fig14}a), which can be difficult to disentangle at low spectral resolution.  The correlation with HX shows an extreme amount of scatter for the Sy 2's, but is still consistent with the same (tighter) correlation we find for
Sy 1's.  We therefore consider the combined H$\alpha$+[\ion{N}{2}] line luminosity a possible ``backup" predictor of non-stellar hard X-rays, which is also independent of Seyfert type:   $L_{HX} \sim 11 L_{([NII] + H\alpha)} $. Since the broad H$\alpha$ and H$\beta$ in Seyfert 1 galaxies vary with time, the non-simultaneity of our spectroscopy with the X-ray observations introduces some artificial scatter into these diagrams. However, the amplitude of variability is usually small enough that this increase in scatter is small compared to what is observed in \cite{R16}.

\subsubsection{Less Reliable Continuum Proxies for the total Nonstellar Luminosity}
\begin{figure*}
\centering
\includegraphics[angle=0,scale=0.47]{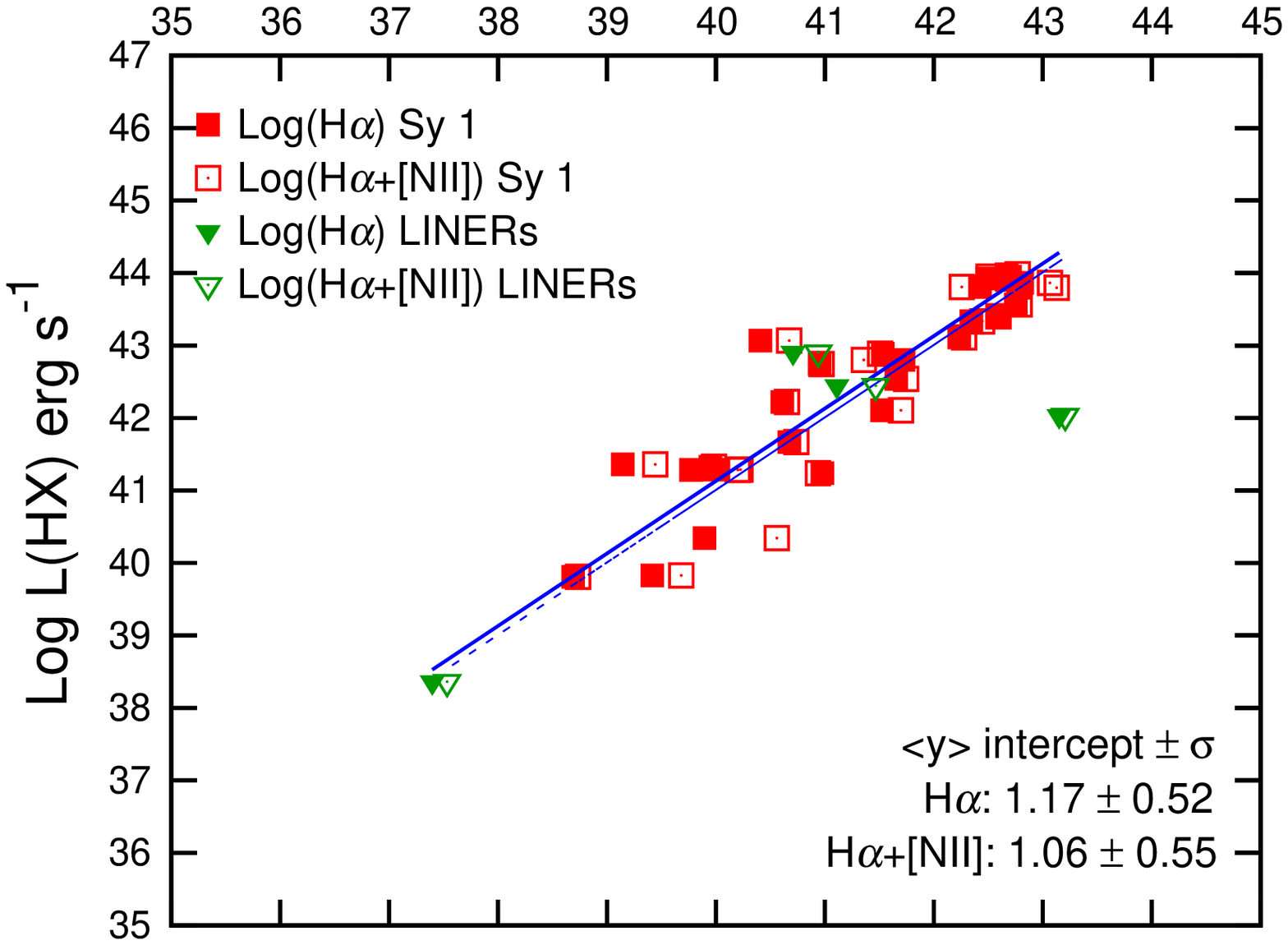}
\includegraphics[angle=0,scale=0.47]{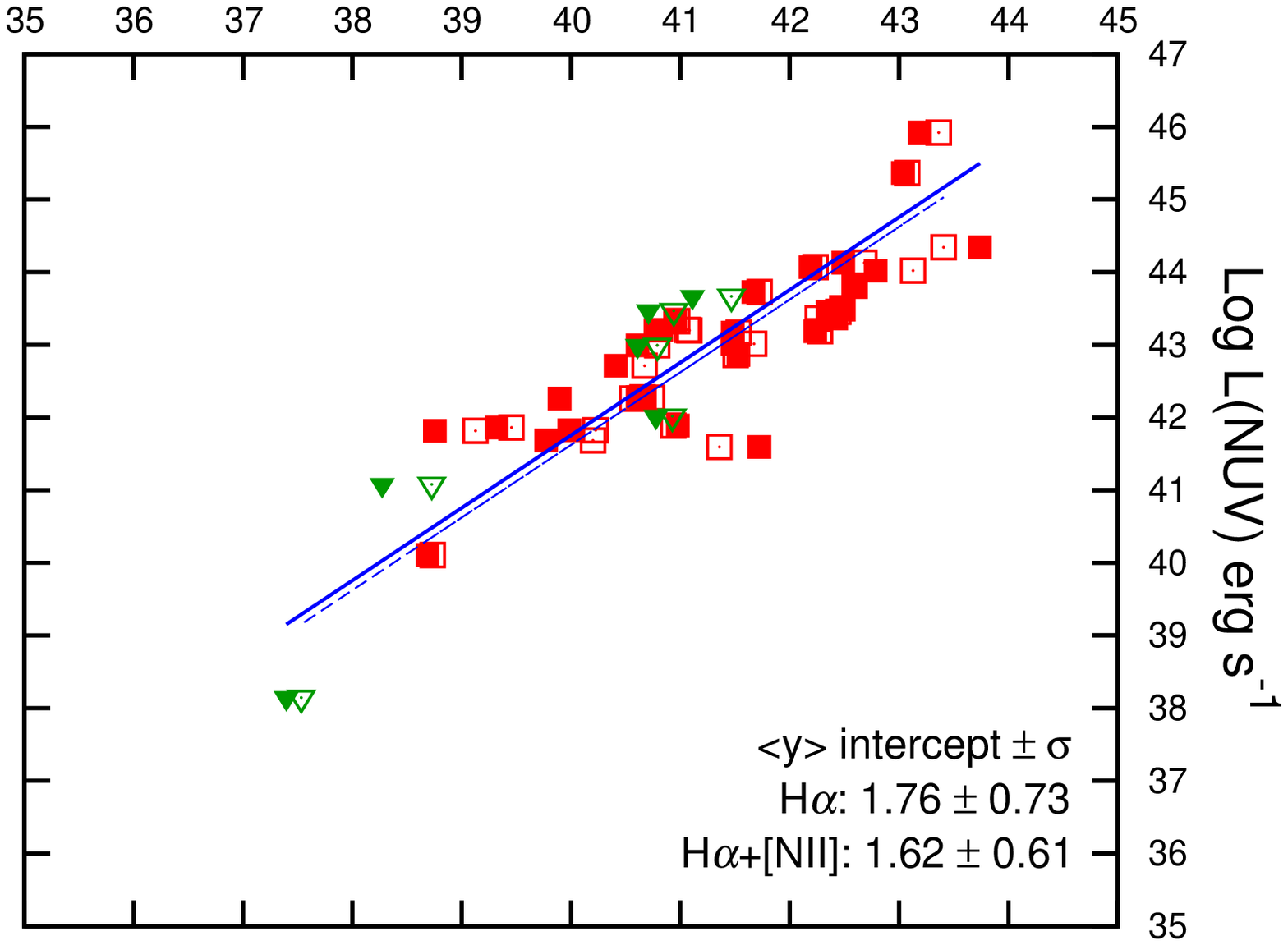}\\
\includegraphics[angle=0,scale=0.47]{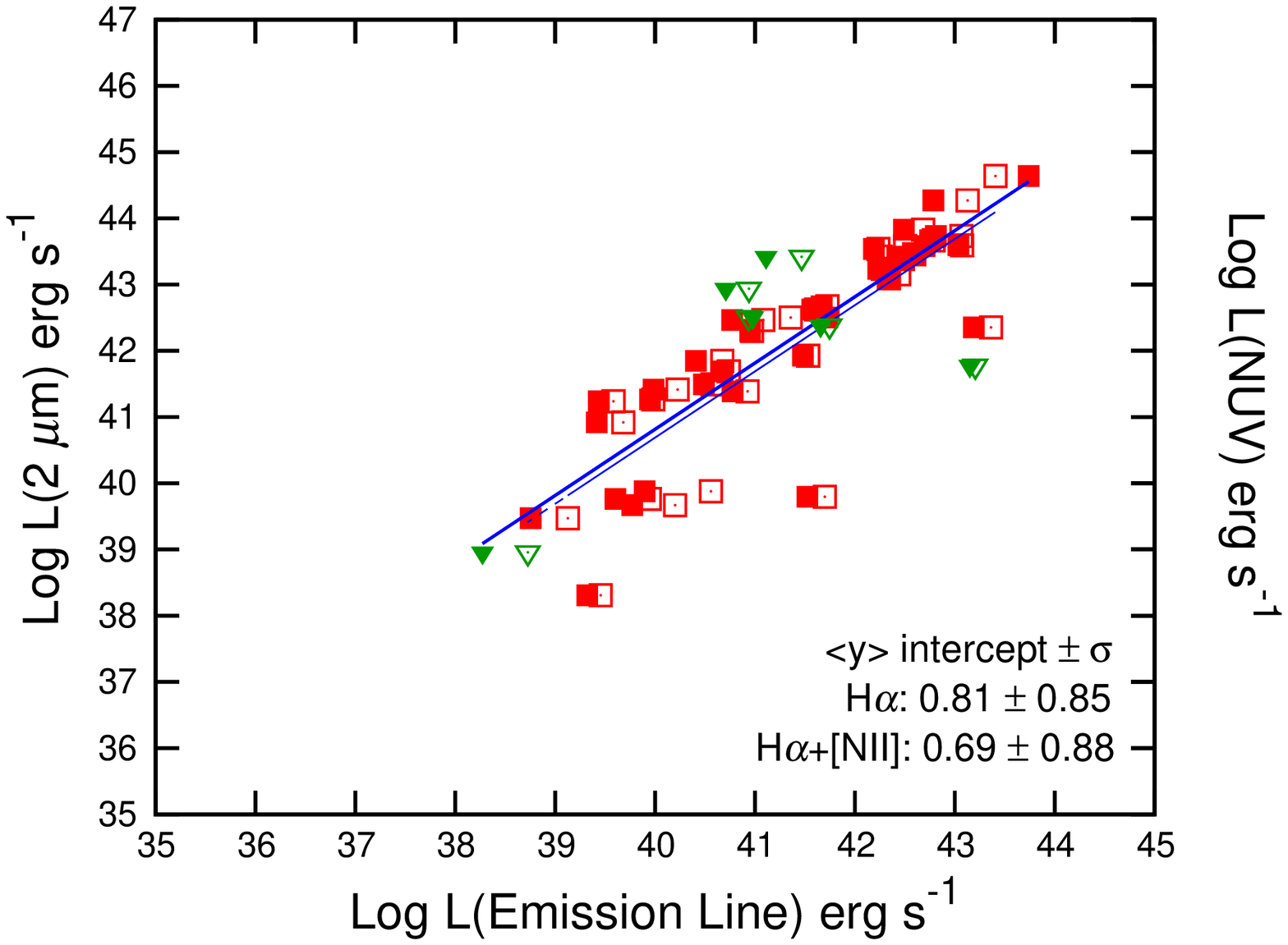}
\includegraphics[angle=0,scale=0.47]{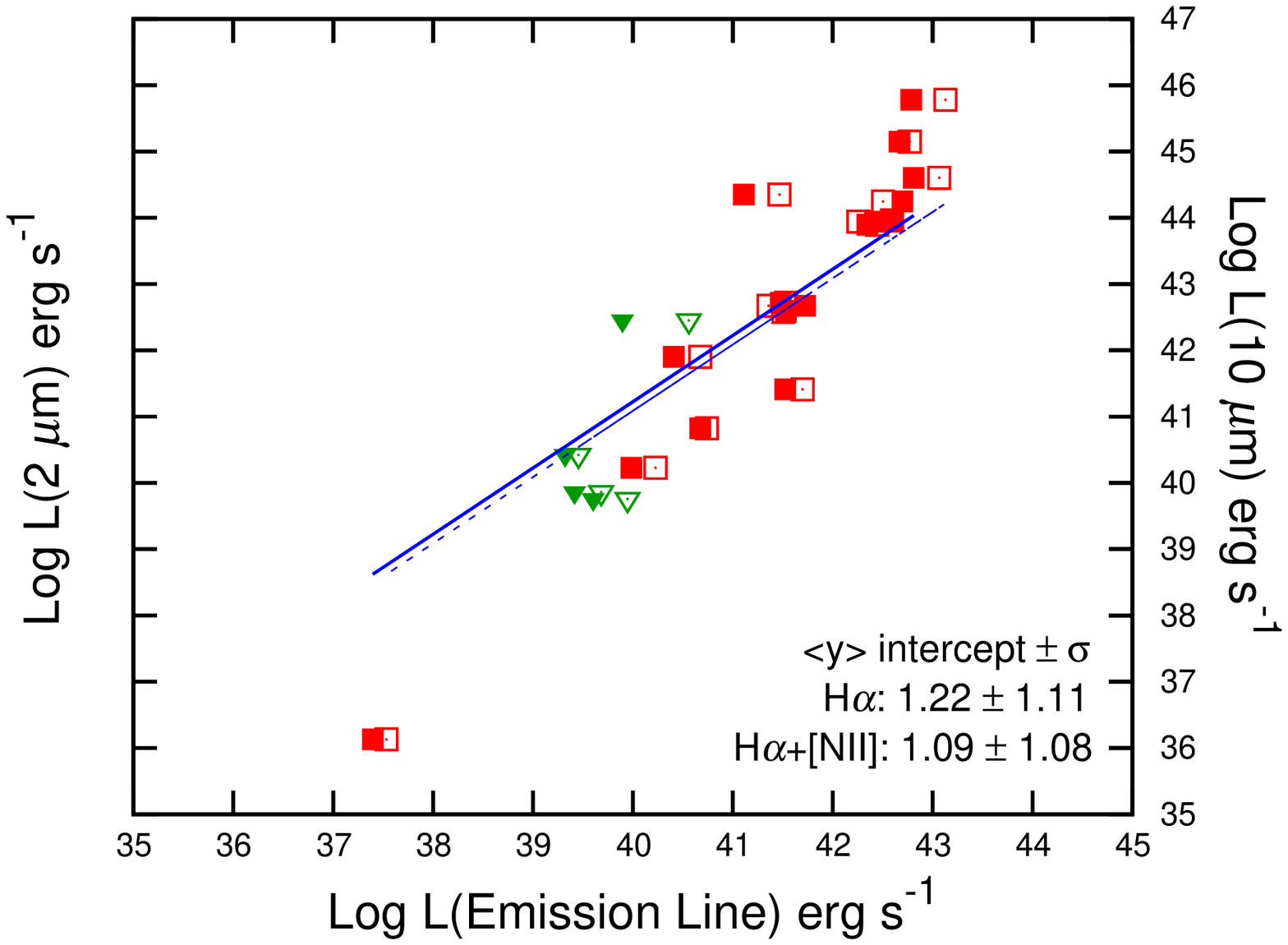}
\caption {Correlation of non-stellar continuum luminosity with emission line luminosity, either with H$\alpha$ (filled in squares) or with H$\alpha$+[\ion{N}{2}] (open squares), for Seyfert 1 galaxies. The filled and open triangles (in green) show observations of 12 $\mu$m Sy1's which were also classified as either LINERs or Starbursts. The plotted lines are fits of $\log(L_{Continuum}) = A\log(L_{Emission\;Line}) + B$ with a fixed slope of $A=1$. The tightest correlation is for the $HX$ luminosities, with  $L_{HX}\sim 15L_{H\alpha}$ for the Sy 1's and $L_{HX}\sim 11.5L_{(H\alpha+[NII])}$, independent of Seyfert type. Those Sy 1 which have secondary classification as LINERs or Starbursts are expected to harbor relative fainter Syfert nuclei. Nonetheless, they do not deviate systematically from the correlation defined by the more AGN-dominated Sy 1's line/continuum.\label{Fig12}}
\end{figure*}

\begin{figure*}
\centering
\includegraphics[angle=0,scale=0.47]{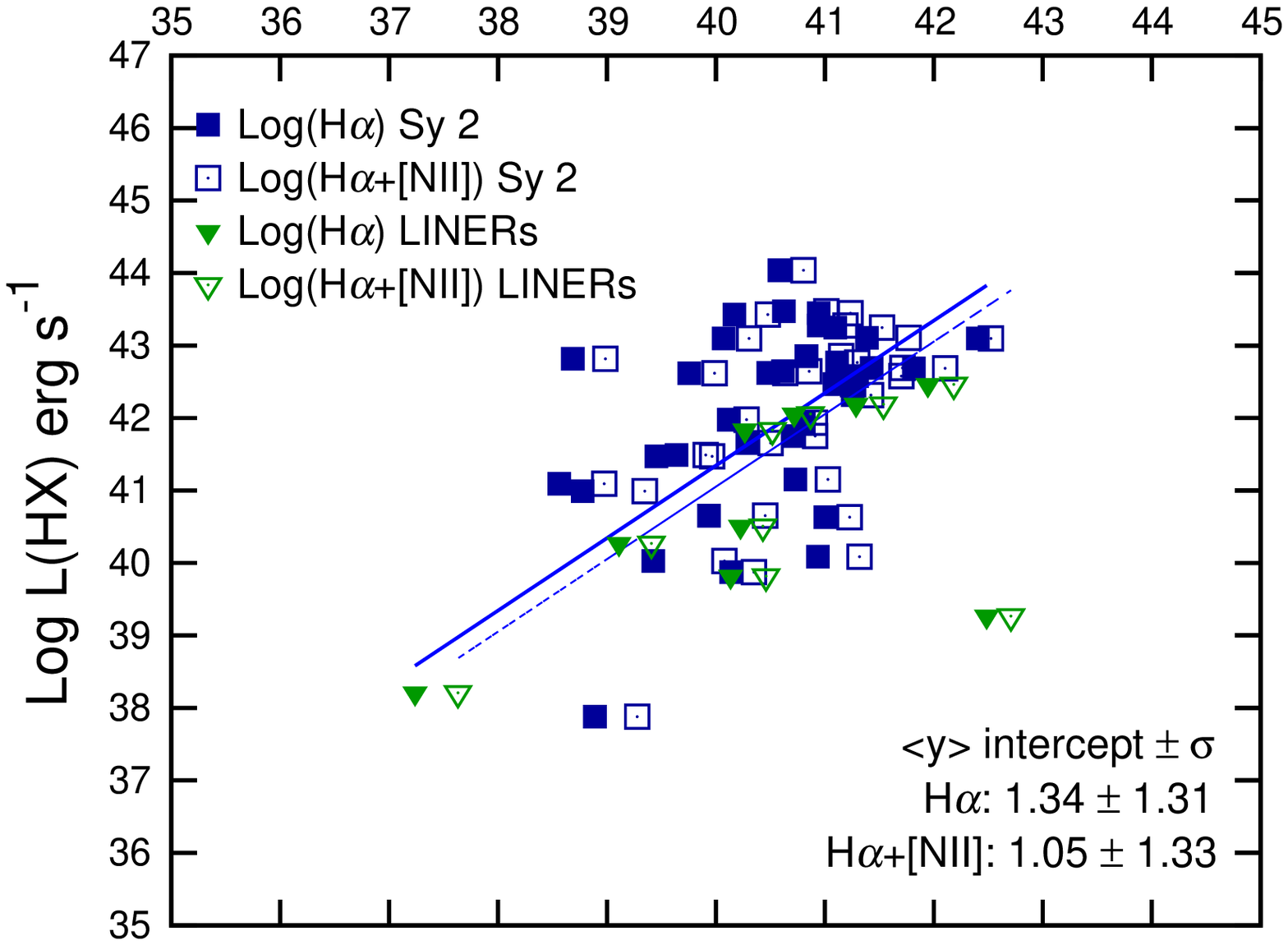}
\includegraphics[angle=0,scale=0.47]{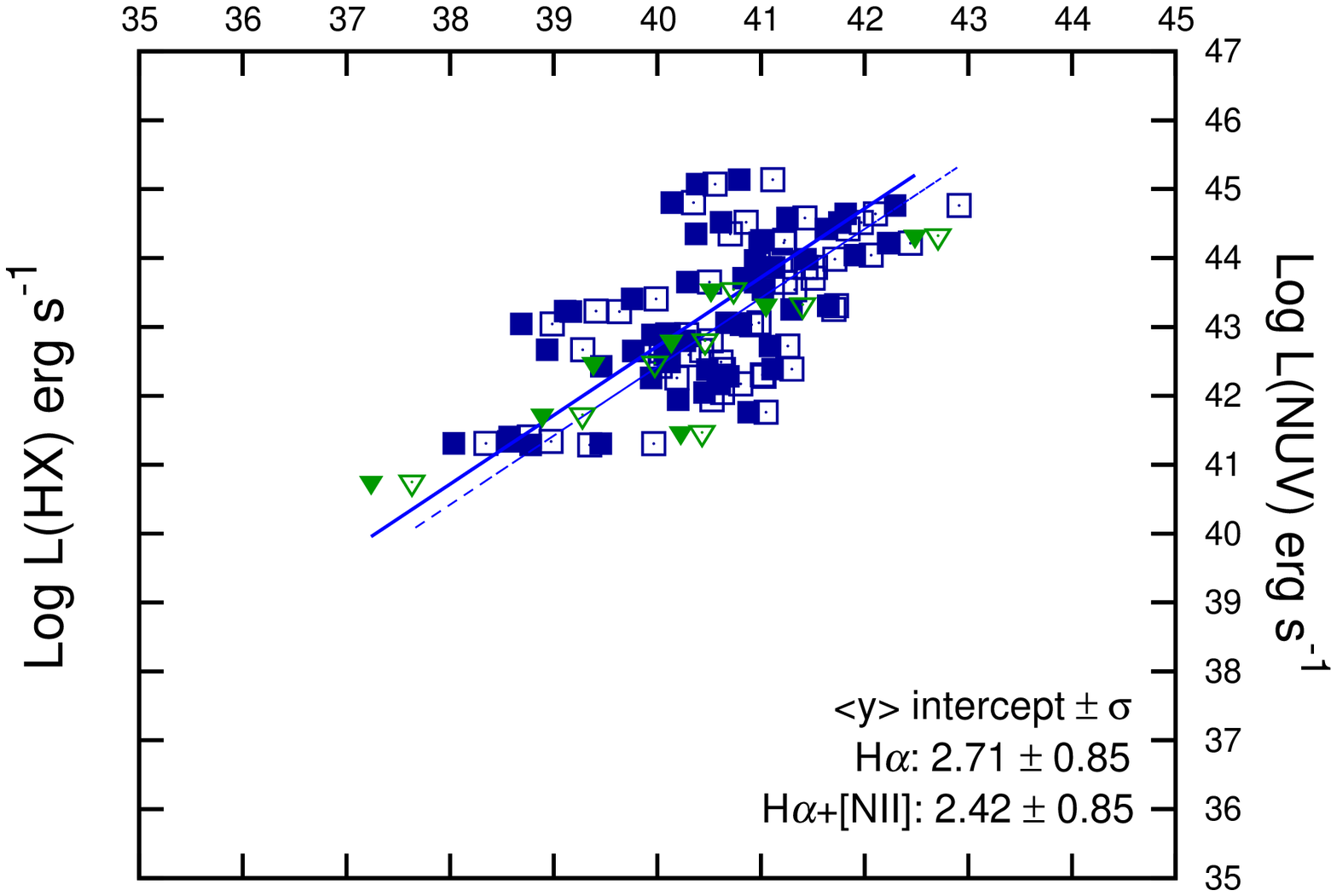}\\
\includegraphics[angle=0,scale=0.47]{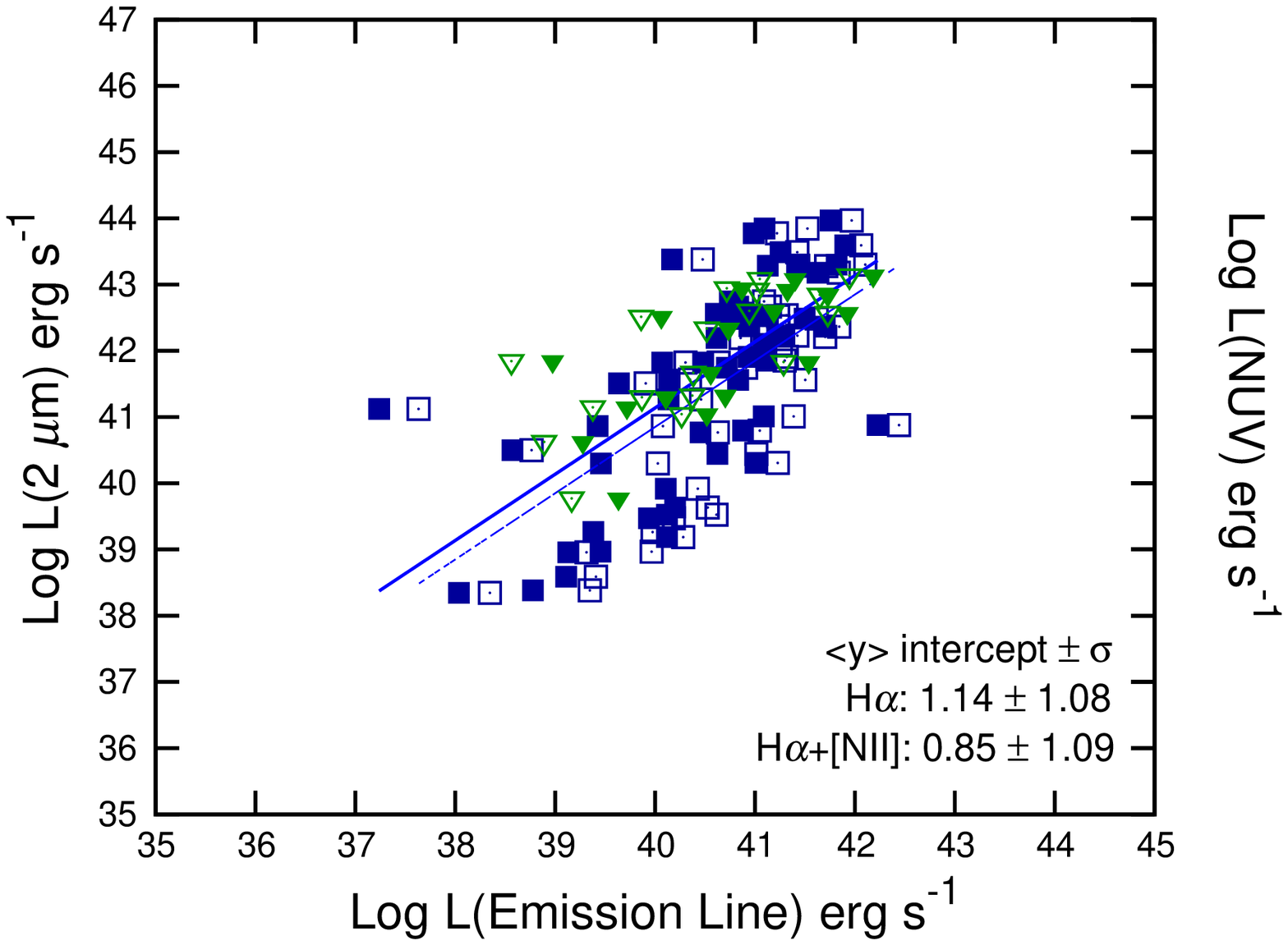}
\includegraphics[angle=0,scale=0.47]{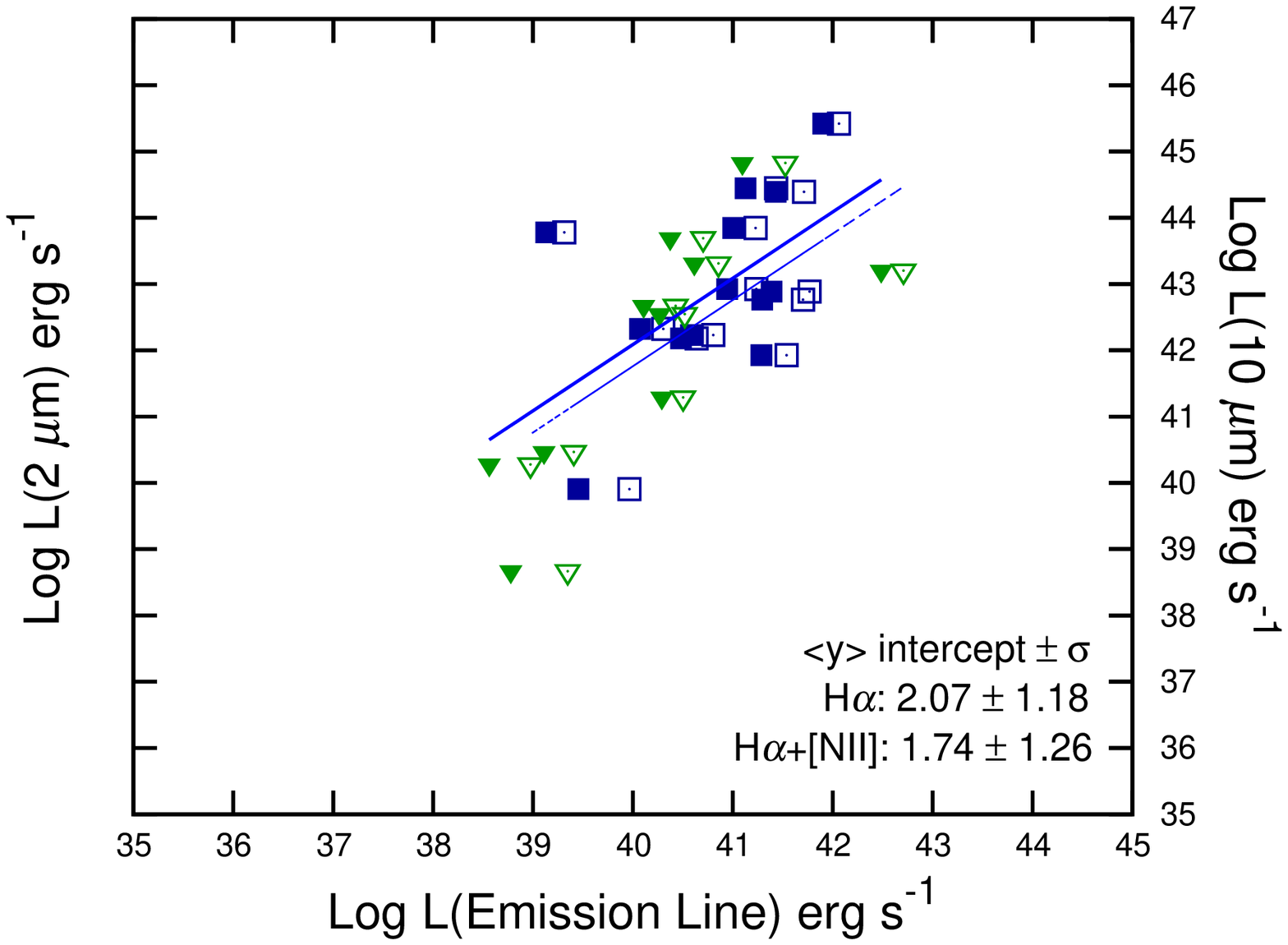}
\caption {Same non-stellar continuum/emission line correlation as shown in Figure~\ref{Fig12}, except the Seyfert 2 galaxies. The tightest correlation is for the $HX$ luminosities, with  $L_{HX}\sim 22L_{H\alpha}$ for the Sy 2's and  $L_{HX}\sim 11.5L_{(H\alpha+[NII])}$, independent of Seyfert type. 
\label{Fig13}}
\end{figure*}

\begin{figure*}
\centering
\includegraphics[angle=0,scale=0.47]{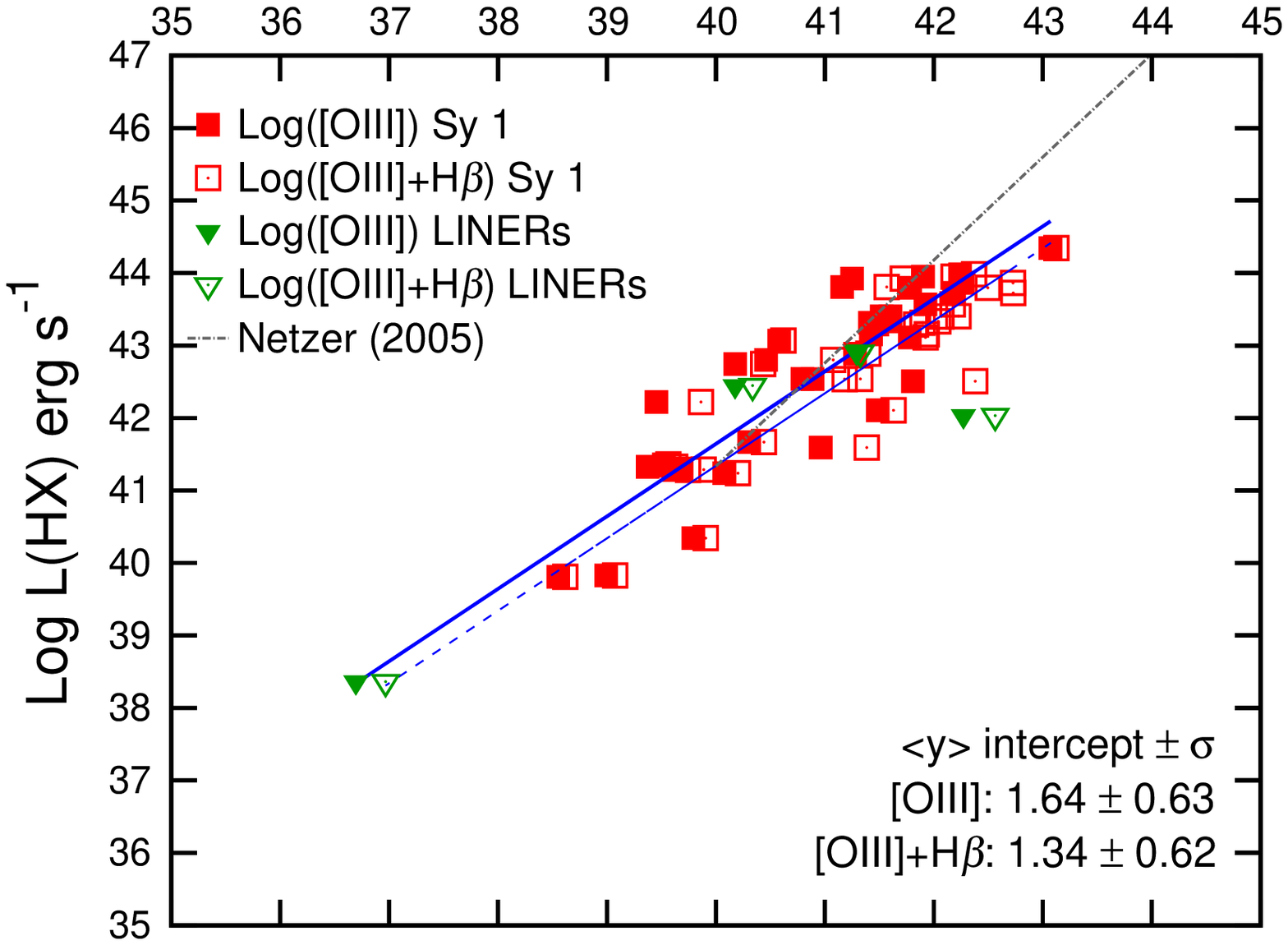}
\includegraphics[angle=0,scale=0.47]{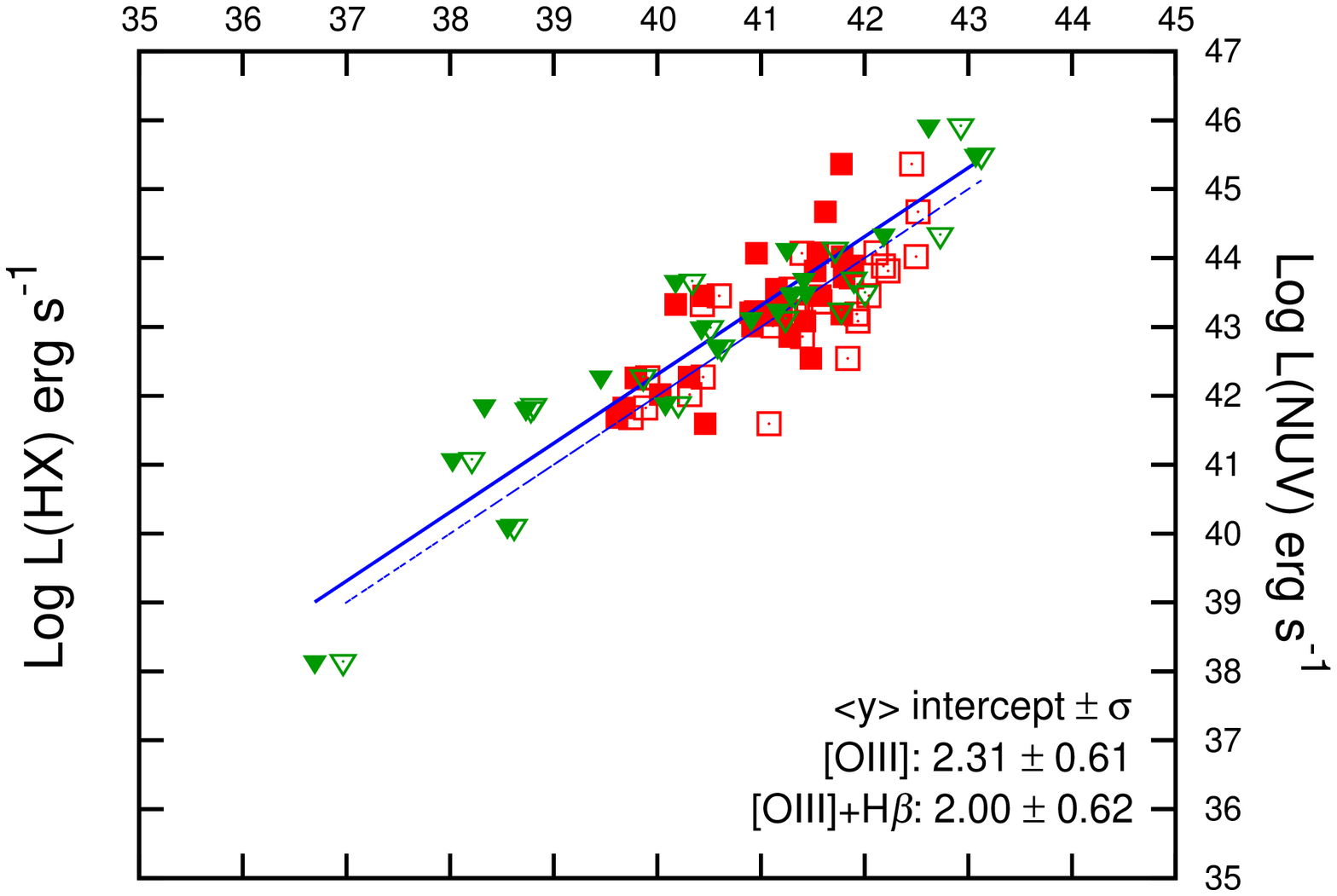}\\
\includegraphics[angle=0,scale=0.47]{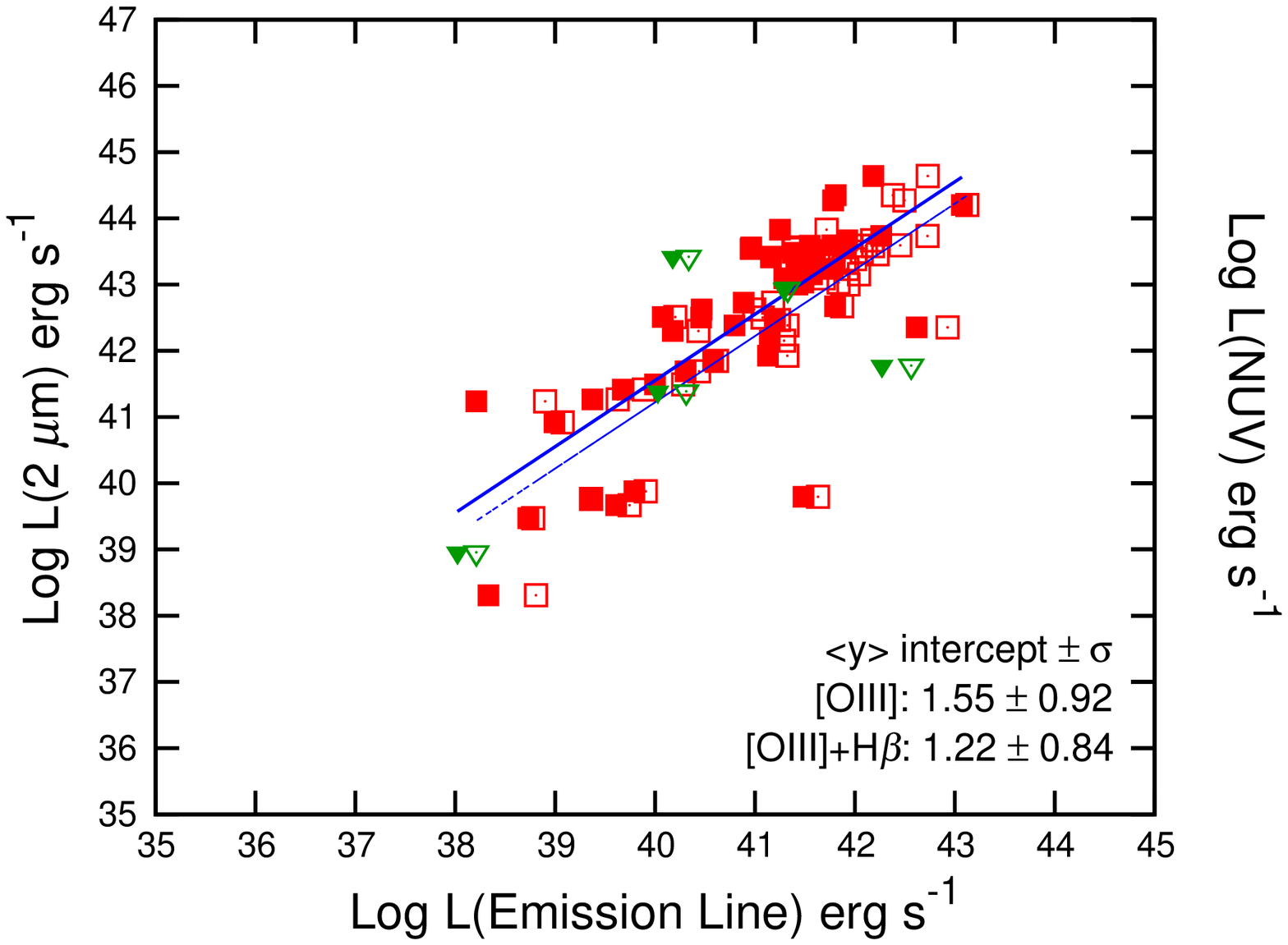}
\includegraphics[angle=0,scale=0.47]{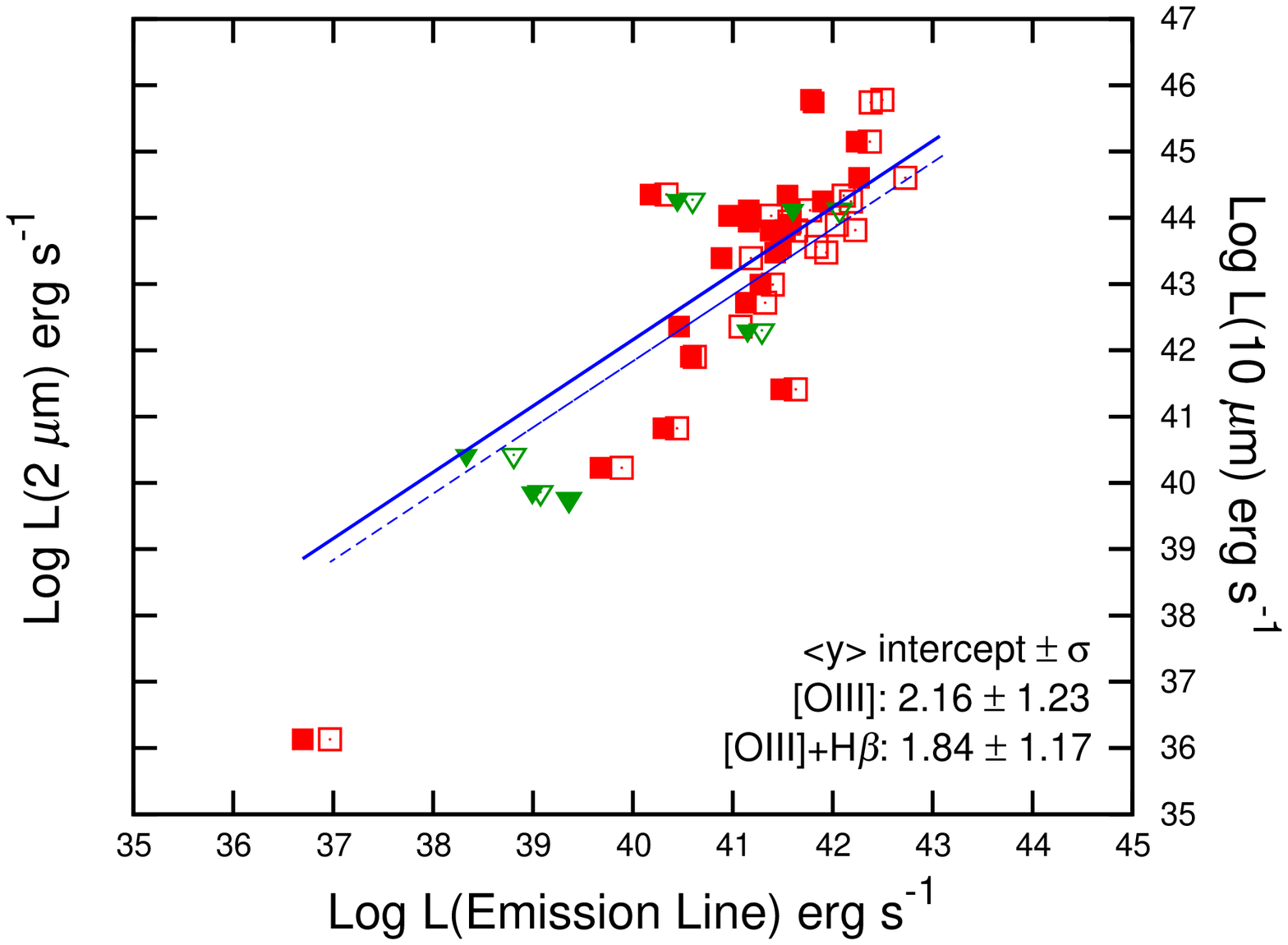}
\caption {Same non-stellar continuum/emission line correlation as shown in Figure~\ref{Fig12}, but for [\ion{O}{3}] and [\ion{O}{3}]+H$\beta$ on the $x$-axis, for Seyfert 1 galaxies. The plotted lines are fits of $\log(L_{Continuum}) = A\log(L_{Emission\;Line}) + B$ with a fixed slope of $A=1$.  The best correlation is for the $HX$ luminosities, with a single relation of $L_{HX}\sim 25L_{([OIII]+H\beta)}$, and with $L_{HX}\sim 44L_{[OIII]}$. The non-linear relation between the $L_{HX}$ and $L_{(Emission\;Line)}$ found by \cite{N06} is indicated by the steep solid line in the upper left panel.
\label{Fig14}}
\end{figure*}

\begin{figure*}
\centering
\includegraphics[angle=0,scale=0.47]{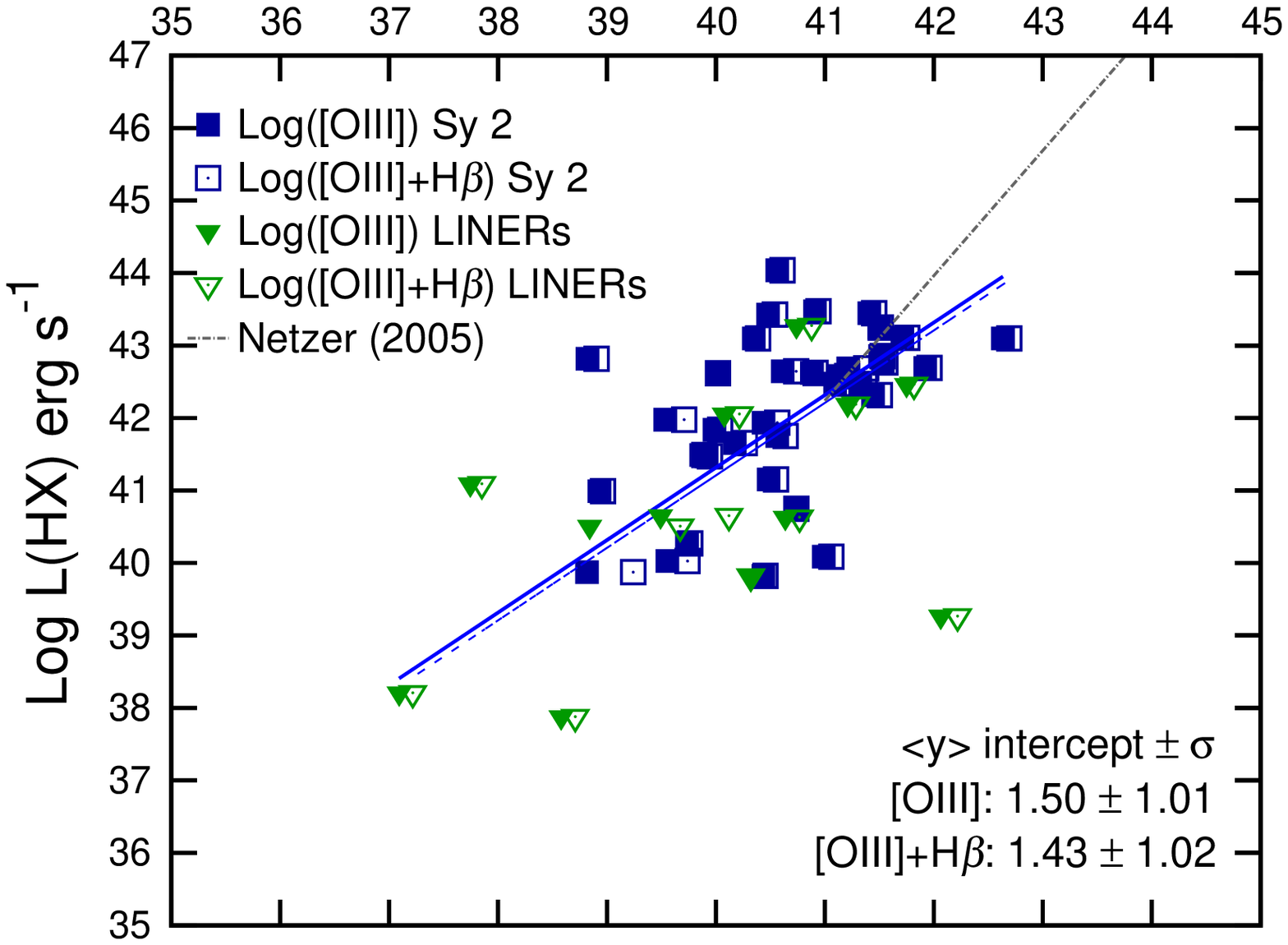}
\includegraphics[angle=0,scale=0.47]{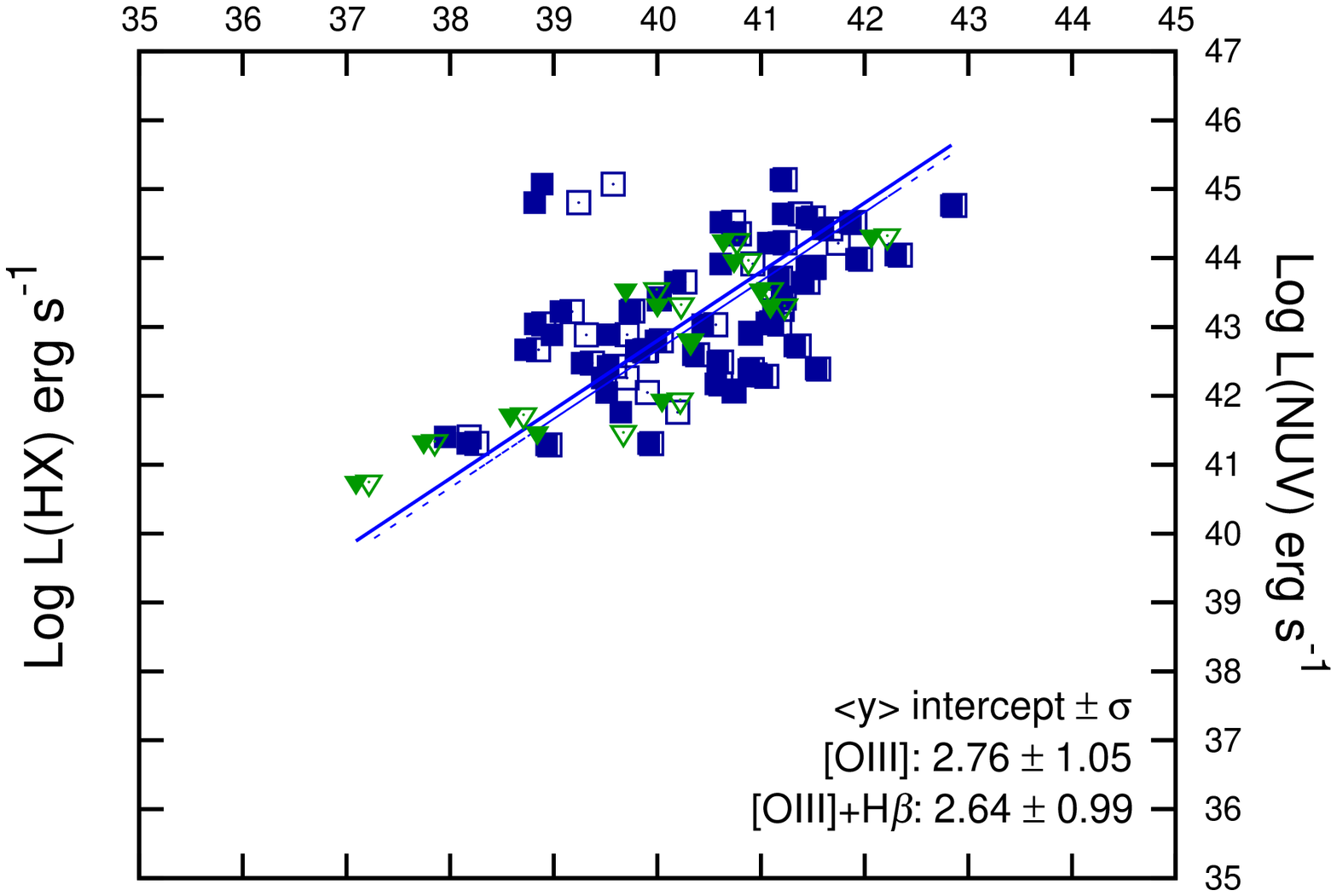}\\
\includegraphics[angle=0,scale=0.47]{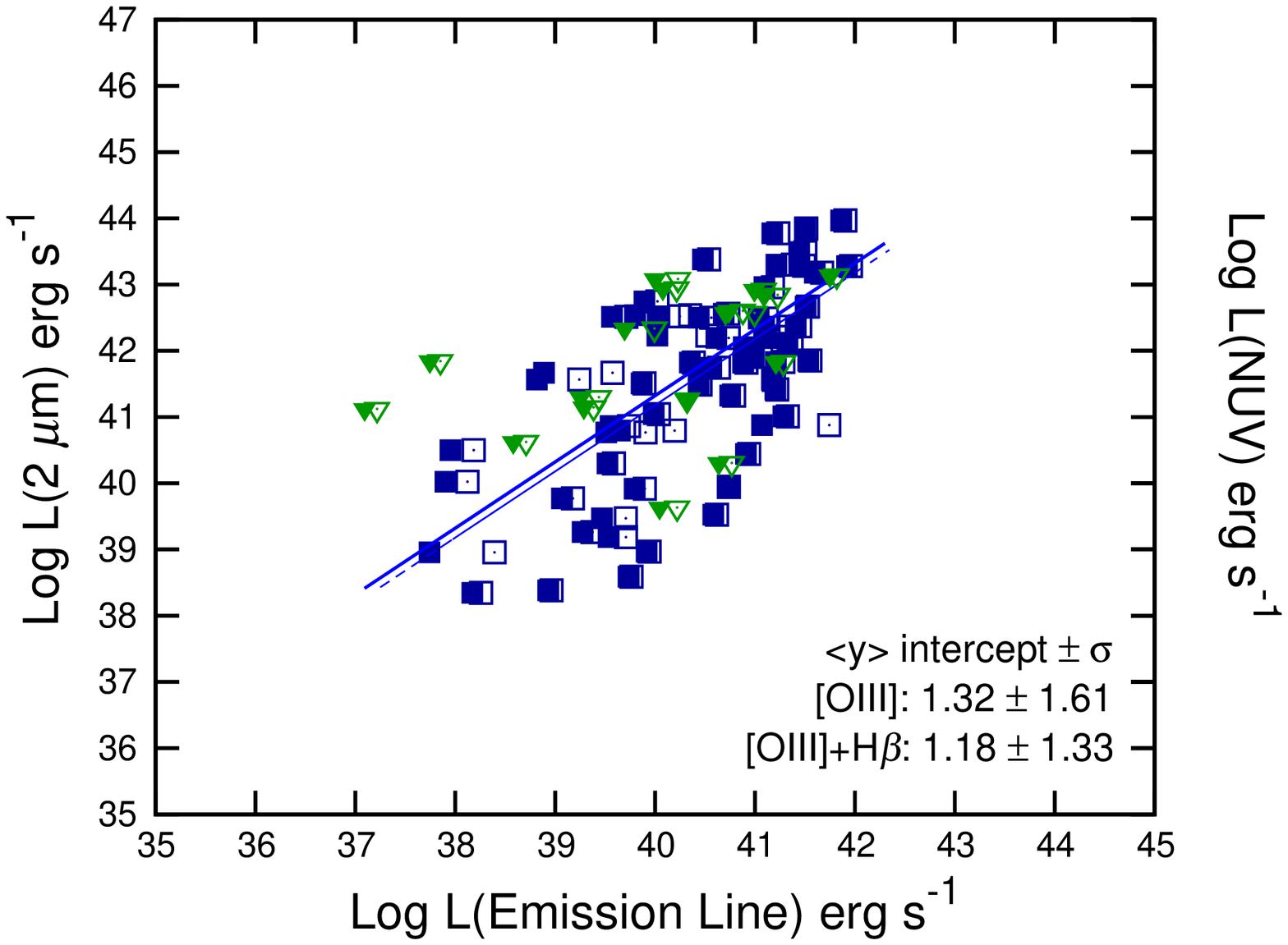}
\includegraphics[angle=0,scale=0.47]{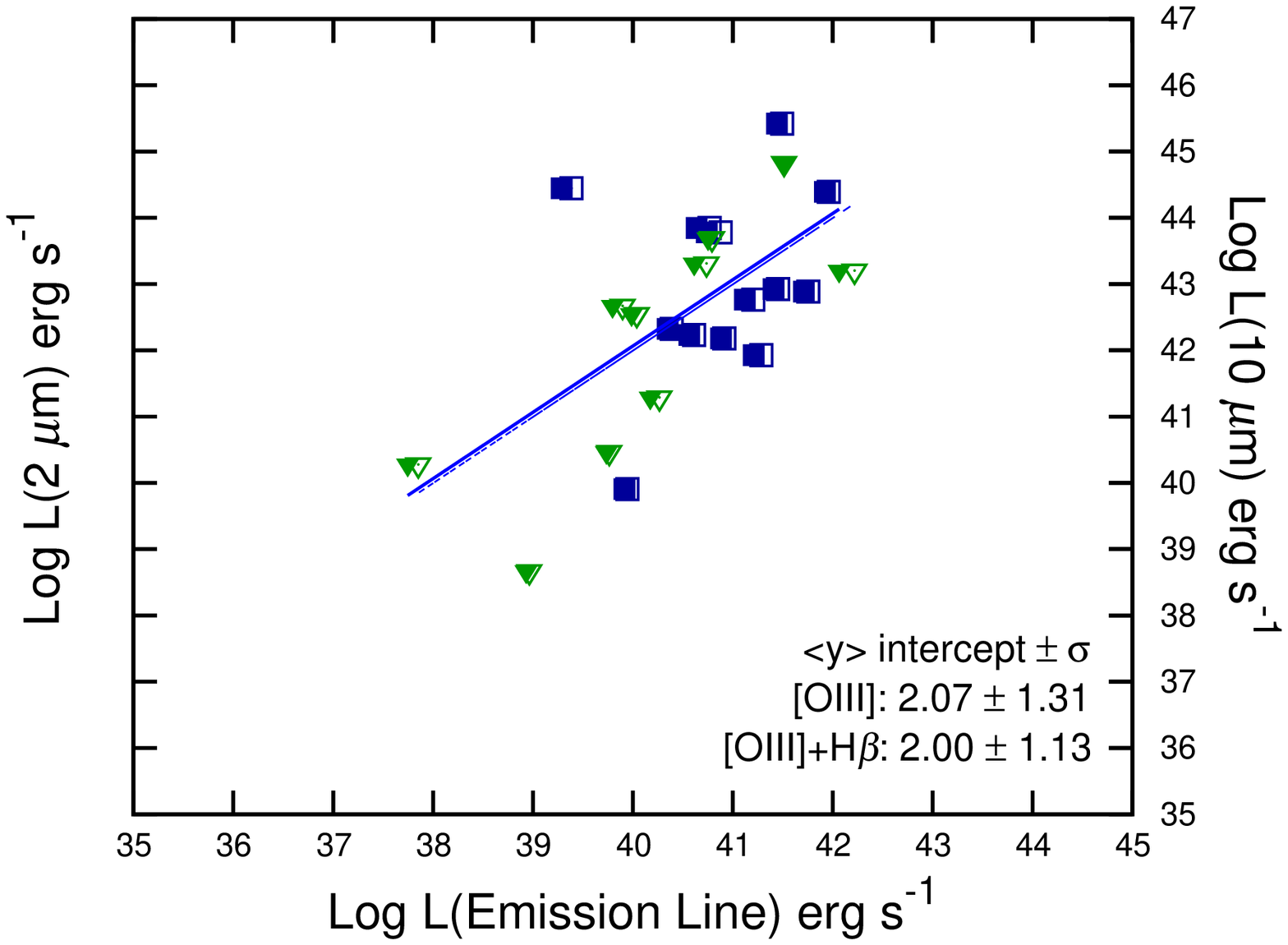}
\caption {Same non-stellar continuum/emission line correlation as shown in Figure~\ref{Fig12}, but for [\ion{O}{3}] and [\ion{O}{3}]+H$\beta$, for Seyfert 2 galaxies. 
The best correlation is for the $HX$ luminosities, with a single relation of $L_{HX}\sim 25L_{([OIII]+H\beta)}$, and $L_{HX}\sim 32L_{[OIII]}$. The steep solid line in the upper left panel is the same as in Figure \ref{Fig14}.
\label{Fig15}}
\end{figure*}

The hard X-rays only carry a minority ($\le 10\%$) of the bolometric luminosity of most AGN, and this fraction tends to be significantly lower in more luminous objects.
We therefore also searched for emission line proxies which could predict non-stellar luminosity at longer wavelengths closer to the bolometric peak output of typical AGN. 
The near-UV luminosity of Sy 1's (around the peak of the ``Big Blue Bump" \citep{MS82}), tends to be about four to five times the HX. As we found for the hard X-rays, the NUV luminosities can also be predicted by our combined [\ion{O}{3}] and [\ion{O}{3}]+H$\beta$ luminosities, with similar scatter ($\pm\, 4$ times for the Sy 1's, and $\pm\,9$ for the Sy 2's), shown in Figures ~\ref{Fig12}b to ~\ref{Fig15}b. Our normalization of $L_{UV} \sim 200 L_{[OIII]} $ and $L_{UV} \sim 100 L_{([OIII] + H\beta)} $ for Sy 1s are similar to the range of models in \cite{N06}, assuming bolometric luminosity is a few times  $L_{UV}$.
 
However, these emission line correlations with NUV luminosity have very different normalizations:
for a given line luminosity, the Sy 2's produce 3 to 8 times {\it more} NUV than the Sy 1's.
Although we are correlating observed luminosities, this discrepancy is not explained by
extinction differences, since we did not find strong Sy 1/Sy 2 differences in optical estimates
of $E(B-V)$, and the non-stellar UV continuum extinction should be even larger. 
Instead, a plausible explanation is that the  UV continuum in Sy 2's is dominated by 
recently formed hot stars which were also included in the large GALEX beam, outshining a possible AGN contribution. 

Next, we made an even less reliable  effort to isolate the non-stellar nuclear continuum,
using the 1.2 $\mu$m, 1.6 $\mu$m,  and 2.2 $\mu$m luminosities of the central pixels in 
2MASS images of each Seyfert nucleus (Figures ~\ref{Fig12}c to ~\ref{Fig15}c).  We decomposed the 
near-IR luminosities of the nucleus into contributions from the AGN and starlight,
both assumed to have constant colors.  The uncertainties in deriving
the 2.2 $\mu$m luminosities of the
weaker AGN are at least a factor of 2 by this so-called ``color given" method
\citep{MF83}.
The correlations of line luminosities with this non-stellar 2 $\mu$m luminosity are even 
weaker than with the HX and NUV, with very large scatter for both Sy 1's and Sy 2's.
The only line luminosity predictor worth possibly considering is the sum of [\ion{O}{3}]+H$\beta$ , which when multiplied by 15 gives a rough prediction for the non-stellar  2.2 $\mu$m luminosity of any Seyfert nucleus, regardless of type.
And our same backup line ratio,  the sum of H$\alpha$+[\ion{N}{2}], when multiplied by by 5.5, gives
a rough prediction, with the Sy 2 nuclei being  on average  40\% brighter at 2.2 $\mu$m .

Our final estimator of the non-stellar AGN luminosity was made from small-beam photometry at 10 $\mu$m, 
from \cite{gorjian04} (Figures ~\ref{Fig12}d to ~\ref{Fig15}d).  The correlations with the line luminosities are too poor to be of much use, with worse than
a factor of ten scatter.  The 10 $\mu$m correlation with [\ion{O}{3}] luminosity is the same for Sy 1's and Sy 2's, perhaps
because both are thought to originate in the NLR.

Five of the broad-line AGN in our sample are LINERs which have been found to exhibit weak broad-line components, usually faint extended wings under H$\alpha$ \citep{1997ApJS..112..391H}. We have included these with the Seyfert 1's in Figure \ref{Fig13}, but plotted with stars and 'x' symbols, not boxes. We were surprised to see that these AGN, which \citeauthor{1997ApJS..112..391H} denotes as ``LINER 1's" also follow roughly the same correlations between emission lines and non-stellar continuum as do the normal Sy 1's.

Similarly, a dozen of the narrow-line AGN in our multi-wavelength sample are now best classified as LINERs, and we included them in the panels of Figures ~\ref{Fig13} and ~\ref{Fig15}, along with the normal Sy 2's, except plotted with star symbols. Within the large scatter of the non-stellar line/continuum luminosity correlations, we again note that the LINERs  lacking any
broad wings overlap entirely with the Sy 2s.  Our results would hardly have changed whether they were included or excluded in the linear fits. Given the large scatter in these emission-line/continuum luminosity correlations, they cannot be used to separate various AGN types, even LINERs.
\end{appendix}


\clearpage


\begin{deluxetable}{lrrccr}
\tabletypesize{\footnotesize}
\tablewidth{0pt}
\tablecolumns{6}
\tablecaption{Seyfert 1 Galaxies \label{tbl1}}
\tablehead
{
\colhead{Name of Galaxy} &
\colhead{RA (J2000)}& 
\colhead{DEC (J2000)} &
\colhead{CfA\tablenotemark{a}}  &
\colhead{Type\tablenotemark{b}} & 
\colhead{Redshift ($z$)}
}
\startdata
MKN 334                       	&	00h03m09.62s	&	+21d57m36.6s	&		$\ast$		&		1.8		&		0.022	\\
MKN 335                       	&	00h06m19.52s	&	+20d12m10.5s	&				&		1.2		&		0.026	\\
ESO 540-G1                      	&	00h34m13.82s	&	-21d26m20.6s	&				&		1.8		&		0.027	\\
ESO 12-G21                       &	00h40m46.26s	&	-79d14m24.2s	&				&		1		&		0.030	\\
IRAS 00488+2907               	&	00h51m35.01s	&	+29d24m04.5s	&		$\ast$		&		1		&		0.036	\\
I ZW 1                        	&	00h53m34.94s	&	+12d41m36.2s	&				&		1/STBRST		&		0.061	\\
NGC 526A                      	&	01h23m54.39s	&	-35d03m55.9s	&		$\ast$		&		1.5		&		0.019	\\
MKN 993                       	&	01h25m31.46s	&	+32d08m11.4s	&		$\ast$		&		1.5		&		0.016	\\
\enddata
\tablenotetext{a}{An asterisk marks if this object was added from the CfA sample.}
\tablenotetext{b}{As listed under `BASIC DATA - Classifications' in Nasa/Ipac Extragalactic Database (NED), https://ned.ipac.caltech.edu.}
\tablecomments{This Table, as are all the data tables, are sorted by increasing RA starting with the Seyfert 1's then doing the same for the Seyfert 2's.}
\end{deluxetable}


\begin{deluxetable}{lrrccr}
\tabletypesize{\footnotesize}
\tablewidth{0pt}
\tablecaption{Seyfert 2 Galaxies\label{tbl2}}
\tablehead
{
\colhead{Name of Galaxy} & 
\colhead{RA (J2000)}& 
\colhead{DEC (J2000)} &
\colhead{CfA\tablenotemark{a}}  &
\colhead{Type\tablenotemark{b}} & 
\colhead{Redshift ($z$)}
}
\startdata
NGC  34                       	&	00h11m06.55s	&	-12d06m26.3s	&				&		2		&		0.020	\\
IRAS F00198-7926                   	&	00h21m53.61s	&	-79d10m07.5s	&				&		2		&		0.073	\\
NGC 262                       	&	00h48m47.14s	&	+31d57m25.1s	&				&		2		&		0.015	\\
IRAS F00521-7054                   	&	00h53m56.15s	&	-70d38m04.2s	&				&		2		&		0.069	\\
ESO 541-IG12                     	&	01h02m17.55s	&	-19d40m08.7s	&				&		2		&		0.057	\\
NGC 424                       	&	01h11m27.63s	&	-38d05m00.5s	&				&		2		&		0.012	\\
NGC 513                       	&	01h24m26.85s	&	+33d47m58.0s	&				&		2		&		0.020	\\
MKN 573                       	&	01h43m57.80s	&	+02d20m59.7s	&		$\ast$		&		2		&		0.017	\\
\enddata

\tablenotetext{a}{An asterisk marks if this object was added from the CfA sample.}
\tablenotetext{b}{As listed under `BASIC DATA - Classifications' in Nasa/Ipac Extragalactic Database (NED), https://ned.ipac.caltech.edu.}
\tablecomments{This Table, as are all the data tables, are sorted by increasing RA starting with the Seyfert 1's then doing the same for the Seyfert 2's.}

\end{deluxetable}

\clearpage

\begin{deluxetable}{llccc}
\tablecaption{Emission Lines\label{tbl3}}
\tabletypesize{\small}
\tablewidth{0pt}
\tablehead{
\colhead{Emission Line} & \colhead{Wavelength} &
\colhead{Ionization Potential} & \colhead{Ionization Potential} & \colhead{Critical Density} \\
\colhead{} & \colhead{(\AA)} &
\colhead{Lower (eV)} & \colhead{Upper (eV)} & \colhead{(cm$^{-3}$)}
}
\startdata
Lyman $\alpha$					&	1216 				& \nodata 	& 13.6 & \nodata\\
\ion{N}{5}							& 1240, 1243 	& 77.5			& 97.9 & \nodata\\
\ion{C}{4}							& 1549, 1551 	& 47.9			& 64.5 & \nodata\\
\ion{C}{3}]							& 1907, 1909 	& 24.4			& 47.9 & 1$\times 10^{10}$\\
\ion{Mg}{2}							& 2796, 2804 	& 7.6				&	15.0 & \nodata\\
$[$\ion{Ne}{5}]					& 3426 				& 97.1			& 126.2 & 1.6$\times 10^{7}$\\
$[$\ion{O}{2}]					& 3727, 3729 	& 13.6 			& 35.1 & 4.5$\times 10^3$\\
$[$\ion{Ne}{3}]					& 3869 				& 41.0			& 63.5 & 9.7$\times 10^6$\\
H$\beta$								& 4861 				& \nodata 	& 13.6 & \nodata\\
$[$\ion{O}{3}]					& 4959 				& 35.1			& 54.9 	& 7.0$\times 10^5$\\
$[$\ion{O}{3}]					& 5007 				& 35.1			& 54.9 	& 7.0$\times 10^5$\\
$[$\ion{O}{1}]					& 6300 				& \nodata 	& 13.6 & 1.8$\times 10^6$\\
$[$\ion{N}{2}]					& 6549, 6583 	& 14.5 			& 29.6 & 8.7$\times 10^4$\\
H$\alpha$								& 6563 				& \nodata		& 13.6 & \nodata\\
$[$\ion{S}{2}]					& 6717, 6734 	& 10.4			& 23.3 & 1.5$\times 10^3$, 3.9$\times 10^3$\\
\enddata

\tablecomments{Emission lines and their corresponding wavelength. The ionization potentials and critical densities are from \citet{do86}, \citet{m03}, \citet{of06}, and SPECTR-W3 (\url{http://spectr-w3.snz.ru/ion.phtml}).}
\end{deluxetable}
\clearpage
\begin{deluxetable}{lrrrrrrrrrrrr}
\tabletypesize{\footnotesize}
\rotate
\tablewidth{0pt}
\tablecaption{Optical Emission Line Data\label{tbl4}}
\tablehead{
\colhead{Name}          & \colhead{[NeV]\tablenotemark{a}}  &
\colhead{[OII]}          & \colhead{[NeIII]}    &
\colhead{H$\beta$}  & \colhead{[OIII] 4959\AA}  &
\colhead{[OIII] 5007\AA} & \colhead{[OI]} &
\colhead{H$\alpha$} & \colhead{[NII]} &
\colhead{H$\alpha$ + [NII]} & \colhead{[SII]} &
\colhead{Ref}
}
\startdata
\textbf{Seyfert 1} &      &       &       &       &       &       &       &           &        &           &       &\\
MKN 334	&	0.71	&	7.03	&	1.23	&	4.48	&	2.38	&	7.87	&	1.29	&	28.35	&	16.1	&	\nodata	&	7.42	&	1,2	\\
MKN 335	&	6.3	&	3	&	11	&	91	&	6.3	&	22.45	&	\nodata	&	\nodata	&	\nodata	&	264.8	&	\nodata	&	1,3	\\
E 540-G1	&	\nodata	&	1.95	&	\nodata	&	2.71	&	\nodata	&	2.59	&	0.4	&	9.69	&	4.72	&	\nodata 	&	2.25	&	4,5	\\
E12-G21	&	\nodata	&	\nodata	&	\nodata	&	14.8	&	\nodata	&	9.95	&	\nodata	&	111	&	\nodata	&	\nodata	&	\nodata	&	6,7	\\
IRAS 00488+2907	&	\nodata	&	\nodata	&	\nodata	&	\nodata	&	\nodata	&	\nodata	&	\nodata	&	\nodata	&	\nodata	&	\nodata	&	\nodata	&		\\
I ZW 1	&	6.3	&	2.7	&	4.3	&	31	&	4.6	&	7.43	&	\nodata	&	74.8	&	11	&	164.8	&	\nodata	&	1,8	\\
NGC 526A	&	\nodata	&	11	&	3.41	&	2.15	&	8.4	&	24.7	&	1.6	&	6.44	&	4.54	&	\nodata	&	3.62	&	3,7	\\
\enddata

\tablenotetext{a}{Emission line fluxes have units of $10^{-14}$ ergs cm$^{-2}$ s$^{-1}$. The wavelengths for these lines can be found in Table \ref{tbl3}.}

\tablerefs{
(1) 	Our data; 
(2) 	\citet{1993ApJ...414..552O}; 
(3) 	\citet{M94}; 
(4) 	\citet{2001MNRAS.327..459L}; 
(5) 	\citet{1996ApJS..106..341M}; 
(6) 	\citet{1997ApJ...486..132B}; 
(7) 	\citet{1992MNRAS.257..677W}; 
(8) 	\citet{1992A&AS...96..389D}; 
(9) 	\citet{2001ApJ...555..650H}; 
(10) 	\citet{1989ApJ...340..713M}; 
(11) 	\citet{2000ApJ...529..157P}; 
(12) 	\citet{1993MNRAS.261..170C}; 
(13) 	\citet{1996MNRAS.281.1206Y}; 
(14) 	\citet{1995ApJS...98..103S}; 
(15) 	\citet{2001ApJS..136...61S}; 
(16) 	\citet{1992ApJ...397..442B}; 
(17) 	\citet{1997ApJS..112..391H}; 
(18) 	\citet{L82}; 
(19) 	\citet{2000ApJS..126...63R}; 
(20) 	\citet{1983ApJ...265...92M}; 
(21) 	\citet{1999ApJ...522..113V}; 
(22) 	\citet{1997ApJS..112..315H}; 
(23) 	\citet{1995ApJS...98..129K}; 
(24) 	\citet{1999PASP..111.1398M}; 
(25) 	\citet{S98}; 
(26) 	\citet{1996ApJ...462..183H}; 
(27) 	\citet{1990AJ.....99.1722K}; 
(28) 	\citet{1994ApJS...94...47C}; 
(29) 	\citet{1996ApJS..106..399P}; 
(30) 	\citet{1995ApJS...98..171V}; 
(31) 	\citet{1989ApJ...347..727A}; 
(32) 	\citet{1996ApJS..105...75C}
.}
\end{deluxetable}

\clearpage


\begin{deluxetable}{lrrrrrrrrrrrrr}
\tabletypesize{\footnotesize}
\rotate
\tablewidth{0pt}
\tablecaption{Ultraviolet Data\label{tbl5}}
\tablehead{
\colhead{Name}& \colhead{Lyman $\alpha$\tablenotemark{a}}    &
\colhead{Ly $\alpha$ EW\tablenotemark{b}}  & \colhead{N V}  &
\colhead{N V EW} & \colhead{CIV} &
\colhead{CIV EW} & \colhead{CIII]} &
\colhead{CIII] EW}  & \colhead{MgII}  &
\colhead{MgII EW} & \colhead{$F_{UV}$\tablenotemark{c}} &
\colhead{$\lambda$ (\AA)} & \colhead{Ref}
}
\startdata
\textbf{Seyfert 1}& &       &        &           &           &           &            &           &           &           &       
&           & \\
MKN 335	&	828	&	109.3	&	280	&	36.45	&	347.48	&	78.97	&	125.9	&	23.17	&	55.79	&	67.55	&	7.89	&	1417.5	&	1,6,9,19	\\
E12-G21	&	22	&	\nodata	&	14	&	\nodata	&	12	&	\nodata	&	3.1	&	\nodata	&	13	&	\nodata	&	0.29	&	1450	&	2	\\
I ZW 1	&	143.05	&	136.5	&	\nodata	&	\nodata	&	44.27	&	24.67	&	70	&	25.39	&	89.83	&	49.4	&	2.35	&	1483.3	&	3,6,7,9	\\
NGC 1097	&	\nodata	&	\nodata	&	\nodata	&	\nodata	&	\nodata	&	\nodata	&	\nodata	&	\nodata	&	\nodata	&	\nodata	&	2.11	&	1507	&	4	\\
F03450+0055	&	17.7	&	101	&	\nodata	&	\nodata	&	7.8	&	73	&	2.13	&	29.3	&	\nodata	&	\nodata	&	0.11	&	1500	&	6	\\
NGC 1566	&	23.1	&	101	&	3.65	&	8	&	28.6	&	92	&	10.95	&	35.5	&	26.26	&	168.5	&	0.23	&	1475	&	2,7	\\
3C 120	&	78.9	&	79	&	\nodata	&	\nodata	&	36.4	&	131	&	\nodata	&	\nodata	&	\nodata	&	\nodata	&	0.74	&	1335	&	1,9	\\
\enddata

\tablenotetext{a}{Emission line fluxes have units of $10^{-14}$ ergs cm$^{-2}$ s$^{-1}$. The wavelengths for these lines can be found in Table~\ref{tbl3}.}
\tablenotetext{b}{Equivalent widths are in \AA.}
\tablenotetext{c}{UV continuum flux is in units of $10^{-14}$ ergs cm$^{-2}$ s$^{-1}$ \AA$^{-1}$.}

\tablerefs{
(1)	\citet{1996ApJ...457..111W}; 
(2)	\citet{1984A&A...131...87C}; 
(3)	\citet{1990A&A...240..247B}; 
(4)	\citet{1995ApJS...98..103S}; 
(5)	\citet{1980MNRAS.192..769C}; 
(6)	\citet{2004ApJS..150..165K}; 
(7)	\citet{2002ApJS..143..257K}; 
(8)	\citet{1983ApJ...265...92M}; 
(9)	\citet{1998ApJ...493....1W}; 
(10)	\citet{1983MNRAS.203..201B}; 
(11)	\citet{1995ApJS...97..331M}; 
(12)	\citet{1984ApJ...281..126T}; 
(13)	\citet{2005ApJ...626...70F}; 
(14)	\citet{1986ApJ...300..658F}; 
(15)	\citet{1988AJ.....95.1371D}; 
(16)	\citet{1994ApJ...435..171K}; 
(17)	\citet{2001ApJ...547..667K}; 
(18)	\citet{1997MNRAS.291..403R}; 
(19)	\citet{1983ApJ...268..591G}
.}
\end{deluxetable}

\clearpage

\begin{deluxetable}{llcc}
\tablecaption{Luminosity Function Parameters\label{tbl6}}
\tablewidth{0pt}
\tablehead{
\colhead{Line} & \colhead{Type} &
\colhead{Log $\Phi_{\star}$} & \colhead{Log $L_{\star}$}
}
\startdata
H$\alpha$	&	Sy 1	&	$-4.18 \pm 0.34$	&	$41.04 \pm 0.13$	\\						
	&	Sy 2	&	$-3.77 \pm 0.30$	&	$39.92 \pm 0.12$	\\						
H$\beta$	&	Sy 1	&	$-4.37 \pm 0.31$	&	$40.72 \pm 0.14$	\\						
	&	Sy 2	&	$-3.93 \pm 0.30$	&	$39.24 \pm 0.12$	\\						
$[$\ion{N}{2}]	&	Sy 1	&	$-4.10 \pm 0.46$	&	$39.98 \pm 0.18$	\\						
	&	Sy 2	&	$-3.76 \pm 0.31$	&	$39.79 \pm 0.12$	\\						
$[$\ion{O}{1}]	&	Sy 1	&	$-4.50 \pm 0.52$	&	$39.38 \pm 0.20$	\\						
	&	Sy 2	&	$-4.21 \pm 0.28$	&	$39.36 \pm 0.13$	\\						
$[$\ion{O}{2}]	&	Sy 1	&	$-4.13 \pm 0.39$	&	$39.58 \pm 0.16$	\\						
	&	Sy 2	&	$-4.27 \pm 0.44$	&	$39.78 \pm 0.17$	\\						
$[$\ion{O}{3}] $\lambda4969$	&	Sy 1	&	$-4.26 \pm 0.37$	&	$39.96 \pm 0.16$	\\						
	&	Sy 2	&	$-4.38 \pm 0.49$	&	$39.75 \pm 0.18$	\\						
$[$\ion{O}{3}] $\lambda5007$	&	Sy 1	&	$-4.11 \pm 0.38$	&	$40.25 \pm 0.15$	\\
	&	Sy 2	&	$-3.99 \pm 0.25$	&	$40.09 \pm 0.11$	\\
$[$\ion{S}{2}]	&	Sy 1	&	$-3.98 \pm 0.45$	&	$39.59 \pm 0.18$	\\
	&	Sy 2	&	$-3.78 \pm 0.31$	&	$39.56 \pm 0.13$	\\
\enddata

\tablecomments{Best fit parameters of $\Phi_{\star}$ and $L_{\star}$ for our luminosity functions, described in Section 2. The assumed luminosity function is a double power law with breaks at $L_*$ is given by Equation \ref{eqn1}. $\Phi_{\star}$ has units of Mpc$^{-3}$ mag$^{-1}$ and $L_{\star}$ has units of ergs s$^{-1}$.}
\end{deluxetable}
\clearpage

\begin{deluxetable}{lcr}
\tablecaption{Ionization Averages\label{tbl7}}
\tablewidth{0pt}
\tablehead{
\colhead{Log(Line-Ratio)} & \colhead{Sy Type} & \colhead{Average}
}
\startdata
$[$\ion{Ne}{3}]\tablenotemark{*}/[\ion{O}{2}]	& 1, 1.2, 1.5 & $-0.11 \pm 0.38$ \\
							& 1.8, 1.9 		& $-0.33 \pm 0.31$ \\
				            & 2				& $-0.48 \pm 0.30$ \\
$[$\ion{O}{3}]/[\ion{O}{2}] 	& 1, 1.2, 1.5 & $0.71 \pm 0.48$ \\
							& 1.8, 1.9 		& $0.38 \pm 0.49$ \\
						    & 2				& $0.43 \pm 0.52$ \\
$[$\ion{Ne}{5}]/[\ion{Ne}{3}] & 1, 1.2, 1.5 & $-0.03 \pm 0.22$ \\
							& 1.8, 1.9 		& $-0.29 \pm 0.27$ \\
							& 2				& $-0.17 \pm 0.40$ \\
\enddata
\tablecomments{This is a list of the line ratio averages divided by type. The Seyfert 1, 1.2, and 1.5's have higher ratios than Seyfert 1.8, 1.9, and 2's.}
\tablenotetext{*}{Although the Seyfert 1s appear to have slightly stronger [\ion{Ne}{3}]$\lambda 3869$ emission, relative to [\ion{O}{3}], as discussed in the text we attribute this to unremoved contamination from the BLR emission of \ion{He}{1}$\lambda 3889$, rather than an intrinsic NLR difference between Sy 1 and Sy 2 galaxies. } 
\end{deluxetable}

\clearpage
\begin{deluxetable}{llrr}
\tablecaption{Fits and Errors for Reddening Ratios\label{tbl8}}
\tablewidth{0pt}
\tablecolumns{7}
\tabletypesize{\footnotesize}
\tablehead
{
\colhead{Ratios}& 
\colhead{Type}&
\colhead{$A$}&
\colhead{$B$}
}
\startdata
H$\alpha$/H$\beta$	vs.	[\ion{S}{2}]/[\ion{O}{2}]	&	Sy	1	&	0.44	$\pm$	0.14	&	0.69	$\pm$		0.03	\\
H$\alpha$/H$\beta$	vs.	[\ion{S}{2}]/[\ion{O}{2}]	&	Sy	2	&	0.44	$\pm$		0.22	&	0.73	$\pm$		0.04	\\
H$\alpha$/H$\beta$	vs.	[\ion{N}{2}]/[\ion{O}{2}]	&	Sy	1	&	0.25	$\pm$		0.13	&	0.61	$\pm$		0.04	\\
H$\alpha$/H$\beta$	vs.	[\ion{N}{2}]/[\ion{O}{2}]	&	Sy	2	&	0.17	$\pm$		0.13	&	0.71$\pm$		0.05	\\
H$\alpha$/H$\beta$	vs.	[\ion{O}{3}]/[\ion{Ne}{3}]	&	Sy	1	&	0.33	$\pm$		0.12	&	0.39$\pm$		0.10	\\
H$\alpha$/H$\beta$	vs.	[\ion{O}{3}]/[\ion{Ne}{3}]	&	Sy	2	&	0.20	$\pm$		0.54	&	0.56	$\pm$		0.53	\\
$[$\ion{S}{2}]/[\ion{O}{2}]	vs.	[\ion{O}{3}]/[\ion{Ne}{3}]	&	Sy	1	&	1.65	$\pm$		0.47	&	-1.40	$\pm$		0.43	\\
$[$\ion{S}{2}]/[\ion{O}{2}]	vs.	[\ion{O}{3}]/[\ion{Ne}{3}]	&	Sy	2	&	2.56	$\pm$		1.26	&	-2.40	$\pm$		1.22	\\
$[$\ion{O}{3}]/[\ion{Ne}{3}]	vs.	[\ion{N}{2}]/[\ion{O}{2}]	&	Sy	2	&	0.59	$\pm$		0.50	&	0.66	$\pm$		0.17	\\
$[$\ion{O}{3}]/[\ion{Ne}{3}]	vs.	[\ion{N}{2}]/[\ion{O}{2}]	&	Sy	2	&	0.27	$\pm$		0.18	&	0.88$\pm$		0.07	\\
$[$\ion{S}{2}]/[\ion{O}{2}]	vs.	[\ion{N}{2}]/[\ion{O}{2}]	&	Sy	1	&	0.87	$\pm$		0.10	&	-0.22	$\pm$		0.05	\\
$[$\ion{S}{2}]/[\ion{O}{2}]	vs.	[\ion{N}{2}]/[\ion{O}{2}]	&	Sy	2	&	0.83	$\pm$		0.10	&	-0.18	$\pm$		0.05	\\
\enddata
\tablecomments{ $A$ and $B$ are the slope and intercept from the Least Squares Orthogonal Fit, respectively.}
\end{deluxetable}
\clearpage
\begin{deluxetable}{lcccclcccc}
\tablecaption{IR, NUV, and HX Luminosities for Seyfert Galaxies\label{tbl10}}
\rotate
\tablewidth{0pt}
\tablecolumns{10}
\tabletypesize{\footnotesize}
\tablehead
{
\colhead{} & 
\colhead{$IR$}&
\colhead{$IR$}&
\colhead{$NUV$}&
\colhead{$HX$} &
\colhead{} & 
\colhead{$IR$}&
\colhead{$IR$}&
\colhead{$NUV$}&
\colhead{$HX$}\\
\colhead{Seyfert 1}&	
\colhead{$10\;\mu m$}&
\colhead{$2\;\mu m$}&
\colhead{2267\AA}&
\colhead{2-10 keV} &
\colhead{Seyfert 2}&	
\colhead{$10\;\mu m$}&
\colhead{$2\;\mu m$}&
\colhead{2267\AA}&
\colhead{2-10 keV}
}
\startdata
MKN 334                       	&	43.81	&	\nodata	&	43.01	&	\nodata	&	NGC  34                       	&	\nodata	&	42.75	&	\nodata	&	\nodata	\\
MKN 355                       	&	\nodata	&	43.46	&	43.81	&	43.40	&	IRAS F00198-7926              	&	\nodata	&	43.49	&	44.58	&	42.32	\\
E 540-G1                      	&	\nodata	&	42.35	&	45.92	&	\nodata	&	M-4-2-18                      	&	\nodata	&	\nodata	&	43.91	&	\nodata	\\
E12-G21                       	&	\nodata	&	43.09	&	\nodata	&	\nodata	&	NGC 262                       	&	42.92	&	42.37	&	43.65	&	43.45	\\
UGC 00524                     	&	\nodata	&	\nodata	&	\nodata	&	\nodata	&	IRAS F00521-7054              	&	\nodata	&	43.97	&	44.52	&	\nodata	\\
IZW1                          	&	45.78	&	44.27	&	44.02	&	43.80	&	E541-IG12                     	&	\nodata	&	43.77	&	44.22	&	\nodata	\\
MKN 993                       	&	\nodata	&	\nodata	&	43.47	&	42.91	&	NGC 424                       	&	\nodata	&	42.96	&	43.04	&	42.47	\\
UGC 01395                     	&	\nodata	&	42.94	&	42.99	&	\nodata	&	NGC 513                       	&	43.31	&	42.19	&	44.52	&	42.64	\\
\enddata
\tablecomments{\\ Luminosities of Seyfert 1 and Seyfert 2 galaxies discussed in Appendix A, units are $Log(Luminosity)$ in erg/s. The Infrared data are obtained from the 2MASS catalogue, the Ultraviolet data are from GALEX, and the X-ray data are from \cite{BN11a, BN11b}, and \cite{P08}.}
\end{deluxetable}
\clearpage
\begin{deluxetable}{lccccc}
\tablecaption{IR, NUV, and HX continuum vs. Optical Emission-Lines\label{tbl9}}
\tablewidth{0pt}
\tablecolumns{6}
\tabletypesize{\small}
\tablehead
{
\colhead{} & 
\colhead{} & 
\colhead{$IR$} &
\colhead{$IR$} &
\colhead{$NUV$} &
\colhead{$HX$}  \\
\colhead{Log L(Emission-Line)}	&
\colhead{Type}	&	
\colhead{$10\;\mu m$} &
\colhead{$2\;\mu m$} &
\colhead{2267\AA} &
\colhead{2-10 keV}
}
\startdata
Log L([\ion{O}{3} ] $ \lambda5007$)		&	Sy 1	&	$2.16\pm1.23$	&	$1.55\pm0.92$	&	$2.31\pm0.61$	&	$1.64\pm0.63$	\\
		&	Sy 2	&	$2.07\pm1.31$	&	$1.32\pm1.61$	&	$2.76\pm1.05$	&	$1.50\pm1.01$	\\
Log L([\ion{O}{3} ] $ \lambda5007+ H\beta$)		&	Sy 1	&	$1.84\pm1.17$	&	$1.22\pm0.84$	&	$2.00\pm0.62$	&	$1.34\pm0.62$	\\
		&	Sy 2	&	$2.00\pm1.30$	&	$1.18\pm1.33$	&	$2.64\pm0.99$	&	$1.42\pm1.02$	\\
Log L($H\alpha$ )		&	Sy 1	&	$1.22\pm1.11$	&	$0.81\pm0.85$	&	$1.76\pm0.73$	&	$1.17\pm0.52$	\\
		&	Sy 2	&	$2.07\pm1.18$	&	$1.14\pm1.08$	&	$2.71\pm0.85$	&	$1.34\pm1.31$	\\
Log L($H\alpha$ +[\ion{N}{2} ])		&	Sy 1	&	$1.09\pm1.08$	&	$0.69\pm0.88$	&	$1.62\pm0.61$	&	$1.06\pm0.55$	\\
		&	Sy 2	&	$1.74\pm1.26$	&	$0.85\pm1.09$	&	$2.42\pm0.85$	&	$1.05\pm1.33$	
\enddata
\tablecomments{Results from least squares fit for $Log(L_{\mathrm{continuum}})=Log(L_{\mathrm{emission-line}})+B$ (discussed in Appendix A), where B is the intercept and the slope = 1. The Infrared data are obtained from the 2MASS catalogue, the Ultraviolet data are from GALEX, and the X-ray data are from \cite{BN11a, BN11b}, and \cite{P08}.}
\end{deluxetable}
\clearpage


\begin{thebibliography}{}
\bibitem[Akritas \& Bershady (1996)]{ak96} Akritas M. G. \& Bershad M. A.1996 \apj, 470, 706
\bibitem[Antonucci (1993)]{ar93} Antonucci, R., 1993, \araa, 31, 473
\bibitem[Armus et al.(1989)]{1989ApJ...347..727A} Armus, L., Heckman,  T.~M., \& Miley, G.~K.\ 1989, \apj, 347, 727
\bibitem[Baldwin(1977)]{b77} Baldwin, J. A. 1977, \apj, 214, 679
\bibitem[Baldwin, Phillips, \& Terlevich(1981)]{bpt81} Baldwin, J. A., Phillips, M. M., Terlevich, R., 1981, \pasp, 93, 5
\bibitem[Barr et al.(1983)]{1983MNRAS.203..201B} Barr, P., Willis, A.~J.,  \& Wilson, R.\ 1983, \mnras, 203, 201
\bibitem[Baskin \& Laor(2004)]{bl04} Baskin, A. \& Laor, A. 2004, \mnras, 350, 31
\bibitem[Bonatto \& Pastoriza(1997)]{1997ApJ...486..132B} Bonatto, C.~J., \& Pastoriza, M.~G.\ 1997, \apj, 486, 132
\bibitem[Bongiorno et al.(2010)]{B10} Bongiorno, A., Mignoli, M., Zamorani, G., Lamareille, F., et al. 2010, \aap, 510, A56
\bibitem[Boroson \& Green(1992)]{bg92} Boroson, T. A. \& Green, R. F.  1992, \apjs, 80, 109
\bibitem[Boroson \& Meyers(1992)]{1992ApJ...397..442B} Boroson, T.~A., \& Meyers, K.~A.\ 1992, \apj, 397, 442
\bibitem[Brightman \& Nandra (2011a)]{BN11a} 
Brightman, M. \& Nandra, K, \mnras, 413, 1206
\bibitem[Brightman \& Nandra (2011b)]{BN11b} Brightman, M. \& Nandra, K, \mnras, 414, 3084
\bibitem[Buson \& Ulrich(1990)]{1990A&A...240..247B} Buson, L.~M. \& Ulrich, M.-H.\ 1990, \aap, 240, 247
\bibitem[Cardelli et al.(1989)]{cardelli} Cardelli, J. A., Clayton, G. C., Mathis, J. S. 1989, \apj, 345, 245
\bibitem[Clavel et al.(1980)]{1980MNRAS.192..769C} Clavel, J., Benvenuti,  P., Cassatella, A., Heck, A., Penston, M.~V., Selvelli, P.~L., Beeckmans,  F., \& Macchetto, F.\ 1980, \mnras, 192, 769
\bibitem[Clavel \& Joly(1984)]{1984A&A...131...87C} Clavel, J., \& Joly, M.\ 1984, \aap, 131, 87
\bibitem[Colbert et al.(1996)]{1996ApJS..105...75C} Colbert, E.~J.~M.,  Baum, S.~A., Gallimore, J.~F., O'Dea, C.~P., Lehnert, M.~D., Tsvetanov,  Z.~I., Mulchaey, J.~S., \& Caganoff, S.\ 1996, \apjs, 105, 75
\bibitem[Coziol et al.(1993)]{1993MNRAS.261..170C} Coziol, R., Pena, M.,  Demers, S., \& Torres-Peimbert, S.\ 1993, \mnras, 261, 170
\bibitem[Cruz-Gonzalez et al.(1994)]{1994ApJS...94...47C} Cruz-Gonzalez,  I., Carrasco, L., Serrano, A., Guichard, J., Dultzin-Hacyan, D.,  \& Bisiacchi, G.~F.\ 1994, \apjs, 94, 47
\bibitem[Dasyra et al. (2008)]{D08} Dasyra, K. M., Ho, L. C., Armus, L., Ogle, P.; Helou, G., Peterson, B. M., Lutz, D., Netzer, H., Sturm, E. 2008, \apj, 674L, 9	
\bibitem[de Grijp et  al.(1992)]{1992A&AS...96..389D} de Grijp, M.~H.~K., Keel, W.~C., Miley, G.~K., Goudfrooij, P., \& Lub, J.\ 1992, \aap, 96, 389
\bibitem[De Robertis \& Osterbrock(1986)]{do86} De Robertis, M. M. \& Osterbrock, D. E., 1986, \apj, 301, 727
\bibitem[De Robertis et al.(1988)]{1988AJ.....95.1371D} De Robertis, M.~M.,  Hutchings, J.~B., \& Pitts, R.~E.\ 1988, \aj, 95, 1371
\bibitem[Dopita et al.(2002)]{dg02} Dopita, M. A., Groves, B. A., Sutherland, R. S., Binette, L., \& Cecil, G. 2002, \apj, 572, 753
\bibitem[Farrah et al.(2005)]{2005ApJ...626...70F} Farrah, D., Surace,  J.~A., Veilleux, S., Sanders, D.~B., \& Vacca, W.~D.\ 2005, \apj, 626, 70
\bibitem[Ferland \& Osterbrock(1986)]{1986ApJ...300..658F} Ferland, G.~J., \& Osterbrock, D.~E.\ 1986, \apj, 300, 658
\bibitem[Freedman et al.(1994)]{f94} Freedman, W. L., Hughes, S. M., Madore, B. F., et al. 1994, \apj, 427, 628
\bibitem[Gallego et al.(1995)]{g95} Gallego, J., Zamorano, J., Aragon-Salamanca, A., \& Rego, M.\ 1995, \apjl, 455, L1
\bibitem[Gallego et al.(1996)]{g96} Gallego, J., Zamorano, J., Rego, M., Alonso, O., \& Vitores, A.~G.\ 1996, \aaps, 120, 323
\bibitem[Gaskell \& Ferland(1984)]{gf84} Gaskell, C. M. \& Ferland, G. J., 1984, \pasp, 96, 393
\bibitem[Gorjian et al.(2004)]{gorjian04} Gorjian, V., Werner, M. W., Jarrett, T. H., Cole, D. M., Ressler, M. E. 2004, \apj, 605, 156
\bibitem[Goto et al.(2011)]{goto11} 	Goto, T., Arnouts, S., Malkan, M., , et al., 2011, \mnras, 414, 1903
\bibitem[Grandi(1983)]{1983ApJ...268..591G} Grandi, S.~A.\ 1983, \apj, 268,  591
\bibitem[Groves et al.(2004)]{gds04} Groves, B. A., Dopita, M.A., Sutherland, R.S. 2004, \apjs, 153, 9
\bibitem[Groves et al.(2004)]{ghk06} Groves, B. A., Heckman, M.H., Kauffman, G. 2006, \mnras, 371, 1559
\bibitem[Hao et al.(2005)]{H05} 	Hao, L., Strauss, M. A., Fan, X., , et al., 2005, \aj, 129, 1795
\bibitem[Heckman(2005)]{hk05} Heckman, T. M, Ptak, A., Hornschemeier, A., Kauffmann, G.  2005, \apj, 634, 161
\bibitem[Ho et al.(1996)]{1996ApJ...462..183H} Ho, L.~C., Filippenko,  A.~V., \& Sargent, W.~L.~W.\ 1996, \apj, 462, 183
\bibitem[Ho et al.(1997a)]{1997ApJS..112..315H} Ho, L.~C., Filippenko,  A.~V., \& Sargent, W.~L.~W.\ 1997, \apjs, 112, 315
\bibitem[Ho et al.(1997b)]{1997ApJS..112..391H} Ho, L.~C., Filippenko,  A.~V., Sargent, W.~L.~W., \& Peng, C.~Y.\ 1997, \apjs, 112, 391
\bibitem[Ho \& Peng(2001)]{2001ApJ...555..650H} Ho, L.~C., \& Peng, C.~Y.\ 2001, \apj, 555, 650
\bibitem[Huchra et al.(1983)]{huchra83} Huchra, J., Davis, M., Latham, D., Tonry, J. 1983, \apjs, 52, 89
\bibitem[Huchra \& Burg(1992)]{hb92} Huchra, J. \& Burg, R., 1992, \apj, 393, 90
\bibitem[Jensen et al.(2016)]{j16} Jensen, T.W., Vivek, M., Dawson, K.S., Anderson, S.F., Bautista, J., et al. 2003, \apj, 833, 199
\bibitem[Kauffmann et al.(2003)]{k03} 	Kauffmann, G., Heckman, T. M., Tremonti, C., et al., 2003, \mnras, 346, 1055
\bibitem[Kewley et al.(2001)]{k01} Kewley, I., Dopita, M.A., Sutherland, R. S., Heiser, C. A., \& Trevena, J. 2001, \apj, 556, 121
\bibitem[Kim et al.(1995)]{1995ApJS...98..129K} Kim, D.-C., Sanders, D.~B.,  Veilleux, S., Mazzarella, J.~M., \& Soifer, B.~T.\ 1995, \apjs, 98, 129
\bibitem[Kirhakos  \& Steiner(1990)]{1990AJ.....99.1722K} Kirhakos, S.~D., \& Steiner, J.~E.\ 1990, \aj, 99, 1722
\bibitem[Kishimoto et al.(2001)]{2001ApJ...547..667K} Kishimoto, M.,  Antonucci, R., Cimatti, A., Hurt, T., Dey, A., van Breugel, W.,  \& Spinrad, H.\ 2001, \apj, 547, 667
\bibitem[Kraemer et al.(1994)]{1994ApJ...435..171K} Kraemer, S.~B., Wu,  C.-C., Crenshaw, D.~M., \& Harrington, J.~P.\ 1994, \apj, 435, 171
\bibitem[Kuraszkiewicz et al.(2002)]{2002ApJS..143..257K} Kuraszkiewicz,  J.~K., Green, P.~J., Forster, K., Aldcroft, T.~L., Evans, I.~N.,  \& Koratkar, A.\ 2002, \apjs, 143, 257
\bibitem[Kuraszkiewicz et al.(2004)]{2004ApJS..150..165K} Kuraszkiewicz,  J.~K., Green, P.~J., Crenshaw, D.~M., Dunn, J., Forster, K., Vestergaard,  M., \& Aldcroft, T.~L.\ 2004, \apjs, 150, 165
\bibitem[Lacy et al.(1982)]{L82} Lacy, J. H., Malkan, M., Becklin, E. E., Soifer, B. T., Neugebauer, G., Matthews, K., Wu, C.-C., Boggess, A., \& Gull, T. R., 1982, \apj, 256, 75
\bibitem[Lumsden et al.(2001)]{2001MNRAS.327..459L} Lumsden, S.~L.,  Heisler, C.~A., Bailey, J.~A., Hough, J.~H.,  \& Young, S.\ 2001, \mnras, 327, 459
\bibitem[Malkan \& Sargent (1982)]{MS82} Malkan, M. A. \& Sargent, W. L. \apj, 254, 22
\bibitem[Malkan (1983)]{m83} Malkan, M. A., 1983, \apjl, 264, L1
\bibitem[Malkan \& Filippenko (1983)]{MF83} Malkan, M. A. \& Fillipenko, A.V. 1983, \apj, 268, 562
\bibitem[Malkan \& Oke(1983)]{1983ApJ...265...92M} Malkan, M.~A., \& Oke, J.~B.\ 1983, \apj, 265, 92
\bibitem[Malkan et al.(1998)]{mgt98} Malkan, M. A., Gorjian, V., \& Tam, R. 1998, \apjs, 117, 25
\bibitem[McQuade et al.(1995)]{1995ApJS...97..331M} McQuade, K., Calzetti,  D., \& Kinney, A.~L.\ 1995, \apjs, 97, 331
\bibitem[Misselt et al.(1999)]{1999PASP..111.1398M} Misselt, K.~A.,  Clayton, G.~C., \& Gordon, K.~D.\ 1999, \pasp, 111, 1398
\bibitem[Moran et al.(1996)]{1996ApJS..106..341M} Moran, E.~C., Halpern,  J.~P., \& Helfand, D.~J.\ 1996, \apjs, 106, 341
\bibitem[Morris  \& Ward(1989)]{1989ApJ...340..713M} Morris, S.~L., \& Ward, M.~J.\ 1989, \apj, 340, 713
\bibitem[Morton(2003)]{m03} Morton, D. C., 2003, \apjs, 149, 205
\bibitem[Mulchaey et al.(1994)]{M94} Mulchaey, J.~S.,  Koratkar, A., Ward, M.~J., Wilson, A.~S., Whittle, M., Antonucci, R.~R.~J.,  Kinney, A.~L., \& Hurt, T.\ 1994, \apj, 436, 586
\bibitem[Nagao, Murayama, \& Taniguchi(2001)]{n01} Nagao, T., Murayama, T., \& Taniguchi, Y. 2001, \pasj, 53, 629
\bibitem[Netzer(2006)]{N06} Netzer, H., 2006, \aap, 453, 525
\bibitem[Netzer(2009)]{N09} Netzer, H., 2009, \mnras, 399, 1907
\bibitem[Oh(2017)]{O17} Oh, K., 2017, \mnras, 464, 1466
\bibitem[Osterbrock(1981)]{o81} Osterbrock, D. E. 1981, \apj, 249, 462
\bibitem[Osterbrock \& Martel(1993)]{1993ApJ...414..552O} Osterbrock, D.~E., \& Martel, A.\ 1993, \apj, 414, 552
\bibitem[Osterbrock \& Ferland(2006)]{of06} Osterbrock, D. E. \& Ferland, G. J., 2006, Astrophysics of Gaseous Nebulae and Active Galactic Nuclei, Second Edition (Sausalito, California: University Science Books)
\bibitem[Panessa et al. (2008)]{P08} 	Panessa, F., Bassani, L., de Rosa, A.,  et al. 2008, \aap, 483, 151
\bibitem[Poggianti  \& Wu(2000)]{2000ApJ...529..157P} Poggianti, B.~M., \& Wu, H.\ 2000, \apj, 529, 157
\bibitem[Polletta et al.(1996)]{1996ApJS..106..399P} Polletta, M., Bassani,  L., Malaguti, G., Palumbo, G.~G.~C., \& Caroli, E.\ 1996, \apjs, 106, 399
\bibitem[Reynolds et al.(1997)]{1997MNRAS.291..403R} Reynolds, C.~S., Ward,  M.~J., Fabian, A.~C., \& Celotti, A.\ 1997, \mnras, 291, 403
\bibitem[Rhee \& Larkin(2005)]{rl05} Rhee, J. H. \& Larkin, J. E., 2005, \apj, 620, 151
\bibitem[Rodr{\'{\i}}guez-Ardila et al.(2000)]{2000ApJS..126...63R}  Rodr{\'{\i}}guez-Ardila, A., Pastoriza, M.~G.,  \& Donzelli, C.~J.\ 2000, \apjs, 126, 63
\bibitem[Runco, J., et al. (2016)]{R16} Runco, J., Cosens, M., Bennert, V. N., et al. 2016, \apj, 821, 33
\bibitem[Rush et al.(1993)]{rms93} Rush, B., Malkan, M. A., Spinoglio, L. 1993, \apjs, 89, 1
\bibitem[Schmitt(1998)]{S98} Schmitt, H.~R.\ 1998, \apj,  506, 647
\bibitem[Simpson (2005)]{S05} Simpson C., 2005, \mnras, 360, 565
\bibitem[Sosa-Brito et al.(2001)]{2001ApJS..136...61S} Sosa-Brito, R.~M.,  Tacconi-Garman, L.~E., Lehnert, M.~D.,  \& Gallimore, J.~F.\ 2001, \apjs, 136, 61
\bibitem[Spinoglio \& Malkan(1989)]{sm89} Spinoglio, L. \& Malkan, M. A., 1989, \apj, 342, 83
\bibitem[Spinoglio \& Malkan(1992)]{sm92} Spinoglio, L. \& Malkan, M. A., 1992, \apj, 399, 504
\bibitem[Stern \& Laor (2012)]{SL12} Stearn, J.\& Laor, A., 2013, \mnras, 426, 2703
\bibitem[Stern \& Laor (2013)]{SL13} Stearn, J.\& Laor, A., 2013, \mnras, 431, 836
\bibitem[Storchi-Bergmann et al.(1995)]{1995ApJS...98..103S}  Storchi-Bergmann, T., Kinney, A.~L., \& Challis, P.\ 1995, \apjs, 98, 103
\bibitem[Theios et al. (2016)]{T16} Theos, R., Malkan, M. \& Ross, N.\ 2016, \apj, 822,  45
\bibitem[Thuan(1984)]{1984ApJ...281..126T} Thuan, T.~X.\ 1984, \apj, 281,  126
\bibitem[Thuan \& Sauvage(1992)]{ts92} Thuan, T. X. \& Sauvage, M. 1992, \aaps, 92, 749
\bibitem[Toba et al.(2014)]{toba14} Toba, Y., Oyabu, S., Matsuhara, H., Malkan, M. A., et al. 2014, \apj, 788, 45
\bibitem[Tommasin et al.(2010)]{tom10} Tommasin, S., Spinoglio, L., Malkan, M. A., Fazio, G.  2010, \apj, 709, 1257
\bibitem[Tsvetanov \& Iankulova(1989)]{ti89} Tsvetanov, Z. I. \& Iankulova, I. M., 1989, \mnras, 237, 707
\bibitem[Veilleux \& Osterbrock(1987)]{vo87} Veilleux, S. \& Osterbrock, D. E., 1987, \apjs, 63, 295
\bibitem[Veilleux et al.(1995)]{1995ApJS...98..171V} Veilleux, S., Kim,  D.-C., Sanders, D.~B., Mazzarella, J.~M.,  \& Soifer, B.~T.\ 1995, \apjs, 98, 171
\bibitem[Veilleux et al.(1999)]{1999ApJ...522..113V} Veilleux, S., Kim,  D.-C., \& Sanders, D.~B.\ 1999, \apj, 522, 113
\bibitem[Wang et al.(1996)]{1996ApJ...457..111W} Wang, T.-G., Zhou, Y.-Y.,  \& Gao, A.-S.\ 1996, \apj, 457, 111
\bibitem[Wang et al.(1998)]{1998ApJ...493....1W} Wang, T.-G., Lu, Y.-J.,  \& Zhou, Y.-Y.\ 1998, \apj, 493, 1 
\bibitem[Winkler(1992)]{1992MNRAS.257..677W} Winkler, H. 1992, \mnras,  257, 677
\bibitem[Xu et al. (1999)]{X99} Xu, C., Livio, M., Baum, S.,	 et al.  \apj, 118, 1169
\bibitem[Yee (1980)]{Y80} Yee, H. K. (1980), \apj, 241, 894
\bibitem[Young et al.(1996)]{1996MNRAS.281.1206Y} Young, S., Hough, J.~H.,  Efstathiou, A., Wills, B.~J., Bailey, J.~A., Ward, M.~J.,  \& Axon, D.~J.\ 1996, \mnras, 281, 1206
\bibitem[Zheng \& Malkan(1993)]{zm93} Zheng, W. \& Malkan, M. A., 1993, \apj, 415, 517
\end{thebibliography}
\end{document}